\documentclass[a4paper,11pt]{article}
\pdfoutput=1 

\usepackage{jheppub} 

\usepackage[T1]{fontenc} 
\usepackage{longtable}
\usepackage{amsmath}
\usepackage{graphicx,slashed,multirow,bbold,mathtools,sidecap,tikz,bm,enumitem,booktabs,array,makecell}
\usepackage[svgnames]{xcolor}
\usepackage{verbatim}
\usepackage[normalem]{ulem}
\usepackage{adjustbox}
\usepackage{tikz}
\usetikzlibrary{positioning,arrows}
\usetikzlibrary{decorations.pathmorphing}
\usetikzlibrary{decorations.markings}
\usepackage{subfigure}
\usepackage{physics}
\usepackage{multirow}
\usepackage{xspace}
\usepackage{subcaption}
\hypersetup{
	colorlinks=true,
	linkcolor=DarkCyan,
	filecolor=magenta,
	urlcolor=DarkCyan,
	citecolor=Maroon,
}

\usepackage[compat=1.1.0]{tikz-feynman}
\newcommand{\beq}{\begin{equation}}
\newcommand{\eeq}{\end{equation}}
\newcommand{\bea}{\begin{eqnarray}}
\newcommand{\eea}{\end{eqnarray}}
\newcommand{\barr}{\begin{array}}
\newcommand{\earr}{\end{array}}

\def \st1{\widetilde t_1}
\def \mst1{m_{\st1}}
\def \sbot1{\widetilde b_1}

\def\bea{\begin{eqnarray}}
	\def\eea{\end{eqnarray}}

\usepackage{color}
\usepackage{colortbl}

\long\def\/*#1*/{}
\usepackage{color}
 \usepackage[normalem]{ulem}
\definecolor{darkgreen}{cmyk}{1,0,1,0.4}
\definecolor{darkred}{cmyk}{0,1,1,0.4}

\newcommand{\ak}[1]{\textcolor{red}{#1}}
 
\definecolor{lime}{HTML}{A6CE39}
\DeclareRobustCommand{\orcidicon}{\hspace{-2.1mm}
\begin{tikzpicture}
\draw[lime, fill=lime] (0,0) circle [radius=0.13] node[white] {{\fontfamily{qag}\selectfont \tiny \,ID}}; \draw[white, fill=white] (-0.0525,0.095) circle [radius=0.007]; 
\end{tikzpicture} \hspace{-3.2mm} }
\foreach \x in {A, ..., Z}{\expandafter\xdef\csname orcid\x\endcsname{\noexpand\href{https://orcid.org/\csname orcidauthor\x\endcsname} {\noexpand\orcidicon}}}

\preprint{\shortstack{IOP-BBSR/2026-11\\IITH-PH-0002/26}}

\title{Light Leptoquarks in a Dark Sector: Scalar Dark Matter, Neutrino Mass, and Collider Signatures}

\author[a]{Priyotosh Bandyopadhyay,}
\author[b, c]{Kirtiman Ghosh,}
\author[d, e]{Anirban Karan,}
\author[a, b, c]{Snehashis Parashar,}
\author[b, c]{Debabrata Sahoo}
\affiliation[a]{\scriptsize
Indian Institute of Technology Hyderabad, Kandi, Sangareddy-502284, Telangana, India}
\affiliation[b]{\scriptsize Institute of Physics, Bhubaneswar, Sachivalaya Marg, Sainik School Post, Bhubaneswar 751005, India}
\affiliation[c]{\scriptsize Homi Bhabha National Institute, Training School Complex, Anushakti Nagar, Mumbai 400094, India}
\affiliation[d]{\scriptsize 
INFN, Gruppo Collegato di Cosenza \& Dipartimento di Fisica, Universit\`a della Calabria,
Arcavacata di Rende, I-87036, Cosenza, Italy}
\affiliation[e]{\scriptsize Departament de F\'isica Te\`orica, Instituto de F\'isica Corpuscular, Universitat de
Val\`encia – Consejo Superior de Investigaciones Cient\'ificas, Parc Cient\'ific,
Catedr\'atico Jos\'e Beltr\'an 2, E-46980 Paterna, Valencia, Spain}

 \emailAdd{bpriyo@phy.iith.ac.in}
\emailAdd{ kirti.gh@gmail.com} 
\emailAdd{kanirban@ific.uv.es}
\emailAdd{snehashis.p@iopb.res.in}
\emailAdd{debabrata.s@iopb.res.in}

\abstract{We investigate the dark matter (DM) phenomenology and subsequent collider signatures of a dark leptoquark (LQ) model containing dark vector-like quarks (VLQs) and a scalar singlet. The dark LQs and VLQs participate in the radiative generation of Majorana neutrino masses at one loop. Since the coloured LQs and VLQs cannot serve as viable DM candidates, a $\mathbb{Z}_2$-odd singlet scalar is introduced as the DM candidate. The $\mathbb{Z}_2$-odd nature of the LQs forbids their conventional decays into a quark and a lepton, allowing them to evade the standard LHC constraints that apply to visible LQ signatures. This framework therefore offers the distinctive possibility of sub-TeV dark LQs coexisting with TeV-scale dark VLQs. These sub-TeV dark LQs help in achieving the observed relic abundance of the singlet through co-annihilation. We analyse the resulting DM phenomenology and assess the prospects for probing this scenario at a future muon collider. }

\keywords{}

\begin{document} 

\tikzset{
  vector/.style={decorate, decoration={snake,amplitude=.4mm,segment length=2mm,post length=1mm}, draw},
  tes/.style={draw=black,postaction={decorate},
    decoration={snake,markings,mark=at position .55 with {\arrow[draw=black]{>}}}},
  provector/.style={decorate, decoration={snake,amplitude=2.5pt}, draw},
  antivector/.style={decorate, decoration={snake,amplitude=-2.5pt}, draw},
  fermion/.style={draw=black, postaction={decorate},decoration={markings,mark=at position .55 with {\arrow[draw=blue]{>}}}},
  fermionbar/.style={draw=black, postaction={decorate},
    decoration={markings,mark=at position .55 with {\arrow[draw=black]{<}}}},
  fermionnoarrow/.style={draw=black},
  scalar/.style={dashed,draw=black, postaction={decorate},decoration={markings,mark=at position .55 with {\arrow[draw=blue]{>}}}},
  scalarbar/.style={dashed,draw=black, postaction={decorate},decoration={marking,mark=at position .55 with {\arrow[draw=black]{<}}}},
  scalarnoarrow/.style={dashed,draw=black},
  electron/.style={draw=black, postaction={decorate},decoration={markings,mark=at position .55 with {\arrow[draw=black]{>}}}},
  bigvector/.style={decorate, decoration={snake,amplitude=4pt}, draw},
  particle/.style={thick,draw=blue, postaction={decorate},
    decoration={markings,mark=at position .5 with {\arrow[blue]{triangle 45}}}},
  gluon/.style={decorate, draw=black,
    decoration={coil,aspect=0.3,segment length=3pt,amplitude=3pt}}
}

\maketitle
\flushbottom

\section{Introduction} 

The Standard Model (SM) of particle physics has been remarkably successful in describing the known fundamental particles and their interactions, with predictions that have been validated to high precision across a wide range of experimental observations. The discovery of the Higgs boson at the Large Hadron Collider (LHC) marked the completion of the SM particle spectrum \cite{ATLAS:2012yve, CMS:2012qbp}. Despite its success, the SM is widely regarded as an effective theory valid up to a certain energy scale, due to its inability to account for several key observations in nature. The most compelling pieces of evidence for physics Beyond the Standard Model (BSM) are the observation of neutrino oscillations and the existence of dark matter. Neutrino oscillation experiments have conclusively demonstrated that neutrinos possess tiny but non-zero masses \cite{Super-Kamiokande:1998kpq, Esteban:2024eli}, directly contradicting the SM, which predicts massless neutrinos due to the absence of right-handed neutrino fields and lepton-number-violating interactions. In addition, overwhelming astrophysical and cosmological evidence points to the existence of dark matter (DM), including galactic rotation curves \cite{Rubin:1970zza, Zwicky:1933gu}, gravitational lensing \cite{Jee:2007nx}, and cosmic microwave background measurements \cite{Planck:2018vyg, ACT:2020gnv}, while the SM contains no viable DM candidate that is simultaneously stable, neutral, non-baryonic, and capable of accounting for the observed relic abundance.

It is therefore particularly appealing to search for BSM extensions in which neutrino masses and dark matter share a common origin. Radiative
neutrino mass models provide a natural setting for such a connection. The smallness of neutrino masses can be associated with loop suppression, while a discrete symmetry acting on the particles inside the loop can simultaneously stabilize the lightest state in the new sector. This idea is exemplified by
scotogenic constructions and their systematic generalizations
\cite{Tao:1996vb, Ma:2006km,Bonnet:2012kz,Restrepo:2013aga, Hirsch:2013ola, Cai:2017jrq, Avila:2019hhv, Karan:2023adm, Avila:2025qsc}. Since the
particles responsible for neutrino mass generation can then lie near the electroweak or TeV scale, these scenarios also offer direct experimental tests at colliders and DM experiments.

Leptoquarks (LQs) are especially interesting candidates for such new states. They are colour-charged bosons that mediate interactions between quarks and leptons and occur in a variety of ultraviolet frameworks, including quark-lepton unification, grand unified theories and composite models. Their simultaneous connection to the quark and lepton sectors gives rise to rich phenomenology involving flavour observables, neutrino masses and collider signatures \cite{Dorsner:2016wpm}. In particular, the scalar LQs $S_1$ and $\widetilde R_2$, respectively singlet and doublet under $SU(2)_L$, can mix after electroweak symmetry breaking (EWSB) and generate Majorana neutrino masses at one loop \cite{Babu:2019mfe,Zhang:2021dgl,Parashar:2022wrd, Dev:2024tto}, providing an attractive coloured realisation of radiative neutrino mass generation. However,  direct-search constraints on conventional LQs are severe. Scalar LQs are efficiently pair-produced through QCD interactions, and searches assuming prompt two-body decays into a quark and a charged lepton or neutrino typically exclude masses in the $1.5$--$1.8$ TeV range, depending on flavour assignments and branching ratios (BRs) \cite{ATLAS:2020dsk, ATLAS:2019qpq}. These limits rely strongly on the assumed decay topology, usually involving a high-$p_T$ jet and lepton, and can be substantially modified when the conventional $\ell q$ or $\nu q$ BRs are suppressed or when the LQ instead undergoes cascade decays involving invisible particles in a compressed spectrum. 

A simple way to realise such non-standard phenomenology is to place the LQs themselves in a dark sector. If an LQ is odd under an exact $\mathbb{Z}_2$ symmetry while all SM particles are even, interactions containing a single LQ and only SM fields are forbidden. Consequently, the LQ cannot undergo its usual two-body decay into a quark and a lepton, nor can it be singly produced through the corresponding Yukawa coupling. It must instead be pair-produced and eventually decay into the lightest $\mathbb{Z}_2$-odd particle, which cannot itself be coloured. Several connections between LQs and DM have previously been explored, including scenarios in which LQs provide a portal between the SM and scalar or fermionic DM, or acquire additional decay modes into a separate dark sector \cite{Queiroz:2014pra,Choi:2018stw,Mandal:2018czf,DEramo:2020sqv,Belanger:2021smw}. A model containing a $\mathbb{Z}_2$-odd scalar LQ, vector-like leptons and a real scalar has also been studied in connection with DM, the muon anomalous magnetic moment and muon collider searches \cite{Ghosh:2023xbj}.

In this work, we introduce a discrete $\mathbb{Z}_2$ symmetry and a minimal set of new fields that transform oddly under it:  two scalar LQs $S_1$ and $\widetilde R_2$, a vector-like quark (VLQ) doublet, a down-type VLQ singlet, and a real scalar singlet $\phi$. This is the most minimal setup within which the two scalar LQs and the two VLQs provide a
coloured realisation of the one-loop T1-2 neutrino mass topology \cite{Bonnet:2012kz}, while the real scalar supplies a neutral endpoint for the decays of the coloured odd states and becomes the stable DM candidate as the lightest $\mathbb{Z}_2$-odd particle. Thus, the same field content connects three phenomenological features: radiative neutrino mass generation, a scalar dark-matter candidate, and non-standard LQ decays that can weaken the usual direct-search limits. The observed DM relic-satisfying region, which currently for a pure inert singlet exists only near $\sim 60$ GeV DM mass or $\gtrsim 5$ TeV~\cite{Goncalves:2025snm} in the light of the recent LUX-ZEPLIN (LZ) direct detection limits~\cite{LZ:2024zvo}, gets significantly altered due to co-annihilation contributions from the new dark sector particles. They also affect the direct detection by supplementing the usual Higgs-portal annihilation of $\phi$ with tree-level VLQ exchange, and loop-induced scalar LQ and VLQ contributions. We specifically consider scenarios where at least one scalar LQ mass eigenstate remains closer in mass to the DM, so that the co-annihilation effects can be realised while simultaneously offering a sub-TeV LQ.  This leads to an LQ-DM system within somewhat of a compressed mass spectrum, where the lightest LQ decays to a soft lepton and quark pair with a $\phi$ via an off-shell VLQ. This invalidates the traditional LQ searches at the LHC \cite{CMS:2018ncu, CMS:2018lab, ATLAS:2021twp, ATLAS:2023uox, CMS:2024bnj}, but makes it susceptible to soft-lepton-based searches \cite{CMS:2018kag, ATLAS:2021twp}, or jets + missing transverse energy (MET) searches \cite{CMS:2019zmd}. The hints for the surviving sub-TeV LQs and their lighter DM partners can be obtained at a TeV-scale lepton collider, utilising visible-energy endpoint and cusp techniques that exploit the known centre-of-mass energy of the collision, a strategy adopted in multiple similar studies \cite{Martyn:2004jc, Han:2009ss, Han:2012nm, Christensen:2014yya}. For our work, we adopt this approach at the 10 TeV muon collider \cite{InternationalMuonCollider:2025sys} for a high-luminosity, high-energy perspective.

In the remainder of this article, we investigate the framework from complementary perspectives of neutrino mass, DM, and collider phenomenology. In \autoref{sec:model}, we motivate the new particle content, present the Lagrangian, and derive the relevant masses and mixing. The resulting Majorana neutrino mass matrix is obtained in \autoref{sec:nm}. In \autoref{sec:DM}, we perform a \texttt{micrOMEGAs}-based scan of the DM parameter space, focusing on scenarios where at least one LQ participates significantly in co-annihilation, and impose applicable collider limits using existing \texttt{CheckMATE} implementations. Then, \autoref{sec:fucol} discusses the future collider phenomenology, with particular emphasis on determining the masses and couplings of the lightest LQ-DM pair at a multi-TeV muon collider. Finally, we summarize our results in \autoref{sec:conc}.

\section{The Model} \label{sec:model}
We consider an extension of the SM with a discrete \( \mathbb{Z}_2 \) symmetry, under which all SM fields are even. The new physics sector, required for the neutrino mass and stable WIMP motivations, minimally requires: two scalar LQ fields, two VLQ fields, and a scalar singlet field, all odd under the \( \mathbb{Z}_2 \) symmetry. The new particle content and their transformation properties under the SM gauge group and the discrete symmetry are summarized in \autoref{tab:particles}. The full symmetry group of the model is
\[
\mathcal{G} = SU(3)_C \times SU(2)_L \times U(1)_Y \times \mathbb{Z}_2.
\]

\begin{table}[h!]
\centering
\begin{tabular}{ccc}
\toprule
Field &  Type & Quantum Numbers under \\
 &   &  \textbf{ \( SU(3)_C \times SU(2)_L \times U(1)_Y \times \mathbb{Z}_2 \)} \\
\midrule
\( \widetilde{R}_2 = (\widetilde{R}_2^{2/3}, \widetilde{R}_2^{-1/3})^T \) & Scalar leptoquark & \( (3,2,+\tfrac{1}{6}, -1) \) \\
\( S_1 \) & Scalar leptoquark & \( (3,1,-\tfrac{1}{3}, -1) \) \\
\( \Psi^q_{L,R} = (\psi^U, \psi^D)_{L,R}^T \) & Vector-like quark & \( (3,2,+\tfrac{1}{6}, -1) \) \\
\( \psi^d_{L,R} \) & Vector-like quark & \( (3,1,-\tfrac{1}{3}, -1) \) \\

\( \phi \) & Scalar WIMP DM & \( (1,1,0,-1) \) \\
\hline
\end{tabular}
\caption{BSM field content and their assigned quantum numbers under the SM gauge group and the discrete \( \mathbb{Z}_2 \) symmetry.}
\label{tab:particles}
\end{table}

Let us summarise the key motivations for introduction of the above fields and the discrete symmetry in our model:

\begin{enumerate}
    \item \textbf{\(\mathbb{Z}_2\)-odd scalar leptoquarks:} The scalar LQs \( \widetilde{R}_2 \) and \( S_1 \), being odd under the discrete symmetry \(\mathbb{Z}_2\), can potentially evade the stringent LHC bounds that typically apply to their \(\mathbb{Z}_2\)-even versions, dropping the $\gtrsim 1.5$ TeV mass bounds on the latter to a few hundred GeV. The singlet-doublet combination is also the most minimal setup for radiative neutrino mass generation with scalar LQs.

    \item \textbf{\(\mathbb{Z}_2\)-odd vector-like quarks:} 
    The radiative neutrino mass generation mechanism that we follow necessitates coloured $\mathbb{Z}_2$-odd fermions in the loop. To this end, we introduce the VLQs, which are naturally anomaly-free. We note that a doublet and a singlet down-type VLQ are the most minimal setup that fulfil the neutrino mass requirement, and their achiral nature allows a down-type singlet to exist without the requirement for an up-type singlet.

    \item \textbf{\(\mathbb{Z}_2\)-odd scalar singlet:} To ensure that all coloured \(\mathbb{Z}_2\)-odd particles decay and do not leave behind stable coloured relics, a \(\mathbb{Z}_2\)-odd scalar singlet \( \phi \) is introduced for the coloured particles to decay into, which becomes our stable WIMP DM. The standalone inert singlet DM is well studied in the literature and faces stringent bounds from direct detection experiments, which may be alleviated by the presence of the additional \(\mathbb{Z}_2\)-odd particles and interactions in our model.
\end{enumerate}

Unlike Ref. \cite{Dey:2025niu}, where neutral components of dark vector-like leptons (VLLs) provide fermionic dark matter alongside dark LQs, our model features dark VLQs. Because both our dark LQs and VLQs are coloured, colour confinement precludes them from being dark matter, requiring a scalar dark matter candidate $\phi$. Crucially, this setup enables a distinct signature of sub-TeV dark LQs (operating alongside TeV-scale VLQs), which is difficult to achieve in the model with dark VLL. Given the three different kinds of fields introduced, in the next subsections we will unpack their interactions, and the relevant mixing and mass eigenstates that they lead to.

\subsection{Scalar Potential}\label{sec:scalpot}

We start with the scalar potential involving the SM Higgs field $H$ and the $\mathbb{Z}_2$-odd scalar fields, which is consistent with the SM gauge symmetry and the discrete \( \mathbb{Z}_2 \) symmetry:
\begin{align}
V(H, \phi, \widetilde{R}_2, S_1) = 
&\ \mu_H^2\, H^\dagger H + \lambda_H (H^\dagger H)^2 
   + \mu_\phi^2\, \phi^2 + \lambda_\phi\, \phi^4 \nonumber \\
&\ + m_{\widetilde{R}}^2\, \widetilde{R}_2^{ \dagger} \widetilde{R}_2 
   + \lambda_{\widetilde{R}} (\widetilde{R}_2^{ \dagger} \widetilde{R}_2)^2 
   + m_{S_1}^2\, S_1^{ *} S_1 + \lambda_{S_1} (S_1^{ *} S_1)^2 \nonumber \\
&\ + \frac{1}{2}\lambda_{H\phi} (H^\dagger H)\,\phi^2
   + \lambda_{H\widetilde{R}} (H^\dagger H)(\widetilde{R}_2^{ \dagger} \widetilde{R}_2) 
   + \lambda_{H S_1} (H^\dagger H)(S_1^{ *} S_1) \nonumber \\
&\ + \lambda_{\phi \widetilde{R}}\, \phi^2\,(\widetilde{R}_2^{ \dagger} \widetilde{R}_2) 
   + \lambda_{\phi S_1} \,\phi^2\,(S_1^{ *} S_1) 
   + \lambda_{S_1 \widetilde{R}} (S_1^{ *} S_1)(\widetilde{R}_2^{ \dagger} \widetilde{R}_2) \nonumber \\
&\ + \widetilde\lambda_{H\widetilde{R}} (H^{\dagger} \widetilde{R}_2)(\widetilde{R}_2^{\dagger} H) + \left[ \kappa\, \widetilde{R}_2^{ \dagger} H\, S_1 + \text{h.c.} \right]
\label{eq:scalar}
\end{align}

The condition \( \mu_H^2 < 0 \) triggers electroweak symmetry breaking via the Higgs vacuum expectation value (VEV), while all other scalars are \( \mathbb{Z}_2 \)-odd and therefore do not acquire VEVs. Furthermore, the trilinear term \( \kappa\, \widetilde{R}_2^{ \dagger} H\, S_1 \) containing the coupling $\kappa$ with mass dimension of one, allowed by both gauge and \( \mathbb{Z}_2 \) symmetries, introduces a nontrivial mixing between the scalar LQs. This term is a crucial ingredient for introducing lepton number violation necessary to generate the Majorana masses for the neutrinos at one loop, with its simultaneous existence alongside the LQ-VLQ Yukawa couplings, which we will see later in \autoref{sec:nm}.

The scalar potential is required to satisfy the usual bounded-from-below (BFB),
perturbativity and perturbative unitarity conditions
\cite{Bandyopadhyay:2016oif,Bandyopadhyay:2021kue,Gonderinger:2009jp}.
For the phenomenological analysis that will follow, all the new scalar quartic couplings are taken
positive and fixed to $\lambda_i=0.01$, which makes the quartic potential manifestly
bounded from below and keeps the couplings safely within the perturbative regime $\abs{\lambda_i} \leq 4\pi$.
We also restrict ourselves to the $\mathbb{Z}_2$- and colour-preserving electroweak vacuum,
$\langle\phi\rangle=\langle\widetilde R_2\rangle=\langle S_1\rangle=0$, with positive
physical scalar mass-squared eigenvalues, as realised by the spectra considered below.

\subsubsection{Masses and mixing of \texorpdfstring{$\mathbb{Z}_2$}{Z2}-Odd Scalars}

After electroweak symmetry breaking, the Higgs doublet acquires a vacuum expectation value,
\begin{align}
H = 
\frac{1}{\sqrt 2}\begin{pmatrix}
\sqrt 2 G^+ \\
v + h + iG^0
\end{pmatrix}, \quad v \approx 246\, \text{GeV},
\end{align}
with $h$ being the physical Higgs boson. The fields $G^\pm$ and $G^0$ are the Goldstone bosons giving masses to $W^\pm$ and $Z$ bosons. The mass of the scalar singlet $\phi$ is given by
\begin{align}
M_\phi^2 = 2\mu_\phi^2 + \frac{1}{2}\lambda_{H\phi} v^2.
\end{align}

The trilinear \( \kappa\, \widetilde{R}_2^{ \dagger} H\, S_1 \) term in \autoref{eq:scalar} facilitates mixing between the component $\widetilde{R}_2^{-1/3}$ and $S_1$, leading to the following mass-squared matrix in the basis $\Phi = (\widetilde{R}_2^{-1/3},\, S_1)^T$:
\begin{equation}
\mathcal{M}^2 = 
\begin{pmatrix}
M_{R}^2  & \delta M^2 \\
\delta M^2 & M_S^2 
\end{pmatrix}, 
\end{equation}
with $ M_R^2 = m_{\widetilde{R}}^2+ \frac{1}{2}\lambda_{H\widetilde{R}}v^2 +\frac{1}{2}\widetilde\lambda_{H\widetilde{R}}v^2$, $M_S^2 = m_{S_1}^2 + \frac{1}{2}\lambda_{H S_1} v^2$, and $\delta M^2 = \frac{\kappa v}{\sqrt{2}}$.

The corresponding mass eigenstates are then given by
\begin{align}
\eta_1 &= \cos\theta_s\, R_2^{-1/3} - \sin\theta_s\, S_1, \\
\eta_2 &= \sin\theta_s\, R_2^{-1/3} + \cos\theta_s\, S_1,
\end{align}
where the mixing angle $\theta_s$ is defined as
\begin{equation}
\tan 2\theta_s = \frac{2 \delta M^2}{M_S^2 - M_R^2}.
\end{equation}
The other component, $\widetilde{ R}_2^{2/3}$, remains unmixed and forms a physical state with mass:
\begin{eqnarray}
    M_{\widetilde{R}_2^{2/3}}^2 = m_{\widetilde{R}}^2 + \frac{1}{2}\lambda_{H\widetilde{R}}v^2.
\end{eqnarray}

\subsection{Yukawa Interactions}

The Yukawa interactions involving the $\mathbb{Z}_2$-odd scalars and fermions, relevant for neutrino mass generation and the decay of the exotic fermions, are given by:
\begin{align}
\mathcal{L}_\text{Yuk} \supset 
 &\ - Y_{\ell R}^{ij}\, \overline{L}_i\, \widetilde{R}_2^{\,C} \, \psi^d_{Rj} 
    - Y_{\ell S}^{ij}\, \overline{\Psi}^{\,q}_{Li} \, {S_1} \, L_{j}^C   \nonumber \\
&\ - Y_{\phi Q}^{ij}\, \overline{\Psi}^{\,q}_{Ri}\, \phi\, Q_{Lj}   
   - Y_{\phi D}^{ij}\, \overline{\psi}^{\,d}_{Li}\, \phi\, d_{Rj}
   + \text{h.c.}
   \label{eq:yukawa}
\end{align}

Here, the SM lepton and quark doublets are denoted by \( L_i = (\nu_i, \ell_i)_L \), \( L_i^C = (-\,\ell_{iL}^C, \nu_{iL}^C) \) and \( Q_i = (u_i, d_i)_L \), respectively. The SM singlet fermions are represented by \( \ell_{Ri}, u_{Ri}, d_{Ri} \). 
 The field \( \widetilde{R}_2^{\prime\,C} \) is defined as  \( \widetilde{R}_2^{\,C} = i\sigma_2 \widetilde R_2^{\,*} \) , where \( \sigma_2 \) is the second Pauli matrix. The couplings \( Y_{AB}^{ij} \) are Yukawa couplings, with \( i \) and \( j \) denoting generation indices. The simultaneous presence of the trilinear term \( \kappa\, \widetilde{R}_2^{ \dagger} H\, S_1 \) in the scalar potential (\autoref{eq:scalar}) and the LQ Yukawa interactions in \autoref{eq:yukawa} allows the lepton number violation necessary for Majorana neutrino mass generation at the one-loop level.\footnote{We do not include baryon number violating terms such as $\overline{(Q_L)^C} \,\psi_L\, \tilde R_2$ to avoid proton decay.}

The exotic fermions, being vector-like, allow for the introduction of Dirac mass terms as well as Yukawa interactions involving the SM Higgs doublet $H$. These Yukawa interactions generate mixings among the $\mathbb{Z}_2$-odd VLQs. The relevant terms in the Lagrangian that contribute to the mixing are:
\begin{align}
\mathcal{L}_\text{mass}^{\rm VLQ} \supset 
&\ - M_Q^{ij}\, \overline{\Psi^q_{Li}}\, \Psi^q_{Rj} 
   - M_D^{ij}\, \overline{\psi^d_{Li}}\, \psi^d_{Rj} 
   - y_H^{ij}\, \overline{\Psi^q_{Ri}}\, H\, \psi^d_{Lj} 
   - \tilde{y}_H^{ij}\, \overline{\Psi^q_{Li}}\, H\, \psi^d_{Rj} 
   + \text{h.c.}
   \label{eq:mass}
\end{align}
\ak{}
The vector-like mass terms involve the \(3\times 3\) complex matrices \(M_Q\) and \(M_D\), while the Yukawa interaction terms involve the \(3\times 3\) complex matrices \(y_H\) and \(\tilde{y}_H\), the most general form of which carry 9 magnitudes and 9 phases. Field redefinitions can be used to remove unphysical parameters: unitary rotations of \(\Psi^q_{L,R}\) and \(\psi^d_{L,R}\) allow \(M_Q\) and \(M_D\) to be diagonalized, while further rephasings can make their diagonal elements real and positive, thereby removing \(3\) phases from each mass matrix. After this basis choice, the remaining physical parameters reside in the Yukawa matrices. However, only three additional phases in total can be absorbed by rephasing the fermion fields simultaneously in the mass eigenstate basis; these rephasings act on both \(y_H\) and \(\tilde{y}_H\), meaning that eliminating phases from one Yukawa matrix will generally reintroduce them into the other.

After electroweak symmetry breaking, the Yukawa interactions generate mixing between the down-type component $\psi^D_i$ (contained in $\Psi^q_i$) and the singlet fermion $\psi^d_j$. The up-type partners $\psi^U_i$ remain unmixed and form Dirac fermions with mass terms determined by $M_Q^{ij}$ alone. Focusing on the charge $-1/3$ fermions, the mass terms can be expressed in the 6-dimensional basis:
\[
\left( \psi^D_{Li},\ \psi^d_{Li} \right)^T \quad \text{and} \quad \left( \psi^D_{Ri},\ \psi^d_{Ri} \right)^T, \quad (i = 1,2,3),
\]
as:
\begin{align}
\mathcal{L}_\text{mass}^{(-1/3)} = 
- \begin{pmatrix}
\overline{\psi^D_{Li}} & \overline{\psi^d_{Li}}
\end{pmatrix}
\begin{pmatrix}
M_Q^{ij} & \tilde{y}_H^{ij} \tfrac{v}{\sqrt{2}} \\
y_H^{ij*} \tfrac{v}{\sqrt{2}} & M_D^{ij}
\end{pmatrix}
\begin{pmatrix}
\psi^D_{Rj} \\
\psi^d_{Rj}
\end{pmatrix}
+ \text{h.c.}
\end{align}

This results in six mass eigenstates after diagonalizing the full $6\times6$ mass matrix. The physical fermions are admixtures of $\psi^D_i$ and $\psi^d_j$, with mixing controlled by the Yukawa matrices $y_H$ and $\tilde{y}_H$, and the vector-like mass matrices $M_Q$ and $M_D$.

\subsubsection{VLQ mass eigenstates and mixing considering one generation} \label{sec:VLQmix}

For the purpose of this article, we consider a simplified scenario with only one generation of \( \mathbb{Z}_2 \)-odd VLQs, where the relevant fermionic fields are:
\begin{itemize}
    \item A vector-like $SU(2)_L$ doublet: \( \Psi^q = (\psi^U, \psi^D) \),
    \item A vector-like $SU(2)_L$ singlet: \( \psi^d \).
\end{itemize}

The mass and mixing terms after electroweak symmetry breaking are given by:
\begin{align}
\mathcal{L}_{\text{mass}}^{\rm VLQ} = 
&\ - M_Q\, \overline{\psi^U_L} \psi^U_R - M_Q\, \overline{\psi^D_L} \psi^D_R 
   - M_D\, \overline{\psi^d_L} \psi^d_R 
   - \tilde{y}_H \frac{v}{\sqrt{2}}\, \overline{\psi^D_L} \psi^d_R 
   - y_H \frac{v}{\sqrt{2}}\, \overline{\psi^D_R} \psi^d_L + \text{h.c.}
\end{align}
which leads to the following mass matrix in the basis \( (\psi^D, \psi^d) \):
\begin{align}
\mathcal{L}_{\text{mass}}^{(-1/3)} = 
- \begin{pmatrix} \overline{\psi^D_L} & \overline{\psi^d_L} \end{pmatrix}
\begin{pmatrix}
M_Q & \tilde{m} \\
m & M_D
\end{pmatrix}
\begin{pmatrix}
\psi^D_R \\
\psi^d_R
\end{pmatrix} + \text{h.c.} 
\end{align}
where we define:
\[
m = y_H^* \frac{v}{\sqrt{2}}, \qquad \tilde{m} = \tilde{y}_H \frac{v}{\sqrt{2}}.
\]

To diagonalize the mass matrix, we perform a bi-unitary transformation:
\begin{align}
U_L^\dagger\, \mathcal{M}\, U_R = \begin{pmatrix}
M_1 & 0 \\
0 & M_2
\end{pmatrix},
\end{align}
where \( M_{1,2} \) are the physical masses. The Hermitian matrix \(\mathcal{M}\,\mathcal{M}^\dagger\) takes the form  
\begin{align}
\mathcal{M}\,\mathcal{M}^\dagger = 
\begin{pmatrix}
M_Q^2 + |\tilde{m}|^2 & M_Q\, m^* + \tilde{m}\, M_D \\
M_Q\, m + \tilde{m}^*\, M_D & |m|^2 + M_D^2
\end{pmatrix},
\end{align}
which is diagonalized by a unitary matrix \(U_L\).  
Similarly, the matrix \(\mathcal{M}^\dagger \mathcal{M}\) is diagonalized by a unitary matrix \(U_R\). The $2\times2$ matrices \(U_L\) and \(U_R\) are  parametrised in terms of mixing angles \(\theta_L\) and \(\theta_R\) (neglecting possible phase factors) as  
\begin{align}
U_{L,R} &=
\begin{pmatrix}
\cos\theta_{L,R} & \sin\theta_{L,R} \\
-\sin\theta_{L,R} & \cos\theta_{L,R}
\end{pmatrix}. 
\end{align}
The mixing angles \( \theta_L \) and \( \theta_R \) for the left-handed and right-handed fermions, respectively, are given by:
\begin{align}
\tan 2\theta_L = \frac{2 |M_Q m + \tilde{m}^* M_D|}{M_Q^2 + |\tilde{m}|^2 - M_D^2 - |m|^2}, \quad \tan 2\theta_R = \frac{2 |M_Q \tilde{m} + m^* M_D|}{M_Q^2 + |m|^2 - M_D^2 - |\tilde{m}|^2}.
\end{align}

After mixing, the two physical VLQ states with charge $-1/3$ are denoted as \( \chi_{1} \) and \( \chi_{2} \), whose left- and right-handed components are related to the gauge basis as: 
\begin{align}
\begin{pmatrix}
\chi_{1L} \\[2pt]
\chi_{2L}
\end{pmatrix}
&= U_L
\begin{pmatrix}
\psi^D_L \\[2pt]
\psi^d_L
\end{pmatrix}, 
&
\begin{pmatrix}
\chi_{1R} \\[2pt]
\chi_{2R}
\end{pmatrix}
&= U_R
\begin{pmatrix}
\psi^D_R \\[2pt]
\psi^d_R
\end{pmatrix},
\end{align}
The up-type component \( \psi^U \) remains unmixed, as a Dirac fermion with mass \( M_Q \). Having established the interactions, masses and mixings of the exotic particles introduced in the model, we now address the neutrino mass generation mechanism in the next section.

\section{Radiative Neutrino Mass Generation} \label{sec:nm}

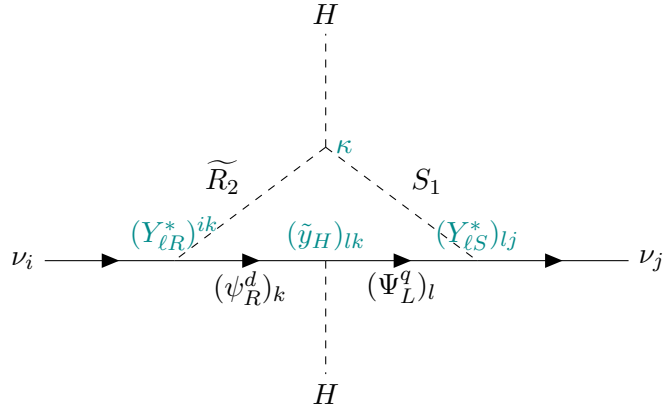
\begin{figure}[h!]
    \centering
\begin{tikzpicture}
  \begin{feynman}
    \vertex (a) {\(\nu_i\)};
    \vertex [right=2cm of a] (b)[label=above:\textcolor{DarkCyan}{\((Y^*_{\ell R})^{ik}\)}];
    \vertex[right=2cm of b] (c)[label=above :\textcolor{DarkCyan}{\((\tilde{y}_H)_{lk}\)}];
    \vertex[right=2cm of c] (d)[label=above :\textcolor{DarkCyan}{\(({Y}^*_{\ell S})_{lj}\)}];
    \vertex[right=2cm of d] (e){\(\nu_j\)};
    \vertex [above=1.5cm of c] (f) [label=right :\textcolor{DarkCyan}{\(\kappa\)}];
    \vertex[above=1.5cm of f] (g){\(H\)};
    \vertex[below=1.5cm of c] (h){\(H\)};

    \diagram* {
      (a) -- [fermion] (b),
      (b) -- [fermion, edge label'=\((\psi^d_R)_k\)] (c),
      (c) -- [fermion, edge label'=\((\Psi^q_L)_l\)] (d),
      (d) -- [fermion] (e),
      (c)--[scalar] (h),
      (f)--[scalar, edge label'=\(\widetilde{R_2}\)](b),
      (f)--[scalar, edge label=\(S_1\)](d),
      (f)--[scalar](g),
    };
  \end{feynman}
\end{tikzpicture}
\caption{Feynman diagram for neutrino mass generation at one-loop as in the T1-2 topology. In the loop, we have VLQs $\Psi^q, \psi^d$ and scalar LQs $\tilde{R_2}, S_1$.} 
    \label{fig:neutrino_mass}
\end{figure}

In this model, neutrino masses are generated radiatively at the one-loop level, following the T1-2 topology classification of Ref.~\cite{Bonnet:2012kz}. The imposed $\mathbb{Z}_2$ symmetry forbids any tree-level neutrino mass. The loop involves $\mathbb{Z}_2$-odd VLQs  ($\Psi^q$, $\psi^d$), and the scalar LQs ($\widetilde{R}_2$, $S_1$). As stated before in \autoref{sec:scalpot}, the presence of the crucial $\kappa-$ term in the Lagrangian of \autoref{eq:scalar}, alongside the Yukawa couplings $ Y_{\ell R}$ and $Y_{\ell S}$, provide the necessary lepton number violation and the LQ mixing, allowing Majorana neutrino mass generation in this topology.
The Feynman diagram responsible for generating the neutrino mass is shown in  \autoref{fig:neutrino_mass}. The resulting contribution to the neutrino mass matrix is:
\begin{equation}
\begin{aligned}
(m_\nu)_{ij} &= \kappa\, v^2 
\sum_{k,\ell} \Big[ 
\big(Y_{\ell S}^{\,il}\big)^*\, M_Q^{l}\, \tilde{y}_H^{\,l k}\, M_D^{k}\, \big(Y_{\ell R}^{\,k j}\big)^* 
+ \big(Y_{\ell R}^{\,i k}\big)^*\, M_D^{k}\, (\tilde{y}_H^{T})^{\,kl}\, M_Q^{l}\, \big(Y_{\ell S}^{\,l j}\big)^*
\Big] \\
&\hspace{2cm} \times I_4\!\left[\left( M_Q^l \right )^ 2,\, \left( M_D^k\right)^2,\, M_{S}^{2},\, M_{R}^{2} \right] \\[4pt]
&\quad + \kappa\, v^2 
\sum_{k,\ell} \Big[
\big(Y_{\ell S}^{\,il}\big)^*\, (y_H)^{\,l k}\, \big(Y_{\ell R}^{\,k j}\big)^* 
+ \big(Y_{\ell R}^{\,i k}\big)^*\, (y_H^{T})^{\,kl}\, \big(Y_{\ell S}^{\,lj}\big)^*
\Big] \\
&\hspace{2cm} \times J_4\!\left[\left( M_Q^l \right )^ 2,\, \left( M_D^k\right)^2,\, M_{S_1}^{2},\, M_{\widetilde{R}_2}^{2} \right] .
\end{aligned}
\label{eq:numass}
\end{equation}
Here, $I_4$ and $J_4$ are the standard four-propagator loop functions, 
with arguments corresponding to the squared masses of the internal states. \cite{Bonnet:2012kz}

\begin{align}
I_4(M_A^2, M_B^2, M_C^2, M_D^2) 
&\equiv 
\int \frac{d^d k}{(2\pi)^d} \, i \,
\frac{1}{(k^2 - M_A^2)(k^2 - M_B^2)(k^2 - M_C^2)(k^2 - M_D^2)}
\nonumber\\
&= -\frac{1}{(4\pi)^2} \Bigg[
\frac{M_A^2 \, \ln\!\left(\frac{M_D^2}{M_A^2}\right)}
{(M_A^2 - M_B^2)(M_A^2 - M_C^2)(M_A^2 - M_D^2)}
\nonumber\\
&\quad+
\frac{M_B^2 \, \ln\!\left(\frac{M_D^2}{M_B^2}\right)}
{(M_B^2 - M_A^2)(M_B^2 - M_C^2)(M_B^2 - M_D^2)}
\nonumber\\
&\quad+
\frac{M_C^2 \, \ln\!\left(\frac{M_D^2}{M_C^2}\right)}
{(M_C^2 - M_A^2)(M_C^2 - M_B^2)(M_C^2 - M_D^2)}
\Bigg] .
\end{align}

\begin{align}
J_4(M_A^2, M_B^2, M_C^2, M_D^2) 
&\equiv 
\int \frac{d^d k}{(2\pi)^d} \, i \,
\frac{k^2}{(k^2 - M_A^2)(k^2 - M_B^2)(k^2 - M_C^2)(k^2 - M_D^2)}
\nonumber\\
&= -\frac{1}{(4\pi)^2} \Bigg[
\frac{M_A^4 \, \ln\!\left(\frac{M_D^2}{M_A^2}\right)}
{(M_A^2 - M_B^2)(M_A^2 - M_C^2)(M_A^2 - M_D^2)}
\nonumber\\
&\quad+
\frac{M_B^4 \, \ln\!\left(\frac{M_D^2}{M_B^2}\right)}
{(M_B^2 - M_A^2)(M_B^2 - M_C^2)(M_B^2 - M_D^2)}
\nonumber\\
&\quad+
\frac{M_C^4 \, \ln\!\left(\frac{M_D^2}{M_C^2}\right)}
{(M_C^2 - M_A^2)(M_C^2 - M_B^2)(M_C^2 - M_D^2)}
\Bigg] .
\end{align}
For TeV-scale internal masses \(M_A~\sim~M_B~\sim~M_C~\sim~M_D \sim~M\) one has the scalings: $J_4 \sim I_4 M^2 \sim \frac{1}{16\pi^2M^2}$. Using these scalings and taking the heavy masses in the loop to be of order a common scale $M$, the two structures in the exact expression in \autoref{eq:numass} scale similarly. Keeping only the leading $\frac{1}{M^2}$ term (and suppressing flavour indices) one finds
\begin{equation}
    (m_\nu)_{ij}\;\sim\;
\kappa\,v^2\;\frac{1}{16\pi^2}\frac{1}{M^2}\;
\Big(
Y_{\ell S}\,\tilde y_H\,Y_{\ell R}
\;+\;
Y_{\ell S}\,y_H\,Y_{\ell R}
\Big)_{ij}. \label{eq:numass_simp}
\end{equation}

In the case with a regular $\mathbb{Z}_2$-even $\widetilde{R}_2 + S_1$ scalar LQ setup \cite{Parashar:2022wrd, Zhang:2021dgl, Babu:2019mfe}, larger $\kappa$ and $\mathcal{}O(1)$ Yukawa couplings can be simultaneously accommodated to generate sub-eV neutrino masses \cite{Elbers:2025vlz, KATRIN:2024cdt} with TeV-scale new physics. However, in our $\mathbb{Z}_2$-odd setup, we find from \autoref{eq:numass_simp} that sub-eV neutrino masses require a more suppressed $\kappa$ coupling, of order $\mathcal{O}(10^{-3})$ GeV, for TeV-scale new physics and $\mathcal{O}(10^{-2} - 1)$ Yukawa couplings, as heavier $M_{D/Q}$ values enter the numerator in \autoref{eq:numass}.
This occurs because the down-type quarks with tiny Yukawa couplings participating in the scotogenic loop of the $\mathbb{Z}_2$-even LQ scenario~\cite{Parashar:2022wrd, Zhang:2021dgl, Babu:2019mfe} are replaced here by VLQs with larger Yukawa couplings ($y_H$ and $\tilde y_H$). Focusing on the subsequent phenomenological implications in the DM sector, we refrain from a detailed analysis of this parameter space in the light of neutrino mass and their mass-squared differences, neutrino mixing angles, and CP-violating phases. The DM phenomenology is primarily sensitive to the overall magnitudes of these couplings, while their flavour structure can be adjusted to reproduce the observed neutrino masses and mixing parameters. With that established, let us now move to the peculiar DM phenomenology that our setup brings forth. 

At the one-loop level, the present framework induces corrections to the muon anomalous magnetic moment, i.e. $(g-2)_\mu$, as well as lepton flavour violating (LFV) transitions such as $\mu \to e \gamma$ and $\mu\text{--}e$ conversion. These contributions proceed via similar diagrams to those discussed in refs.~\cite{Parashar:2022wrd,Zhang:2021dgl}, with the standard LQs and quarks inside the loops replaced by their $\mathbb{Z}_2$-odd dark LQ and VLQ counterparts; the resulting amplitudes depend on the Yukawa couplings $Y_{\ell S}$, $Y_{\ell R}$ and the respective BSM masses. Concurrently, the model impacts the $Z \to b\bar b$ vertex and quark flavour violating processes—including $b\to s\gamma$, $B_s \to \mu^+\mu^-$, and $B_{(s)}-\bar B_{(s)}$ mixing—at one-loop governed by the $Y_{\phi Q}$ and $Y_{\phi D}$ couplings alongside the VLQ and dark scalar $\phi$ mass scales. Furthermore, atomic parity violation constraints are modified at one loop via four-fermion box diagrams and oblique corrections ($S$ and $T$ parameters), the latter of which are sensitive to the custodial symmetry breaking mass splittings within the BSM doublets. While these tight flavour and precision electroweak bounds can be successfully circumvented by a combination of a decoupled BSM mass spectrum, small Yukawa couplings, and nearly degenerate doublet components, an exhaustive numerical scanning of this parameter space is deferred to future work. In this paper, we restrict our focus to the dark matter phenomenology and collider signatures of the model.

\section{Dark Matter Phenomenology} \label{sec:DM}

The decay of the real singlet scalar, being the lightest particle odd under the \( \mathbb{Z}_2 \) symmetry, is forbidden, and hence the particle remains stable in the theory. Since it is a neutral gauge singlet, it can only interact with the SM particles through the Yukawa couplings given in \autoref{eq:yukawa} or through the scalar potential couplings in \autoref{eq:scalar}. Consequently, the singlet scalar \(\phi\) is a viable dark matter candidate, provided it satisfies the relic density constraints and the bounds from dark matter direct detection experiments, primarily LUX-ZEPLIN \cite{LZ:2024zvo}. A detailed discussion of the direct detection cross section and the contribution to the relic density is presented in the following section.

\subsection{Direct Detection of the Scalar Dark Matter $\phi$} \label{sec:dd-1}

\begin{figure}[h]
    \centering
    \subfigure[]{\includegraphics[width=0.3\linewidth]{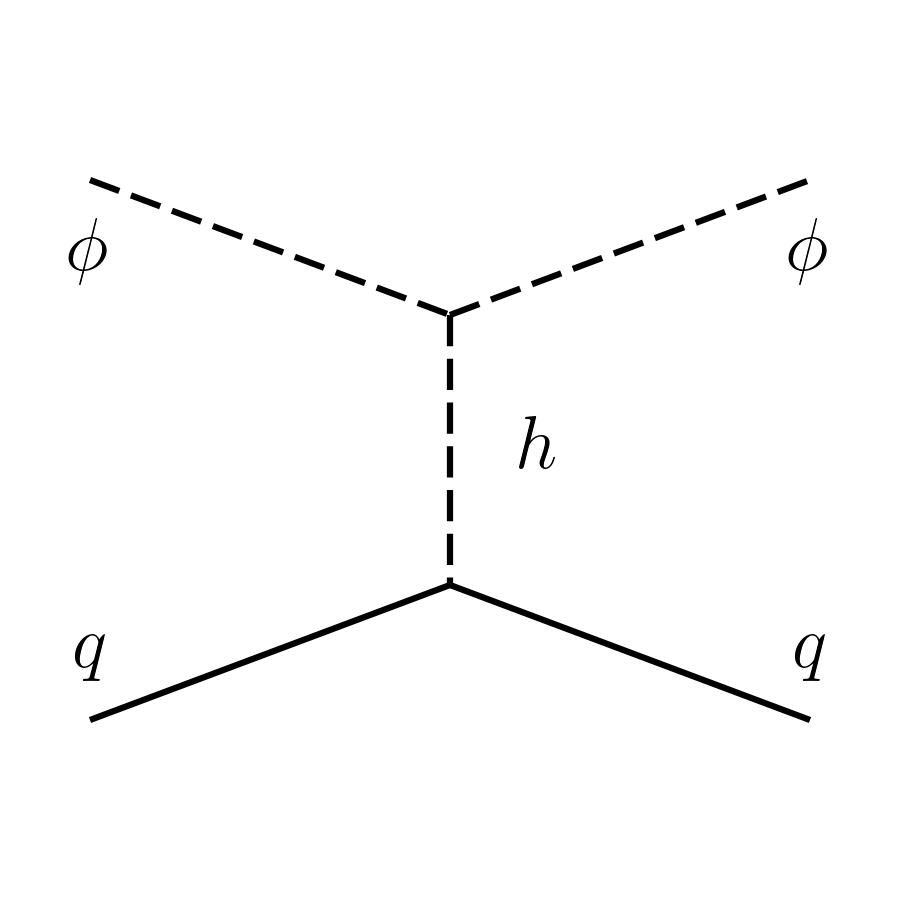}}
    \hspace{2em}
    \subfigure[]{\includegraphics[width=0.3\linewidth]{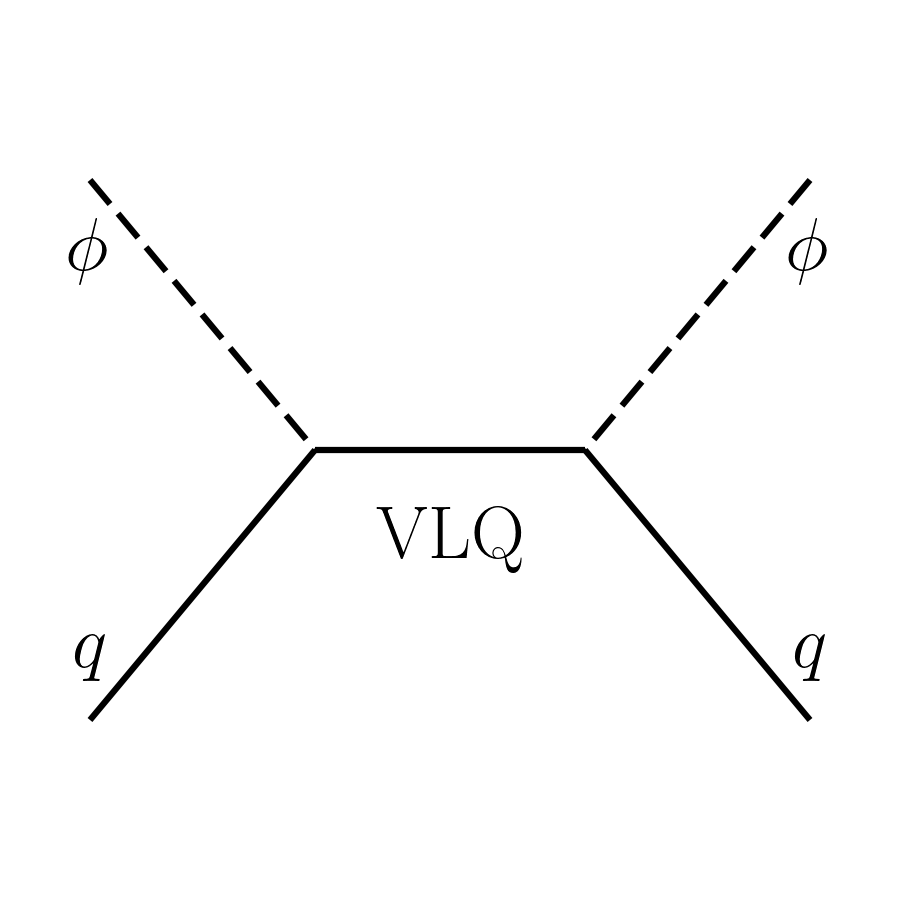}}

    \subfigure[]{\includegraphics[width=0.3\linewidth]{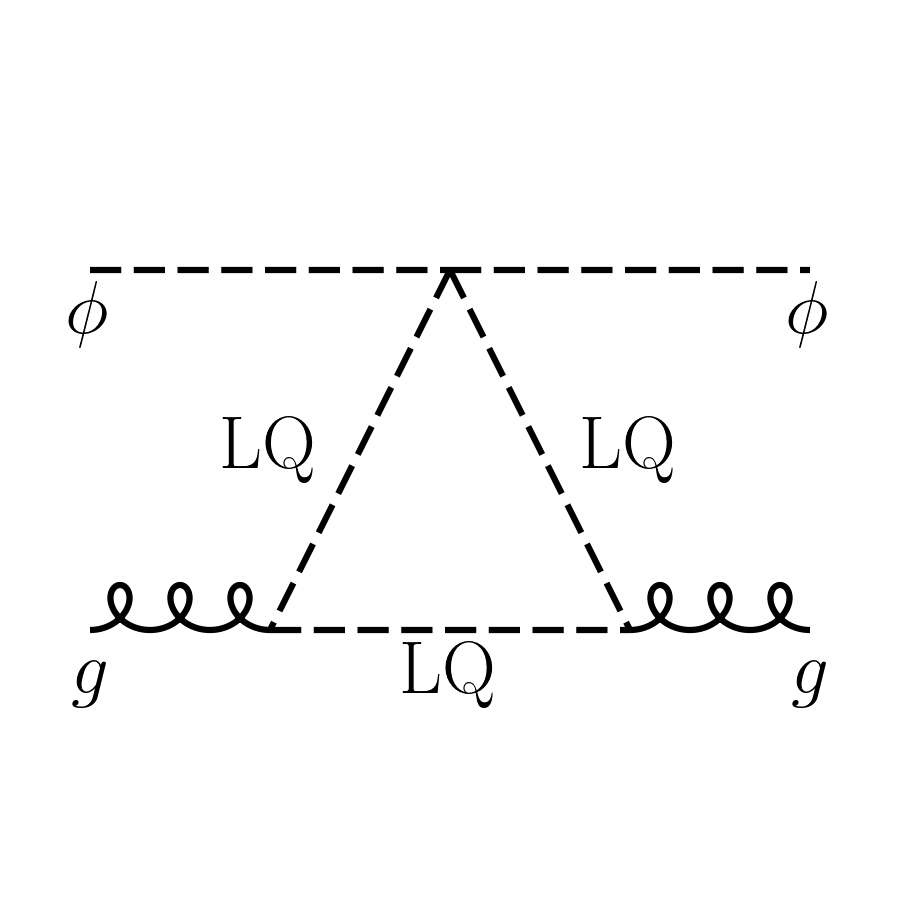}}
    \subfigure[]{\includegraphics[width=0.3\linewidth]{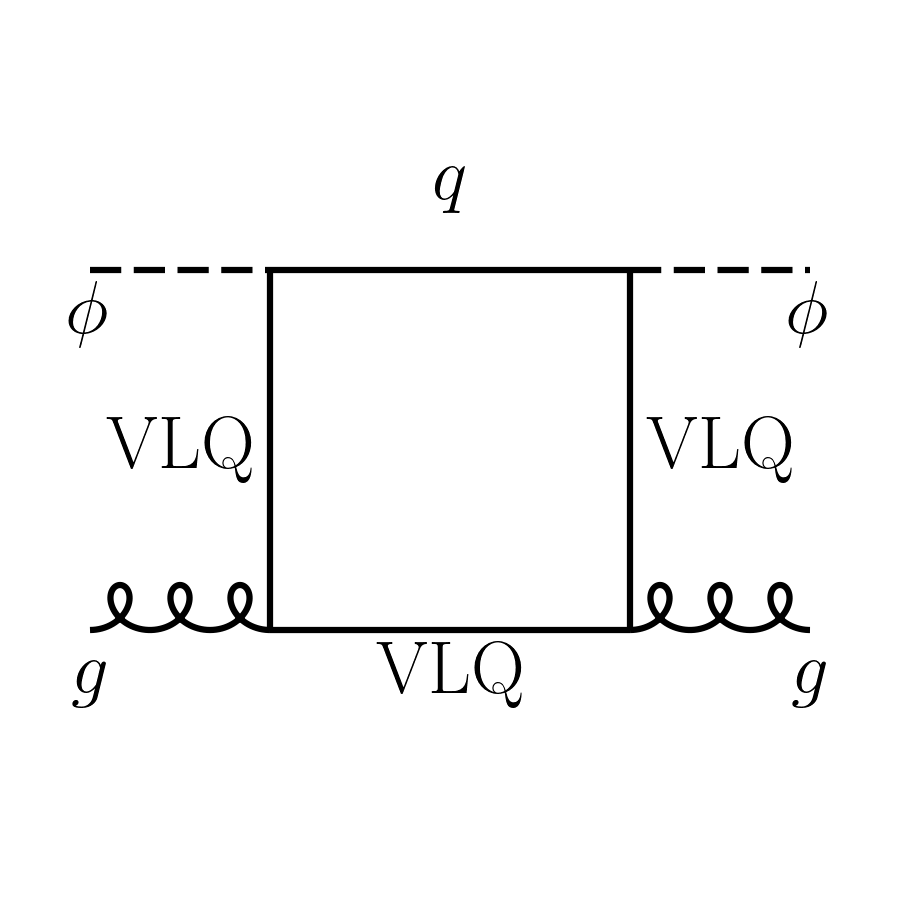}}
    \subfigure[]{\includegraphics[width=0.3\linewidth]{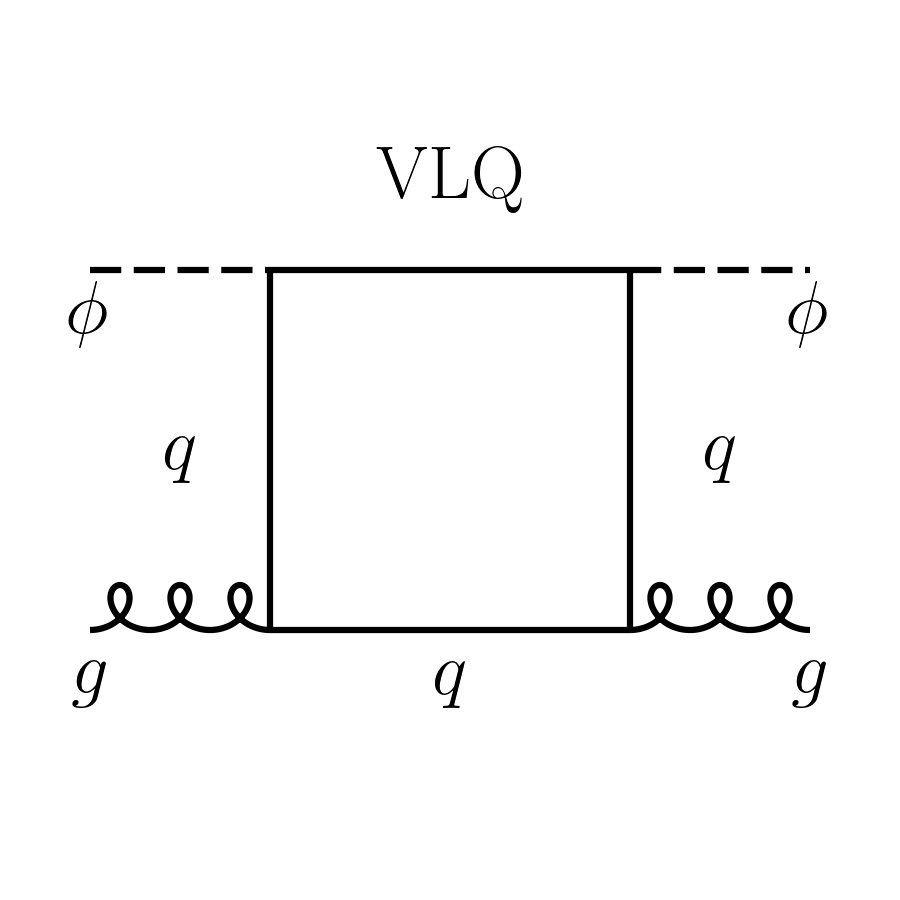}}
    \caption{(a) Tree-level Higgs-portal contribution to DM-nucleon scattering. (b) Tree-level VLQ-mediated contribution to DM-nucleon scattering. (c) Loop-induced contribution to DM-nucleon scattering from scalar leptoquark triangles. (d,e) Loop-induced contributions to DM-nucleon scattering from VLQ boxes.}
    \label{fig:dd}
\end{figure}

The scalar singlet $\phi$ is stable due to the imposed $\mathbb{Z}_2$ symmetry and can
scatter elastically off nuclei through both tree-level and loop-level interactions, receiving relevant contributions from Higgs exchange through the scalar portal coupling $\lambda_{H\phi}$, tree-level exchange of the VLQs, and
loop-induced couplings to gluons generated by scalar LQ and VLQ
loops. Since direct-detection experiments probe small momentum transfer, the heavy
mediators can be integrated out. The low-energy interactions relevant for
spin-independent scattering may therefore be written as
\begin{equation}
\mathcal{L}_{\rm eff}
=
\frac{1}{2}\sum_q C_q\,\phi^2 \bar q q
+
\frac{1}{2}C_g\,\phi^2\frac{\alpha_s}{12\pi}
G^a_{\mu\nu}G^{a\mu\nu},
\label{eq:DD_eff_lag}
\end{equation}
where the Wilson coefficients $C_q$ and $C_g$ receive contributions from both the Higgs portal and the VLQ-mediated processes, as well as the LQ and VLQ loops. The dimension-6 quark operator of the form $\phi^2 \bar q \slashed \partial q$ vanishes for a real scalar  \cite{Cirelli:2013ufw, DEramo:2020sqv}. 

\subsubsection{Wilson Coefficients for Direct Detection}

\autoref{fig:dd}(a) shows the Higgs-portal contribution to direct detection, the relevant term for which, following \autoref{eq:scalar}, reads
\begin{equation}
V\supset \frac12 \lambda_{H\phi}\phi^2 H^\dagger H.
\end{equation}

From here, one obtains the following contribution to the scalar Wilson coefficient:
\begin{equation}
C_q^{(h)}
=
\frac{\lambda_{H\phi}m_q}{m_h^2}.
\label{eq:Cq_higgs}
\end{equation}

The VLQ-portal contribution, as shown in \autoref{fig:dd}(b), follows from the Yukawa interactions connecting $\phi$, the VLQs, and the SM quarks:

\begin{equation}
\mathcal{L}\supset 
-Y_{\phi Q}\,\overline{\Psi^q_{R}}\,\phi\,Q_{L}
\;
-Y_{\phi D}\,\overline{\psi^d_{L}}\,\phi\,d_{R}
\;+\;\text{h.c.},
\label{eq:VLQ-yuk}
\end{equation}
In the mass basis, following the discussion in \autoref{sec:VLQmix} the scalar Wilson coefficient can be written as
\begin{equation}
C_q^{({\rm VLQ})}
=
\sum_{a=1}^{2}
\frac{
Y_{\phi Q}(U_R)_{1a}
\left[Y_{\phi D}(U_L)_{2a}\right]^*
}{M_a}
+{\rm h.c.}
\label{eq:Cq_VLQ_exact}
\end{equation}
For small VLQ mixing and real Yukawa couplings that we consider, this reduces to
\begin{equation}
C_d^{({\rm VLQ})}
\simeq
2 \frac{\tilde m}{M_QM_D}
Y_{\phi Q}Y_{\phi D},
\qquad
\tilde m=\frac{\tilde y_Hv}{\sqrt2}.
\label{eq:Cq_VLQ_smallmix}
\end{equation}
Thus, at tree level, the VLQ-mediated direct-detection constraint is controlled
by the product $Y_{\phi Q}Y_{\phi D}\tilde y_H$. If either $Y_{\phi Q}$,
$Y_{\phi D}$, or $\tilde y_H$ vanishes, this scalar operator is absent at leading
order.

The loop-induced gluon operator receives contributions from both scalar
LQ triangles (\autoref{fig:dd}(c)) and VLQ boxes (\autoref{fig:dd}(d,e)). The total Wilson coefficient is the sum of these two contributions:
\begin{equation}
C_g=C_g^{(\triangle{\rm LQ})}+C_g^{(\Box{\rm VLQ})}.
\end{equation}
For the LQ contribution, the general expression after scalar mixing is
\begin{equation}
C_g^{(\triangle{\rm LQ})}
=
\frac14
\left[
\frac{\lambda_{\phi \widetilde{R}}}{M_{\widetilde{R}_2^{2/3}}^2}
\mathcal{A}_0\left(\frac{M_\phi^2}{4M_{\widetilde{R}_2^{2/3}}^2}\right)
+
\sum_{i=1}^{2}
\frac{\lambda_{\phi\eta_i}}{M_{\eta_i}^2}
\mathcal{A}_0\left(\frac{M_\phi^2}{4M_{\eta_i}^2}\right)
\right],
\label{eq:Cg_LQ_general}
\end{equation}
where
\begin{align}
\lambda_{\phi\eta_1}
&=
\lambda_{\phi \widetilde{R}}\cos^2\theta_s
+
\lambda_{\phi S_1}\sin^2\theta_s,
\\
\lambda_{\phi\eta_2}
&=
\lambda_{\phi \widetilde{R}}\sin^2\theta_s
+
\lambda_{\phi S_1}\cos^2\theta_s.
\end{align}
The loop function is normalized as $\mathcal{A}_0(0)=1$:
\begin{equation}
\mathcal{A}_0(\tau)
=
3\tau^{-2}\left[f(\tau)-\tau\right],
\qquad
f(\tau)=
\begin{cases}
\arcsin^2\!\sqrt{\tau}, & \tau\le1,\\[3pt]
-\tfrac14\!\left[\ln\!\dfrac{1+\sqrt{1-\tau^{-1}}}{1-\sqrt{1-\tau^{-1}}}-i\pi\right]^2, & \tau>1,
\end{cases}.
\label{eq:A0_norm}
\end{equation}
In the limit of small LQ mixing, considering $M_{\widetilde{R}_2^{2/3}} \approxeq M_{\eta_2^{1/3}} = M_{R_2}$ and $M_{\eta_1^{1/3}} = M_{S_1}$, \autoref{eq:Cg_LQ_general} simplifies to
\begin{equation}
C_g^{(\triangle{\rm LQ})}
\simeq
\frac14
\left[
2\frac{\lambda_{\phi \widetilde{R}}}{M_{R_2}^2}\mathcal{A}_R
+
\frac{\lambda_{\phi S_1}}{M_{S_1}^2}\mathcal{A}_S
\right],
\label{eq:Cg_LQ_nondeg}
\end{equation}
with
\begin{equation}
\mathcal{A}_R=
\mathcal{A}_0\left(\frac{M_\phi^2}{4M_{R_2}^2}\right),
\qquad
\mathcal{A}_S=
\mathcal{A}_0\left(\frac{M_\phi^2}{4M_{S_1}^2}\right).
\end{equation}

For the VLQ box contribution, in the small mixing limit, one obtains:
\begin{align}
C_g^{(\Box{\rm VLQ})}
\simeq
\frac{1}{2}
\Bigg[
\sum_{q=u,c,t}\frac{|Y_{\phi Q}^{\,q}|^{2}}{M_Q^{2}-M_\phi^{2}}
+\sum_{q=d,s,b}\!\left(
\frac{|Y_{\phi Q}^{\,q}|^{2}}{M_Q^{2}-M_\phi^{2}}
+\frac{|Y_{\phi D}^{\,q}|^{2}}{M_D^{2}-M_\phi^{2}}
\right)
\Bigg]
+\mathcal{O}(\theta_{L,R}^{2})\;.
\label{eq:Cg_VLQ_smallmix}
\end{align}

\subsubsection{Matching to nucleons and spin-independent cross-sections}

Having established the effective quark- and gluon-level interactions in \autoref{eq:DD_eff_lag}, we now connect them to hadronic observables relevant for direct detection experiments. The key step is to express the quark and gluon operators in terms of nucleon matrix elements. For light quarks, the scalar bilinears are parameterized as
\begin{align}
m_q\langle N|\bar q q|N\rangle &= m_N f_{Tq}^{(N)}, 
& q &= u,d,s,
\label{eq:lightq}
\end{align}
while heavy-quark loops and gluon operators are related by the QCD trace anomaly,
\begin{align}
\langle N| \tfrac{\alpha_s}{12\pi} G_{\mu\nu}^a G^{a\mu\nu} |N\rangle
&= -\frac{2}{27}\,m_N\,f_{TG}^{(N)},
&
f_{TG}^{(N)} &= 1 - \sum_{q=u,d,s} f_{Tq}^{(N)}.
\label{eq:gluonmatrix}
\end{align}
Here, $f_{Tq}^{(N)}$ denote the light-quark scalar form factors, while $f_{TG}^{(N)}$ represents the gluon contribution to the nucleon mass. Combining the quark and gluon contributions from eq.~\eqref{eq:DD_eff_lag}, the nucleon-level interaction can be written as
\begin{equation}
\mathcal{L}_{\rm eff}
\;\supset\;
\frac{1}{2}f_N^{\rm tot}\,\phi^2\,\bar N N,
\label{eq:LeffN}
\end{equation}
with the effective scalar coupling
\begin{equation}
f_N^{\rm tot}
=
m_N
\left[
\sum_{q=u,d,s} C_q^{\rm tot}\,\frac{f_{Tq}^{(N)}}{m_q}
-\frac{2}{27}\,C_g^{\rm eff}\,f_{TG}^{(N)}
\right].
\label{eq:fNtot}
\end{equation}
The first term accounts for the Higgs- and VLQ-mediated contact interactions with light quarks, while the second term captures the loop-induced coupling to gluons arising from both scalar LQ and VLQ loops, also incorporating the heavy-quark contributions through the trace anomaly relation in \autoref{eq:gluonmatrix}. The corresponding spin-independent cross section 
\begin{equation}
\sigma_{\rm SI}^{(N)}
=
\frac{m_N^2 (f_N^{\rm tot})2}{4\pi(M_\phi+m_N)^2}
\simeq
\frac{m_N^2 (f_N^{\rm tot})^2}{4\pi M_\phi^2},
\label{eq:sigmaSI_general}
\end{equation}
where the final expression holds for $M_\phi\gg m_N$. For our evaluation, we use the following numerical values for the scalar form factors~\cite{Belanger:2013oya}:
\begin{align}
&f_{Tu}^{(p)} = 0.0153,\quad f_{Td}^{(p)} = 0.0191,\quad f_{Ts}^{(p)} = 0.0447,~~\Longrightarrow ~~ f_{TG}^{(p)} = 1 - \sum_{q=u,d,s} f_{Tq}^{(p)}= 0.9209
\label{eq:formfactors}
\end{align}

Using $m_N=0.939~{\rm GeV}$, $m_h=125.25~{\rm GeV}$,
$m_d(2~{\rm GeV})=4.67\times10^{-3}~{\rm GeV}$, and the scalar form factors
quoted above, the Higgs-portal contribution gives
\begin{equation}
\sigma_{\rm SI}^{(N)}\big|_h
\simeq
7.887\times10^{-39}
\frac{\lambda_{H\phi}^2}{(M_\phi/{\rm GeV})^2}
~{\rm cm}^2 .
\label{eq:sigma_higgs}
\end{equation}
Therefore,
\begin{equation}
\lambda_{H\phi}
\lesssim
\left(
\frac{\sigma_{\rm lim}}
{7.887\times10^{-39}~{\rm cm}^2}
\right)^{1/2}
\left(\frac{M_\phi}{\rm GeV}\right).
\label{eq:lambdaHphi_bound_general}
\end{equation}
Here, $\sigma_{\rm lim}$ denotes the experimental limit on the spin-independent cross-section. For the representative value $\sigma_{\rm lim}=10^{-47}~{\rm cm}^2$ \cite{LZ:2024zvo}, this becomes
\begin{equation}
\lambda_{H\phi}
\lesssim
3.6\times10^{-5}
\left(\frac{M_\phi}{\rm GeV}\right).
\label{eq:lambdaHphi_bound}
\end{equation}

Similarly, the tree-level VLQ contribution gives
\begin{equation}
\sigma_{\rm SI}^{(N)}\big|_{\rm VLQ}
\simeq
1.613\times10^{-27}
\frac{1}{(M_\phi/{\rm GeV})^2}
\left(\frac{\tilde m}{M_QM_D}\right)^2
|Y_{\phi Q}Y_{\phi D}|^2
~{\rm cm}^2 .
\label{eq:sigma_VLQ_tree}
\end{equation}
Using $\tilde m=\tilde y_Hv/\sqrt2$, this implies
\begin{equation}
|Y_{\phi Q}Y_{\phi D}\tilde y_H|
\lesssim
\frac{1}{246}
\left(
\frac{2\sigma_{\rm lim}}
{1.613\times10^{-27}~{\rm cm}^2}
\right)^{1/2}
\left(\frac{M_\phi}{\rm GeV}\right)
\left(\frac{M_Q}{\rm GeV}\right)
\left(\frac{M_D}{\rm GeV}\right),
\label{eq:VLQ_tree_bound_general}
\end{equation}
and for the same representative value $\sigma_{\rm lim}=10^{-47}~{\rm cm}^2$, this gives
\begin{equation}
|Y_{\phi Q}Y_{\phi D}\tilde y_H|
\lesssim
4.53\times10^{-13}
\left(\frac{M_\phi}{\rm GeV}\right)
\left(\frac{M_Q}{\rm GeV}\right)
\left(\frac{M_D}{\rm GeV}\right).
\label{eq:VLQ_tree_bound}
\end{equation}

For non-degenerate scalar LQs, the LQ-triangle contribution gives
\begin{equation}
\sigma_{\rm SI}^{(N)}\big|_{\triangle{\rm LQ}}
\simeq
7.01\times10^{-33}
\frac{1}{(M_\phi/{\rm GeV})^2}
\left[
2\frac{\lambda_{\phi \widetilde{R}}}{(m_R/{\rm GeV})^2}\mathcal{A}_R
+
\frac{\lambda_{\phi S_1}}{(m_{S_1}/{\rm GeV})^2}\mathcal{A}_S
\right]^2
~{\rm cm}^2 .
\label{eq:sigma_LQ_nondeg}
\end{equation}
Thus, direct detection constrains the combination
\begin{equation}
\left|
2\frac{\lambda_{\phi \widetilde{R}}}{(m_R/{\rm GeV})^2}\mathcal{A}_R
+
\frac{\lambda_{\phi S_1}}{(m_{S_1}/{\rm GeV})^2}\mathcal{A}_S
\right|
\lesssim
\left(
\frac{\sigma_{\rm lim}}
{7.01\times10^{-33}~{\rm cm}^2}
\right)^{1/2}
\left(\frac{M_\phi}{\rm GeV}\right),
\label{eq:LQ_nondeg_bound_general}
\end{equation}
which for the representative value $\sigma_{\rm lim}=10^{-47}~{\rm cm}^2$ becomes
\begin{equation}
\left|
2\frac{\lambda_{\phi \widetilde{R}}}{(m_R/{\rm GeV})^2}\mathcal{A}_R
+
\frac{\lambda_{\phi S_1}}{(m_{S_1}/{\rm GeV})^2}\mathcal{A}_S
\right|
\lesssim
3.78\times10^{-8}
\left(\frac{M_\phi}{\rm GeV}\right).
\label{eq:LQ_nondeg_bound}
\end{equation}
To simplify this representation, we can consider degenerate LQs $M_{R_2} = M_{S_1} = M_{\rm LQ}$, which reproduces
\begin{equation}
\lambda_{\phi \widetilde{R}},\lambda_{\phi S_1}
\equiv \lambda
\lesssim
1.26\times10^{-8}
\left(\frac{M_\phi}{\rm GeV}\right)
\left(\frac{M_{\rm LQ}}{\rm GeV}\right)^2 .
\label{eq:LQ_deg_bound}
\end{equation}
For TeV-scale scalar LQs, this is a weak constraint, possibly allowing for $\lambda_{\phi \widetilde{R}},\lambda_{\phi S_1}$ in the $\mathcal{O}(1)$ range. However, we choose to be conservative in our numerical scan and consider $\lambda_{\phi R},\lambda_{\phi S_1}$ in $\mathcal{O}(10^{-2})$, as these couplings do not affect other aspects of the phenomenology.

Finally, assuming flavour-universal couplings  $Y_{\phi Q}^q=Y_{\phi D}^q\equiv y$ and degenerate VLQ masses $M_Q=M_D=M_{\rm VLQ}$, from \autoref{eq:Cg_VLQ_smallmix} one obtains a simplified expression for the VLQ-box contribution:
\begin{equation}
\sigma_{\rm SI}^{(N)}\big|_{\Box{\rm VLQ}}
\simeq
2.270\times10^{-30}
\frac{1}{(M_\phi/{\rm GeV})^2}
\left[
\frac{y^2}
{(M_{\rm VLQ}/{\rm GeV})^2-(M_\phi/{\rm GeV})^2}
\right]^2
~{\rm cm}^2 .
\label{eq:sigma_VLQ_box}
\end{equation}
Therefore,
\begin{equation}
y
\lesssim
\left(
\frac{\sigma_{\rm lim}}
{2.270\times10^{-30}~{\rm cm}^2}
\right)^{1/4}
\left(\frac{M_\phi}{\rm GeV}\right)^{1/2}
\left[
\left(\frac{M_{\rm VLQ}}{\rm GeV}\right)^2
-
\left(\frac{M_\phi}{\rm GeV}\right)^2
\right]^{1/2}.
\label{eq:VLQ_box_bound_general}
\end{equation}
For $\sigma_{\rm lim}=10^{-47}~{\rm cm}^2$ and $M_{\rm VLQ}\gg M_\phi$, this
simplifies to
\begin{equation}
y
\lesssim
4.6\times10^{-5}
\left(\frac{M_{\rm VLQ}}{\rm GeV}\right)
\left(\frac{M_\phi}{\rm GeV}\right)^{1/2}.
\label{eq:VLQ_box_bound}
\end{equation}
This limit on the VLQ Yukawa couplings becomes relevant only when $Y_{\phi Q}=Y_{\phi D} \sim \mathcal{O}(1)$. But as we will see later, our parameter space evaluation keeps one of them to be substantially smaller than the other, depending on the mass hierarchy of $\mathbb{Z}_2$-odd particles. Hence, they do not impose additional constraints on the parameter space beyond those already derived from the tree-level processes in \autoref{eq:VLQ_tree_bound}. In the numerical scan that will follow in \autoref{sec:dmpheno}, we use the direct-detection constraints mainly to motivate
conservative choices for the scalar and Yukawa couplings. In particular, we reiterate the requirements for our DM parameter scan:
\begin{align}
\lambda_{H\phi}
&\lesssim
3.6\times10^{-5}
\left(\frac{M_\phi}{\rm GeV}\right),
\label{eq:scan_lambdaHphi}
\\
|Y_{\phi Q}Y_{\phi D}\tilde y_H|
&\lesssim
4.53\times10^{-13}
\left(\frac{M_\phi}{\rm GeV}\right)
\left(\frac{M_Q}{\rm GeV}\right)
\left(\frac{M_D}{\rm GeV}\right),
\label{eq:scan_VLQ_tree}
\end{align}
and choose $\lambda_{\phi \widetilde{R}},\lambda_{\phi S_1}$ in the $\mathcal{O}(10^{-2})$
range, which is safely below the LQ-loop direct-detection bound for the heavy LQs used in the scan. It is noteworthy that here we have considered these limits from individual contributions separately, whereas the different contributions add coherently in $f_N^{\rm tot}$, so cancellations are possible in principle. Since the quark piece enters with \(+\sum_{q} C_q^{\rm tot} f_{Tq}^{(N)}/m_q\) while the gluon piece enters with a \emph{negative} coefficient \(-\tfrac{2}{27} C_g f_{TG}^{(N)}\) [eq.~\eqref{eq:fNtot}], destructive interference among the Higgs portal, VLQ tree, and gluon terms can take place. This opens a fine-tuned region where sizeable $Y_{\phi Q}$, $Y_{\phi D}$ and/or scalar quartics \(\lambda_{H\phi},\lambda_{\phi\widetilde R},\lambda_{\phi S_1}\) still yield a suppressed \(f_N^{\rm tot}\) and hence a small \(\sigma_{\rm SI}^{(N)}\). However, the above limits are imposed
without relying on such cancellations, making our estimate more conservative.

\subsection{Evaluation of DM parameter space}\label{sec:dmpheno}

Recognising the multitude of free parameters in the model, we intend to perform a scan over a subset that best represents the goals of this work, primarily to accommodate a much lighter LQ than its current $\sim 1.5$ TeV lower limit from LHC searches, owing to its $\mathbb{Z}_2$-odd nature. We also want the next-to-lightest $\mathbb{Z}_2$-odd particle to affect the relic density of the inert singlet $\phi$ via co-annihilation, so that relic-satisfying/underabundant DM masses in a few hundred GeVs can be realised without exclusion from the DD experiments. In that regard, we devise two mass hierarchies to follow in the scan, allowing either the singlet or the doublet LQ to be closer in mass to the DM, and essentially decoupling the VLQs to a high-TeV scale.

\begin{table}[h]
    \centering
    \begin{tabular}{ccc}
    \toprule
        Mass range & Hierarchy 1 & Hierarchy 2 \\  
    \midrule
    {[400, 900]} GeV & $M_\phi$ & $M_\phi$ \\
    {[$M_{\phi} + 10, M_{\phi} + 150$]} GeV & $M_{S_1}$ & $M_{R_2}$ \\
    1.5 TeV & $M_{R_2}$ & $M_{S_1}$ \\
    2.0 TeV & $M_D$ & $M_D$ \\
    2.5 TeV & $M_Q$ & $M_Q$ \\
    \bottomrule
    \end{tabular}
    \caption{Chosen mass hierarchies and respective masses for the $\mathbb{Z}_2$-odd particles.}
    \label{tab:hierarchy}
\end{table}

\autoref{tab:hierarchy} shows the two aforementioned hierarchies, keeping $S_1$ and $\widetilde{R}_2$ as the lightest $\mathbb{Z}_2$-odd coloured particle, in Hierarchy 1 (H1) and Hierarchy 2 (H2), respectively. For a decoupled inert singlet DM model in light of the LZ 2024 data, the observed relic is only satisfied in a mass close to $\sim 60$ GeV \cite{Yu:2024xsy}, with the remaining parameter space up to $\sim 20$ TeV being ruled out \cite{EscuderoAbenza:2025cfj}. However, with the inclusion of additional co-annihilating particles, we can look at heavier masses for $\phi$, keeping the next-to-lightest $\mathbb{Z}_2$-odd particle within 10-150 GeV of its mass, helping us realise a lighter LQ as mentioned.

\subsection{DM phenomenology for Hierarchy 1} \label{sec:dm-h1}

\begin{figure}[h]
    \centering
    \subfigure[]{\includegraphics[width=0.25\linewidth]{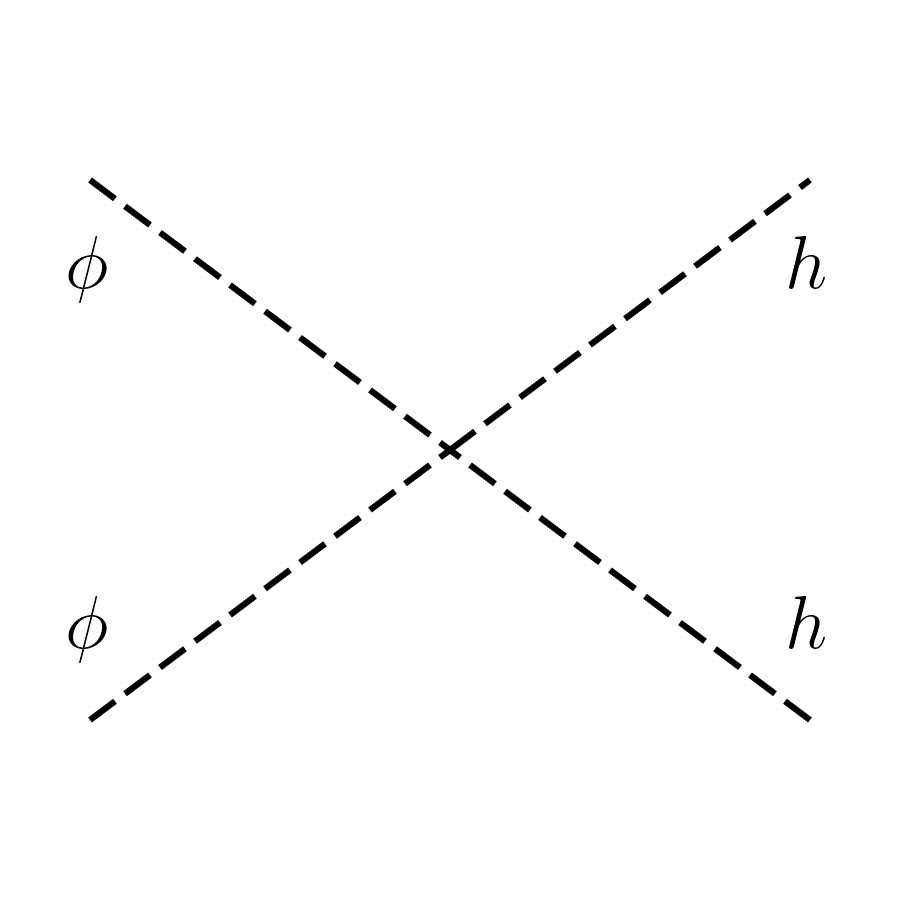}}
    \subfigure[]{\includegraphics[width=0.25\linewidth]{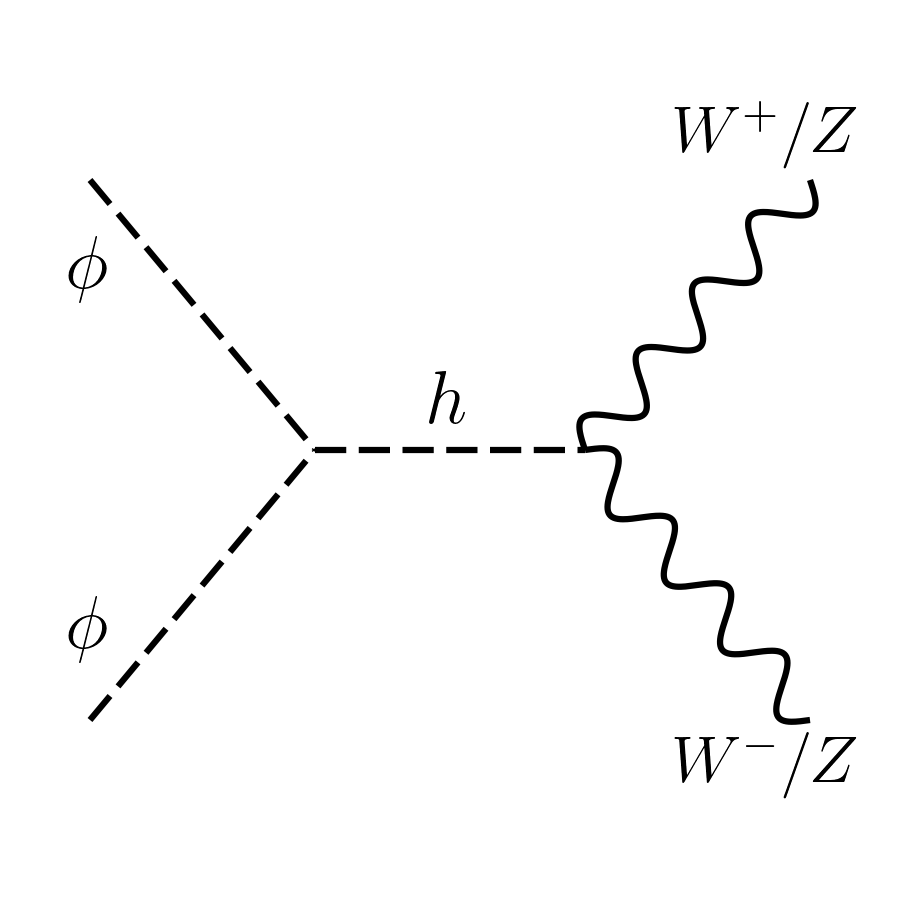}}
    \subfigure[]{\includegraphics[width=0.25\linewidth]{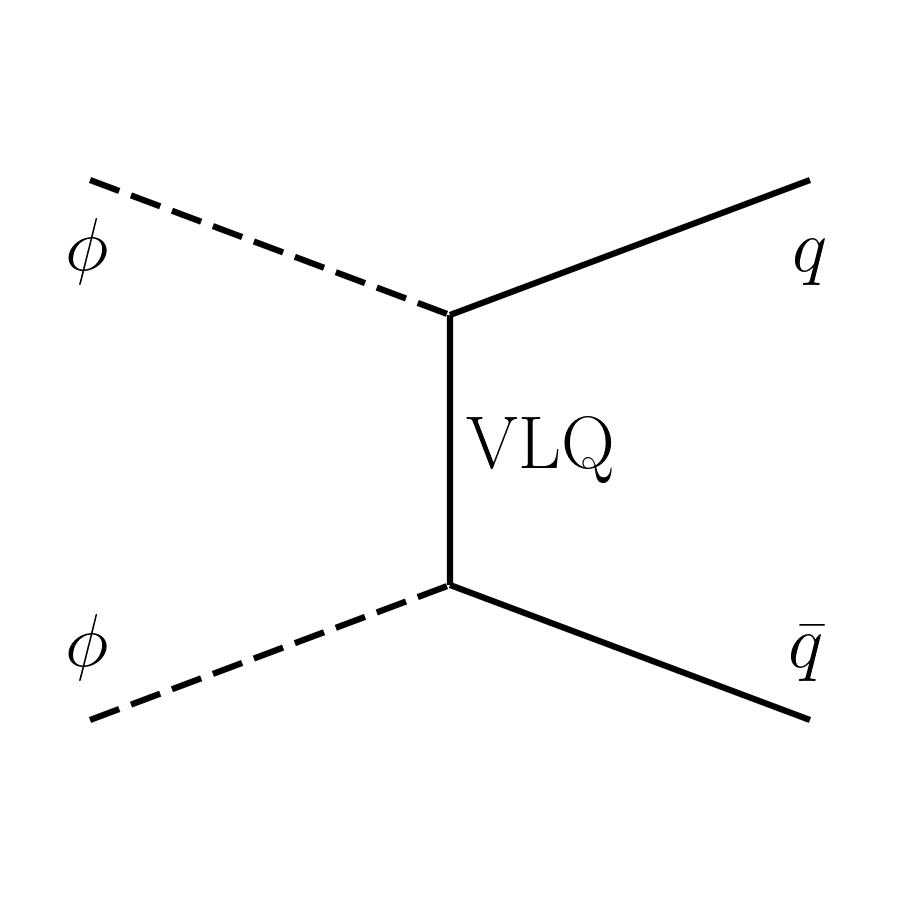}}
    \subfigure[]{\includegraphics[width=0.25\linewidth]{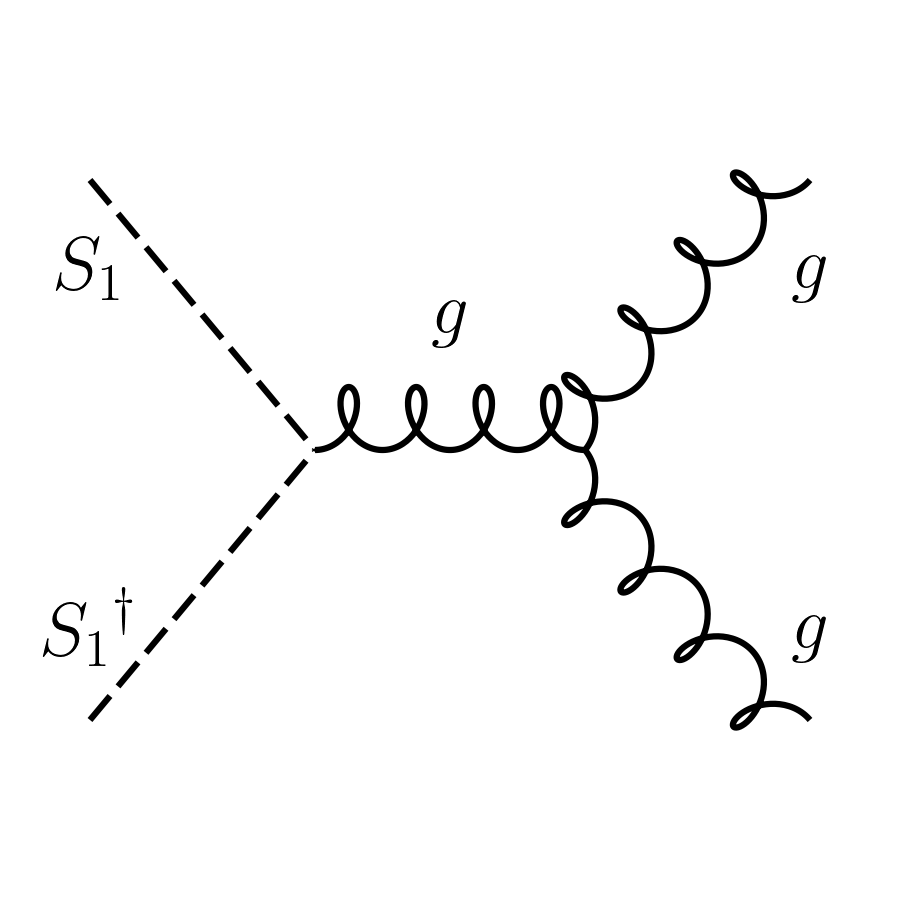}}
    \subfigure[]{\includegraphics[width=0.25\linewidth]{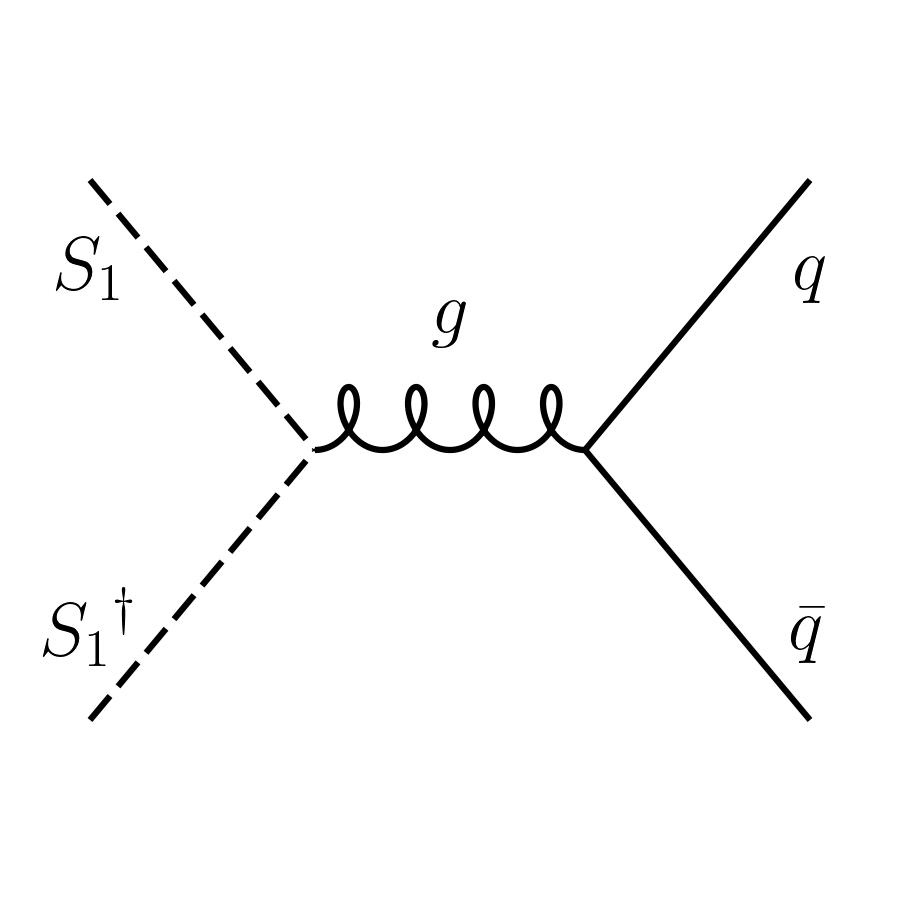}}
    \subfigure[]{\includegraphics[width=0.25\linewidth]{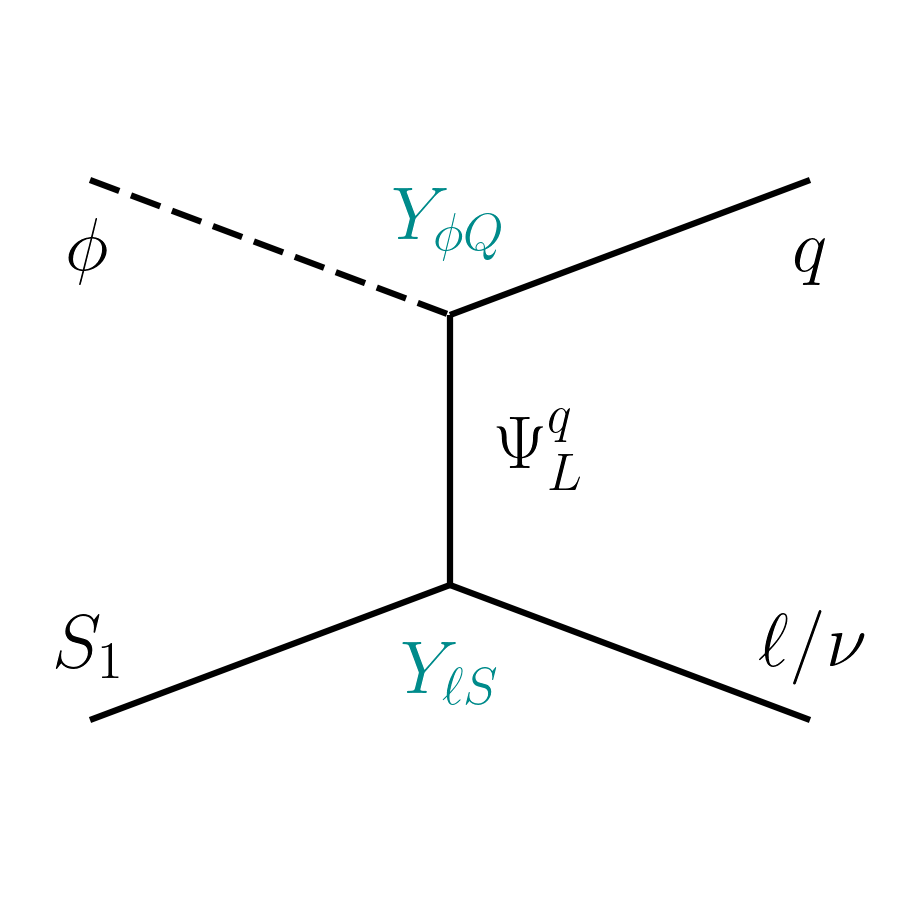}}
    \caption{(a)-(c) show $\phi-\phi$ annihilation channels, (d)-(e) show $S_1 - S_1$ co-annihilation channels, (f) shows $\phi-S_1$ co-annihilation channel.}
    \label{fig:h1-annihilation}
\end{figure}

\autoref{fig:h1-annihilation}(a)-(c) show the dominant annihilation channels, and (d)-(f) show the dominant co-annihilation channels for the H1 case. Choosing TeV-scale masses for particles in the neutrino mass loop in \autoref{eq:numass} requires $\kappa \sim 10^{-3}$ GeV if we want to work with Yukawa couplings in $\mathcal{O}(0.01-1)$, and for such small mixing, $S_1$ remains almost completely singlet. The same is true for $\widetilde{R_2}^{1/3}$ which remains almost purely doublet. Hence, from this section onwards, we drop the $\eta_{1,2}$ nomenclature for them and keep the gauge basis notations only. The diagram in \autoref{fig:h1-annihilation}(f) is especially intriguing, as both $Y_{\phi Q}$ and $Y_{\ell S}$ contribute to it. The same couplings are responsible for the 3-body decay of  $S_1$, which we will see later. The Higgs-portal phenomenology of $\phi$ is well studied in decoupled singlet scenarios, which is less intriguing here. Considering all of these, we fix all the new quartic couplings $\lambda_i = 0.01$ for the scans, which respects the direct detection limits from \autoref{eq:lambdaHphi_bound} and \autoref{eq:LQ_deg_bound}. Next, referring to \autoref{eq:VLQ_tree_bound}, we fix $y_H = \tilde y_H = 0.05$ and $Y_{\phi D} = 0.01$. Even though it does not impact the DM evolution, we fix $Y_{\ell R} = 0.01$ for uniformity. Then, fixing $M_{R_2}, M_D, M_Q$ at the values from \autoref{tab:hierarchy}, we pick six values of $M_\phi \in [400,900]$ GeV with 100 GeV intervals, and vary $M_{S_1}$ accordingly. We also randomly vary $Y_{\phi Q}, Y_{\ell S} \in [0.01, 1.0]$, and scan the points using {\tt micrOMEGAs} \cite{Alguero:2023zol}.\footnote{To obtain the neutrino masses and $\Delta m_{ij}$ for neutrinos, we can tune the components of $Y_{\ell R}$ as it does not affect the DM phenomenology in H1.} The loop-induced contributions to the spin-independent cross-section, as discussed in \autoref{sec:dd-1} are already implemented in {\tt micrOMEGAs} following the prescription of ref. \cite{Hisano:2015bma}. 

\begin{figure}[h]
    \centering
    \subfigure[]{\includegraphics[width=\linewidth]{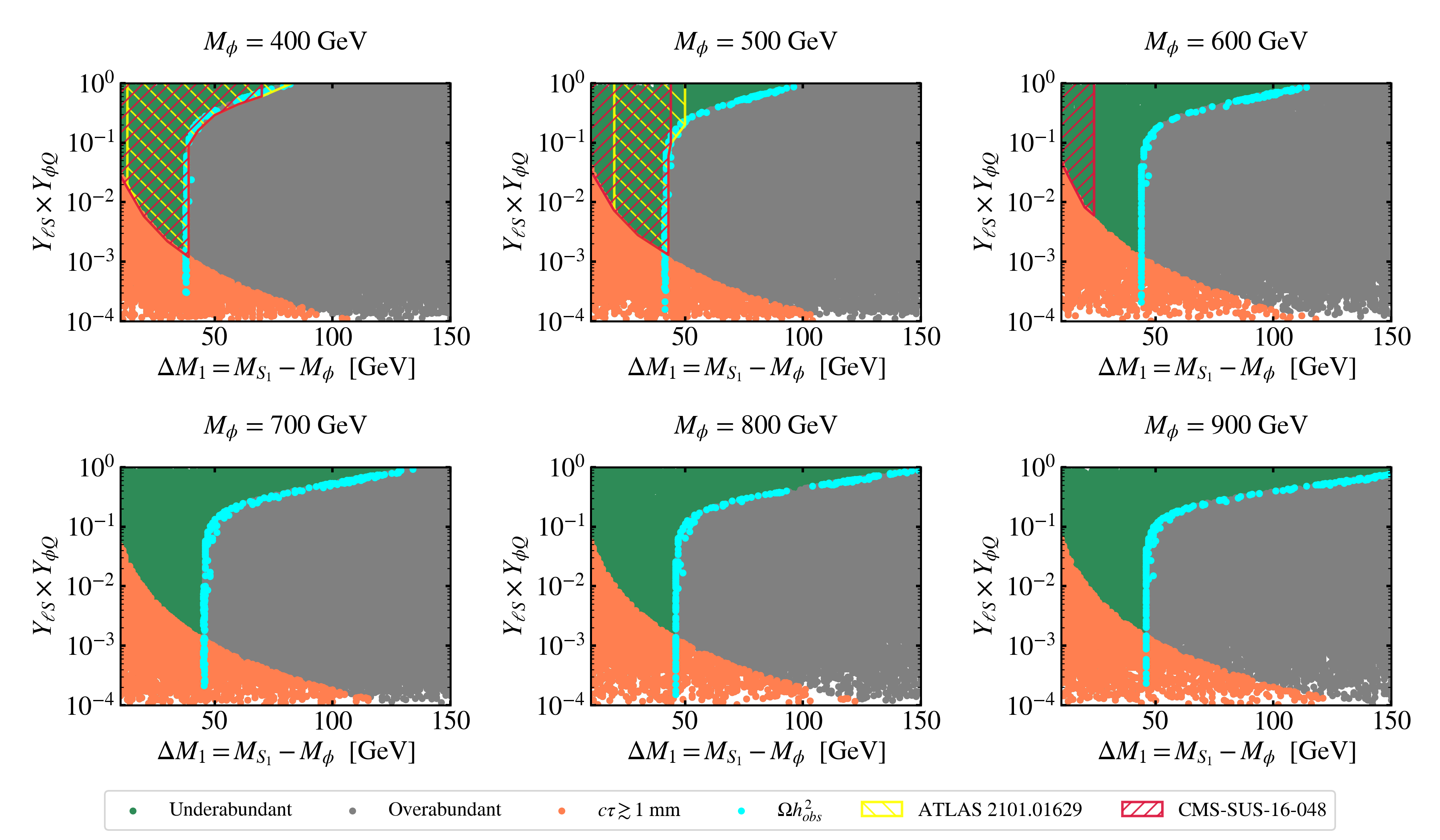}}
    \subfigure[]{\includegraphics[width=0.6\linewidth]{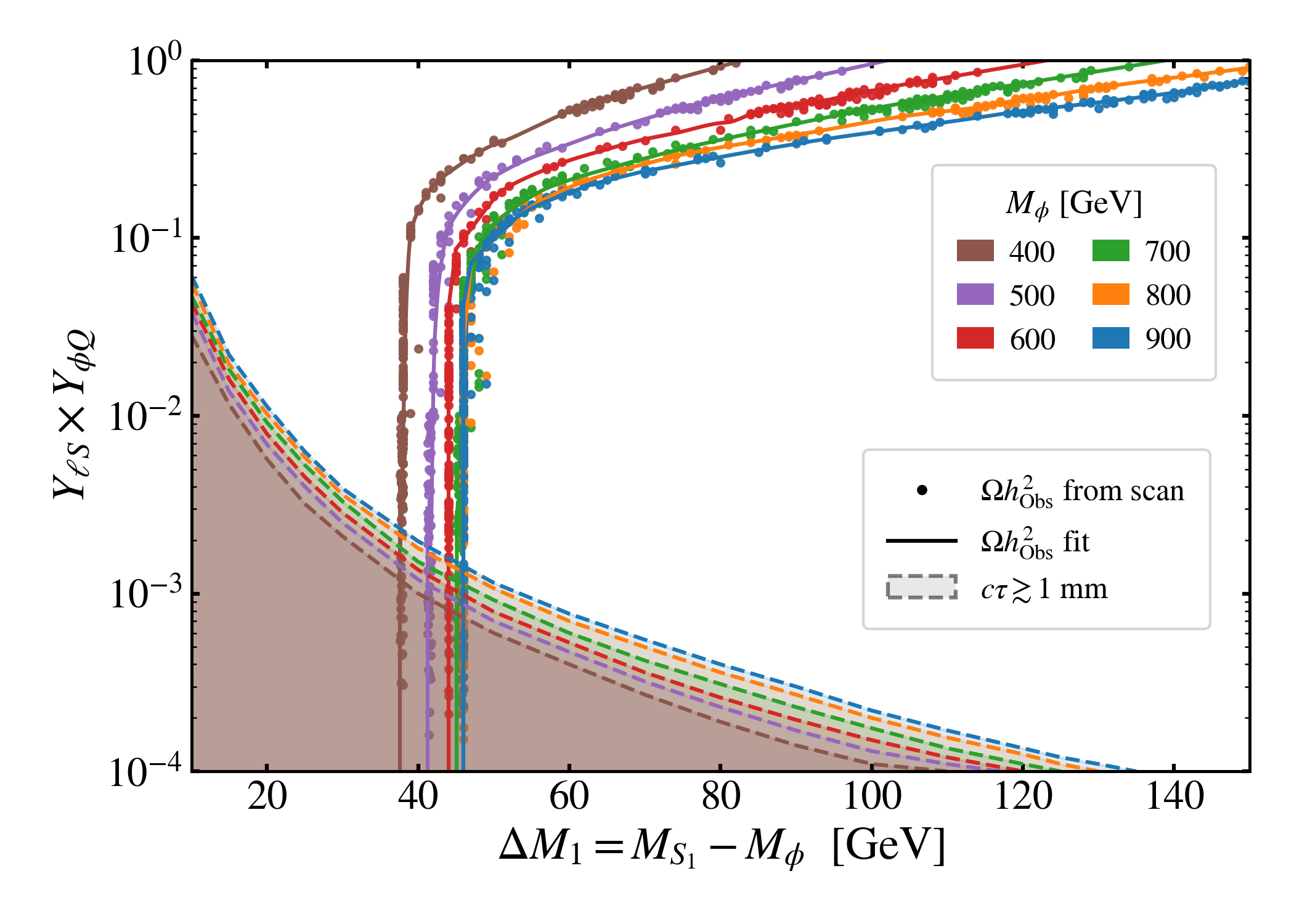}}
    \hfil
    \subfigure[]{\includegraphics[width=0.3\linewidth]{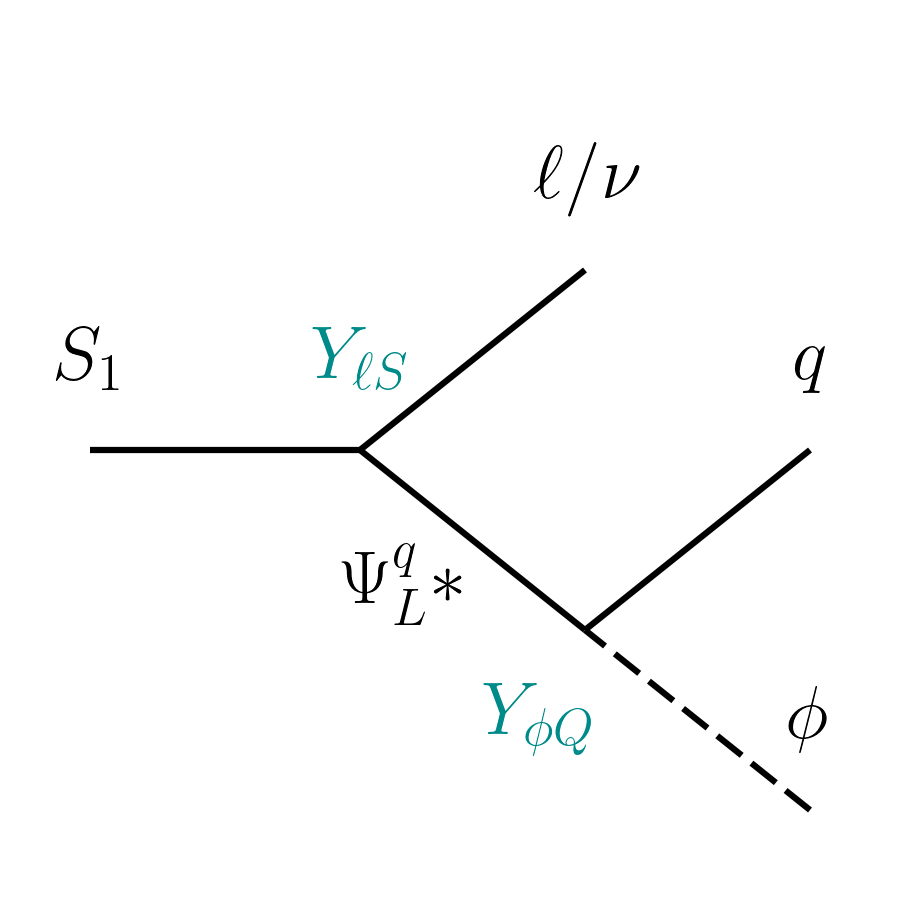}}
    \caption{(a) Scatter plots of $\Delta M_1 = M_{S_1} - M_\phi$ vs $Y_{\ell S} \times Y_{\phi Q}$ for six different $M_\phi$ choices. The grey and green points are overabundant and underabundant in DM respectively, while the cyan points satisfy $\Omega h^2_{\rm Obs} =0.1198 \pm 0.0012$. The orange points lead to displaced decay of $S_1$ at detectors. The red and yellow hatched regions mark points excluded by CMS-SUS-16-048 \cite{CMS:2018kag} and ATLAS 2101.01629 \cite{ATLAS:2021twp} analyses, as per \texttt{CheckMATE} \cite{Dercks:2016npn}. (b) $\Delta M_{1} = M_{S_1} - M_\phi$ vs $Y_{\ell S}\times Y_{\phi Q} $, showing the trends of $c \tau \gtrsim 1$ mm limits (shaded) and the $\Omega h^2_{\rm Obs}$ points for the six DM masses considered. (c) 3-body decay Feynman diagram for $S_1$.}
    \label{fig:h1_combo_masses}
\end{figure}

The key ingredients in the context of the DM relic in this scenario are the masses $M_{\phi}, M_{S_1}$, the mass splitting $\Delta M_1 = M_{S_1} - M_\phi$, and the product of the Yukawa couplings $Y_{\ell S}\times Y_{\phi Q} $ for the region of low-mass splitting of our interest. We choose a twofold way to present their correlation. Firstly, in \autoref{fig:h1_combo_masses}(a) we present scatter plots of $\Delta M_{1} = M_{S_1} - M_\phi$ vs $Y_{\ell S}\times Y_{\phi Q} $ for the six chosen DM masses. The grey and green points signify DM overabundance and underabundance respectively, while the cyan points satisfy the observed relic $\Omega h_{Obs}^2 = 0.1198 \pm 0.0012$~\cite{Planck:2018vyg}. Clearly, the DM becomes overabundant when the co-annihilating $S_1$ essentially decouples, and for a majority of the parameter space, the relic is satisfied with $\Delta M_{1} \sim 38-48$ GeV, which tends to shift towards larger values for higher DM masses. Smaller mass gaps allow more co-annihilation and lead to underabundance. Larger values of $Y_{\ell S}\times Y_{\phi Q}  \gtrsim 0.2$ are capable of enhancing the co-annihilation, allowing for more prominent mass splittings, up to 80-150 GeV, to satisfy the relic. The same two couplings affect the 3-body decay of $S_1 \to (\Psi_L^d)^* \ell/\nu \to \phi q \ell/\nu$ from \autoref{fig:h1_combo_masses}(c), and the off-shell nature and the small mass splitting can lead to decay widths small enough that can cause $S_1$ to have a displaced decay vertex in the detectors. However, dealing with a $\mathbb{Z}_2$-odd displaced coloured object is tricky, as they must form bound states with SM quarks/gluons (similar to $R$-hadrons in some supersymmetric theories), the treatment of which is beyond the scope of this paper. Hence, we keep the points that ensure prompt decays at the detectors, demanding a mean proper decay length $c\tau \lesssim 1$ mm, corresponding to $\Gamma(S_1) \gtrsim 10^{-13}$ GeV. The orange points in \autoref{fig:h1_combo_masses}(a) mark the points that do not satisfy this promptness criterion. For a clearer picture of how the DM relic and promptness shift with DM mass, in \autoref{fig:h1_combo_masses}(b) we plot the relic-satisfying points for each $M_{\phi} = $ 400 GeV (brown), 500 GeV (purple), 600 GeV (red), 700 GeV (green), 800 GeV (orange), and 900 GeV (blue), fitting them with solid curves. As the DM and consequently the LQ get heavier, the $S_1 S_1 \to \text{SM SM}$ co-annihilation-dominated, almost vertical region shifts to larger $\Delta M_1$ values, but the distance between them reduces; in fact, $M_{\phi} = 800$ and $900$ GeV lines almost overlap. However, at higher $Y_{\ell S}\times Y_{\phi Q} $ values, heavier DM masses (and correspondingly heavier LQ mass) allow larger $\Delta M_1$ values to satisfy the relic, which is a consequence of the enhanced co-annihilation rate from \autoref{fig:h1-annihilation}(f). Reduced Boltzmann suppression for the heavier LQs in the initial state of the co-annihilation, via the term $e^{-x\Delta}$ with $\Delta = (M_{S_1} - M_\phi)/M_\phi$, enhances the allowed $\Delta M_{1}$ as DM mass increases \cite{Baker:2015qna}. The shaded regions with dashed-line boundaries correspond to the displaced vertex regions for each DM mass, which shift slightly towards larger $Y_{\ell S}\times Y_{\phi Q} $ and $\Delta M_{1}$ values as $M_{\phi}$ increases.

\subsubsection{LHC constraints for Hierarchy 1}

The introduction of light LQs imperatively calls for the LHC bounds on them to be considered. The current limits on $S_1$ from pair production at the LHC are around $\sim 1.5$ TeV, but they assume a 100\% branching ratio to a charged lepton and a quark, which is not the case in our scenario due to the 3-body decay with large MET contributions from $\phi$. Hence, we need to check for other LHC searches that can be sensitive to our signal. We use {\tt CheckMATE} \cite{Dercks:2016npn} to identify the ATLAS search for squarks and gluinos at the 13 TeV LHC via monolepton + jets + MET signature \cite{ATLAS:2021twp} (denoted as ATLAS 2101.01629), and the CMS search for events with two soft opposite-sign leptons and MET at 13 TeV \cite{CMS:2018kag} (denoted as CMS-SUS-16-048), as the most significant limits on our model space. The excluded regions from these searches at 95\% CL from {\tt CheckMATE} are superimposed as yellow (ATLAS 2101.01629) and red (CMS-SUS-16-048) hatched regions, respectively in \autoref{fig:h1_combo_masses}(a). The CMS search is sensitive for lower mass splitting values compared to the ATLAS one because of softer leptons being preferred. The 400 GeV DM mass region is completely excluded by both searches.\footnote{The corresponding signal region (SR) for the CMS search is the stop-specific SR with MET $> 300$ GeV in ref. \cite{CMS:2018kag}, and for the ATLAS search is the one containing 2 jets with a $b$-veto and effective mass between 700-1300 GeV, from ref. \cite{ATLAS:2021twp}.} Considering a 500 GeV DM, the lightest $\mathbb{Z}_2$-odd singlet LQ mass that is still allowed from collider searches is $\sim 550$ GeV. The CMS search rules out a small region in the 600 GeV DM mass case, but allows LQ masses $\gtrsim 630$ GeV. The Higher DM mass regions and subsequent LQ mass considerations are unaffected by the searches included in {\tt CheckMATE}. This drives the proof-of-concept for our original intention: to have a LQ setup that allows at least one mass eigenstate to be much lighter than the current LHC limits of $\sim 1.5$ TeV.

\subsection{DM Phenomenology for Hierarchy 2}

\begin{figure}[h]
    \centering
    \subfigure[]{\includegraphics[width=\linewidth]{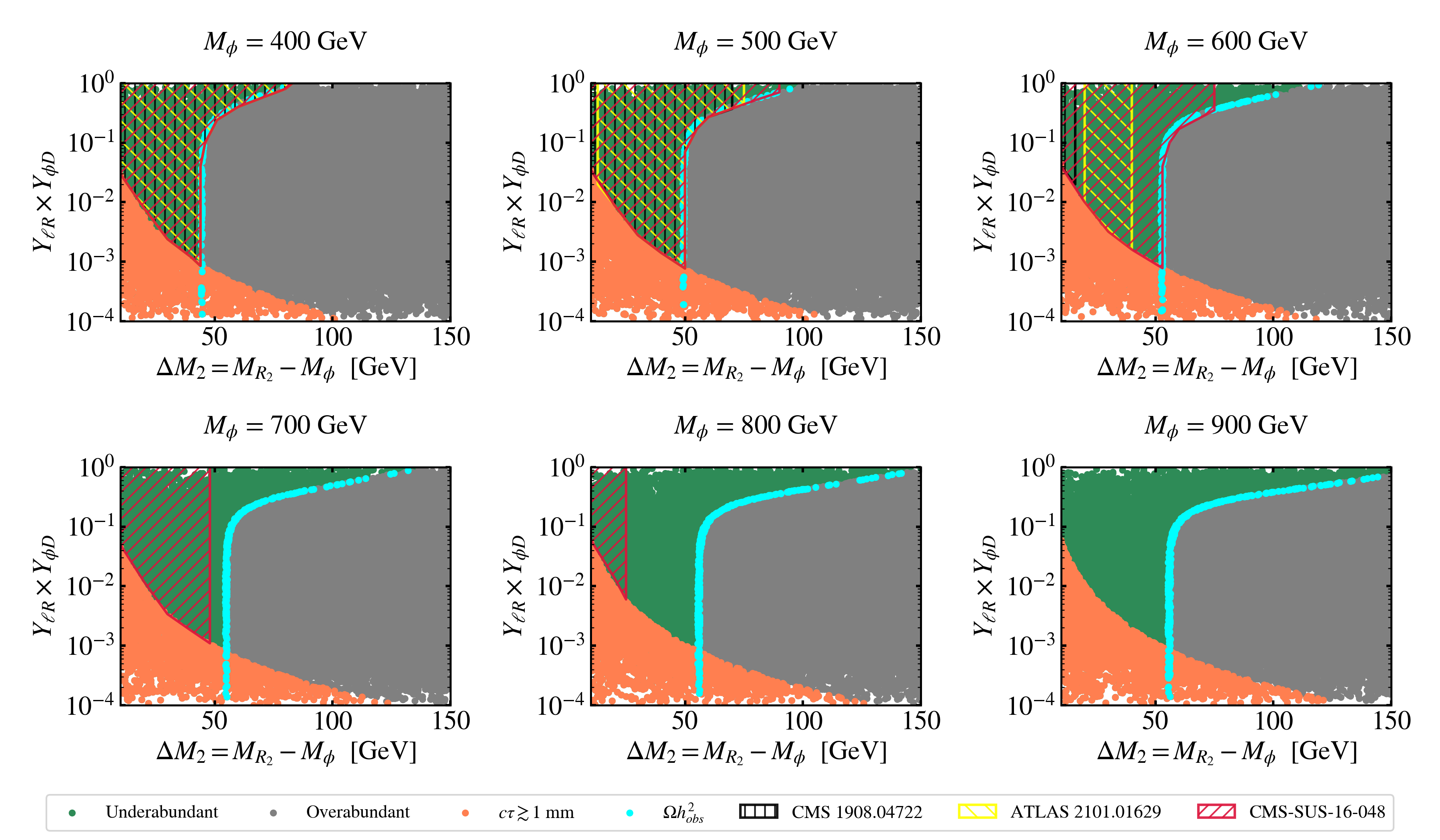}}
    \subfigure[]{\includegraphics[width=0.6\linewidth]{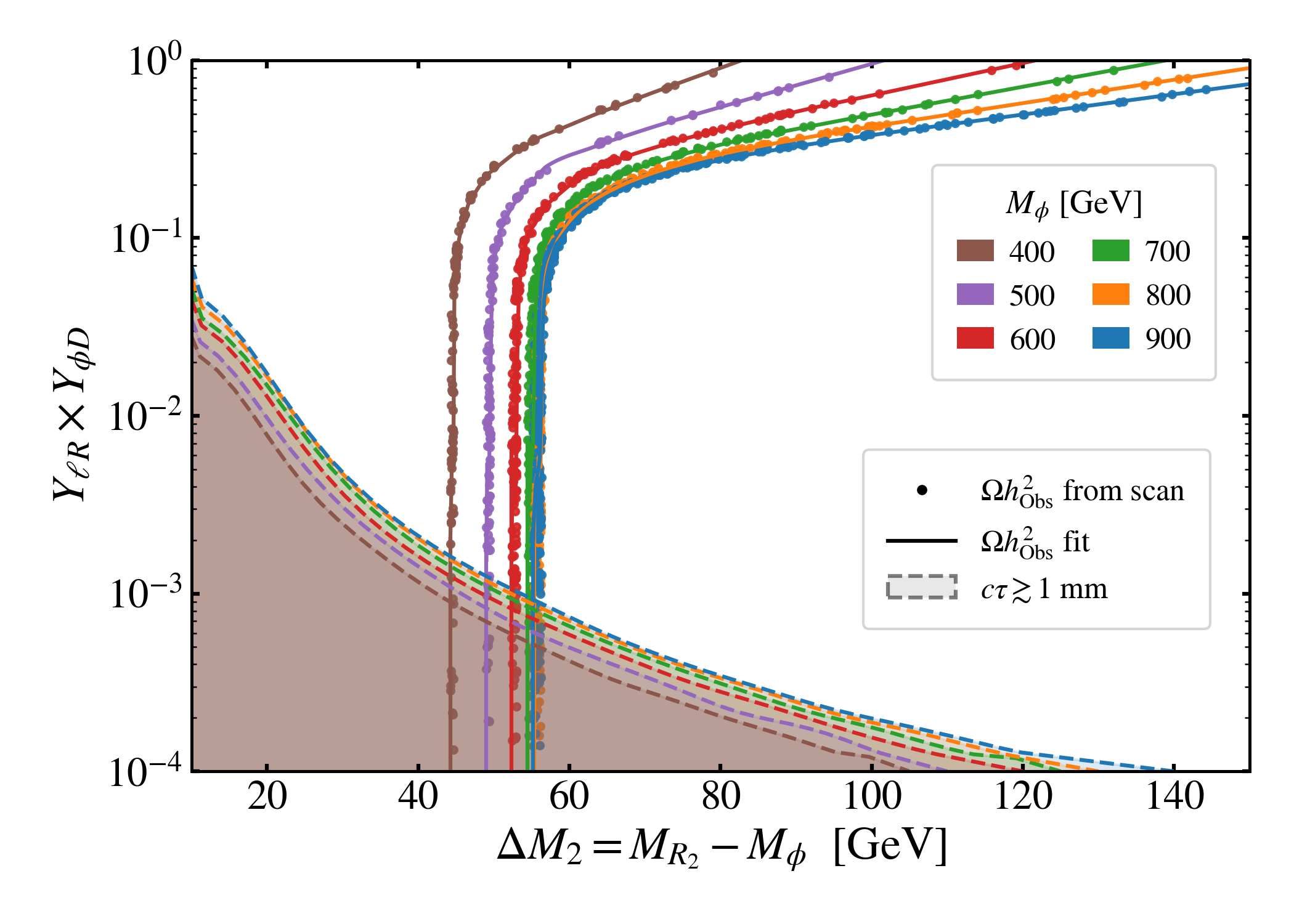}}
    \hfil
    \subfigure[]{\includegraphics[width=0.3\linewidth]{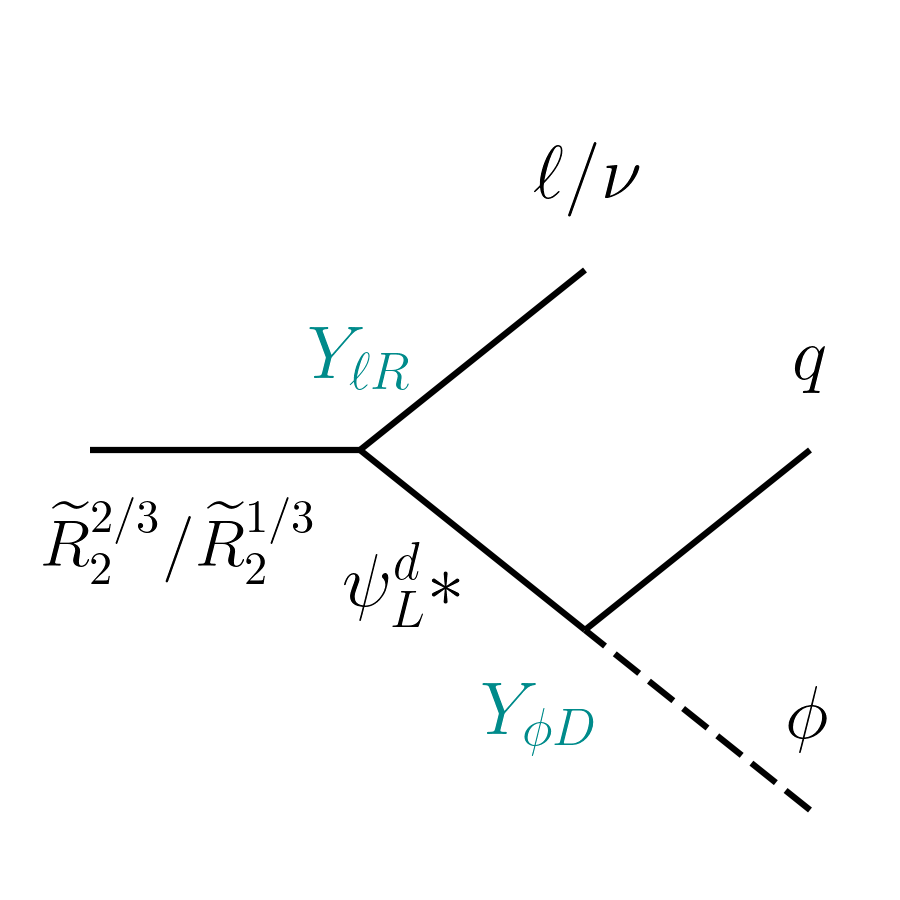}}
     \caption{(a) Scatter plots of $\Delta M_2 = M_{R_2} - M_\phi$ vs $Y_{\ell R} \times Y_{\phi D}$ for six different $M_\phi$ choices. The grey and green points are overabundant and underabundant in DM respectively, while the cyan points satisfy $\Omega h^2_{\rm Obs} =0.1198 \pm 0.0012$. The orange points lead to displaced decay of $\widetilde{R}_2$ at detectors. The black, red and yellow hatched regions mark points excluded by CMS 1908.04722 \cite{CMS:2019zmd}, CMS-SUS-16-048 \cite{CMS:2018kag} and ATLAS 2101.01629 \cite{ATLAS:2021twp} analyses, as per \texttt{CheckMATE} \cite{Dercks:2016npn}. (b) $\Delta M_{2} = M_{R_1} - M_\phi$ vs $Y_{\ell R}\times Y_{\phi D} $, showing the trends of $c \tau \gtrsim 1$ mm limits (shaded) and the $\Omega h^2_{\rm Obs}$ points for the six DM masses considered. (c) 3-body decay Feynman diagram for $\widetilde{R}_2^{2/3}/\widetilde{R}_2^{1/3}$.}
    \label{fig:h2_combo_masses}
\end{figure}

For Hierarchy 2, the annihilation and co-annihilation channels remain more-or-less the same as H1 (\autoref{fig:h1-annihilation}), but with $\widetilde{R}_2$ instead of $S_1$. The key difference is that the Yukawa couplings responsible for the DM relic and the 3-body decay of $\widetilde{R}_2$ are now $Y_{\ell R}$ and $Y_{\phi D}$. We fix $y_H = \tilde{y_H} = 0.05,\, Y_{\phi Q} = 0.01,\, Y_{\ell S} = 0.01$, keep $M_{S_1}, M_D, M_Q$ at the values from \autoref{tab:hierarchy}, and again vary $M_{R_2}$ in accordance with the same six $M_{\phi}$ values as the previous case, randomly varying $Y_{\phi D}, Y_{\ell R} \in [0.01, 1.0]$, and scanning the points using {\tt micrOMEGAs}.

Following the same approach as in H1, \autoref{fig:h2_combo_masses}(a) shows the scatter plots of $\Delta M_2 = M_{R_2} - M_\phi$ vs $Y_{\ell R} \times Y_{\phi D}$ for the six chosen DM masses. In this case, considering minimal mixing, we note that $M_{R_2^{2/3}} \approxeq M_{\eta_2^{1/3}}$, and hence we use $M_{R_2}$ as the uniform representative variable for the doublet LQ mass. We also use $\widetilde{R}_2^{1/3}$ instead of $\eta_2^{1/3}$ in this section, as the mixing is tiny and it remains almost completely doublet. The grey and green points signify DM overabundance and underabundance respectively, while the cyan points satisfy the observed relic $\Omega h_{Obs}^2 = 0.1198 \pm 0.0012$. As two LQ eigenstates offer more co-annihilation possibilities, the relic is satisfied for larger values of $\Delta M_2$ compared to H1 when the product of the Yukawa couplings is small, pushing the range to $\sim 45-55$ GeV. Owing to the near-degeneracy, both components of the doublet LQ have very close total decay widths. The orange points mark the points that do not satisfy the prompt decay criterion for $\widetilde{R}_2$, demanding a mean proper decay length $c\tau \lesssim 1$ mm, corresponding to $\Gamma(R_2) \gtrsim 10^{-13}$ GeV. Following on from H1, in \autoref{fig:h2_combo_masses}(b) we again fit the relic-satisfying points for $M_{\phi} = $ 400 GeV (brown), 500 GeV (purple), 600 GeV (red), 700 GeV (green), 800 GeV (orange), and 900 GeV (blue), with solid curves. The trends for these fits remain identical to H1, with only a shift in the $\Delta M_2$ values as mentioned. The shaded regions with dashed-line boundaries correspond to the displaced vertex regions for each DM mass with similar trends to H1 as well.

\subsubsection{LHC Constraints on Hierarchy 2}

The noteworthy difference here is that, while in H1 the singlet LQ could decay into both charged leptons and neutrinos with comparable branchidng ratios, $\widetilde{R}_2^{1/3}$ decays only to neutrinos and $\widetilde{R}_2^{2/3}$ decays only to charged leptons, with jets and MET contributions from $\phi$ in both cases, as seen in \autoref{fig:h2_combo_masses}(c). Hence, from the {\tt CheckMATE} analyses, we also obtain constraints from the CMS 2-jet + MET search \cite{CMS:2019zmd} (denoted as CMS 1908.04722) which exclusively comes from the $\widetilde{R}_2^{1/3}$ pair production, alongside the CMS-SUS-16-048 and ATLAS 2101.01629 analyses which are sensitive to the $\widetilde{R}_2^{2/3}$ pair production, similar to H1. The excluded regions from these searches at 95\% CL from {\tt CheckMATE} are superimposed in \autoref{fig:h2_combo_masses}(a) as black (CMS 1908.04722), red (CMS-SUS-16-048), and yellow (ATLAS 2101.01629) hatched regions, respectively. The CMS-SUS-16-048 bound appears much stronger in this case compared to H1, owing to enhanced cross-sections for the doublet LQ, as well as larger branching ratios to charged leptons from the  $\widetilde{R}_2^{2/3}$ pair production. The CMS 1908.04722 search\footnote{The corresponding signal region is $b$-vetoed 2-jets with $H_T$ and $H_T^{\rm miss} > 600$ GeV, from \cite{CMS:2019zmd}.} is sensitive to lower mass splitting values similar to the  CMS-SUS-16-048 one, but restricts a smaller region of the parameter space, being even weaker than the ATLAS 2101.01629 search. The 400 GeV DM mass region is again completely excluded by both searches. For the 500 GeV DM mass, a minute region with $M_{R_2} \gtrsim 590$ GeV is still allowed, but the rest of the parameter space is excluded. For $M_\phi = 600, 700,$ and 800 GeV, the allowed $M_{R_2}$ masses become $\gtrsim 675, 750,$ and 825 GeV respectively. The 900 GeV DM mass region and the corresponding $M_{R_2} \gtrsim 910$ GeV remain unconstrained from the {\tt CheckMATE} analyses.

It is important to note that this work serves as a proof-of-concept of having at least one sub-TeV LQ mass eigenstate in a scenario that can generate neutrino mass and offer a viable singlet scalar DM that evades direct detection limits via co-annihilation. The existing {\tt CheckMATE} analyses help us demonstrate this faithfully, while leaving room for dedicated recasts of more recent analogous searches from CMS for compressed SUSY spectra using the full Run-2 data \cite{CMS:2021edw, CMS:2025ttk}, which can be explored in future work. 

\section{Future Collider Prospects} \label{sec:fucol}

The discussion around the existing LHC limits on the sub-TeV dark LQ parameter space already reveals prospects for exclusion or discovery hints at the HL-LHC from pair-production and three-body decay final states involving soft leptons, jets and large MET, which can be transferred to the FCC-hh as well. The remaining unstable dark-sector particles of the model, i.e. the heavier LQ ($\widetilde{R}_2/S_1$ in H1/H2) and the VLQ states, can undergo peculiar cascade decays that always result in $\phi$ along with multiple leptons and jets. While a detailed signal-to-background analysis of such states is beyond the scope of this work, we briefly discuss their prospects at future colliders. To begin with, let us look at the decay cascades and final states from the pair production channels.

\begin{table}[h]
    \centering

    \renewcommand{\arraystretch}{1.2}
    \begin{tabular}{cccc}
    \toprule
    Hierarchy & pair production & key decays & final state \\
    \toprule
    \multirow{6}{*}{H1}&$S_1^{+1/3} S_1^{-1/3}$ & $S_1^{1/3} \to \ell/\nu \, q \,\phi$ &\makecell{ $(0-2)\ell + 2j +$MET \\ (soft leptons/jets)} \\
    \cmidrule{2-4}
    &$\widetilde{R}_2^{+1/3} \widetilde{R}_2^{-1/3}$ & \makecell{ $\widetilde{R}_2^{1/3} \to \nu \, q \, \phi$ \\ $\widetilde{R}_2^{1/3} \to h S_1^{1/3}$ (BR $\sim 2$\%) } & \makecell{$2j+$MET \\ $4b + (0-2)\ell + 2j+$ MET}\\
    \cmidrule{2-4}
    &$\widetilde{R}_2^{+2/3} \widetilde{R}_2^{-2/3}$ & $\widetilde{R}_2^{2/3} \to \ell \, q \, \phi$ & $2\ell + 2j+$MET \\
    \cmidrule{2-4}
    &$\Psi^{U}\,\overline{\Psi}^{U}$ &
    $\Psi^{U} \to q_u \, \phi$ &
    $2j + $MET, $2t+$MET \\
    \cmidrule{2-4}
    &$\Psi^{D}\,\overline{\Psi}^{D}$ &
    $\Psi^{D} \to \nu\,\widetilde R_2^{-1/3}$&
    $2j +$ MET \\
    \cmidrule{2-4}
    &$\psi^{d}\,\overline{\psi}^{d}$ &
    $\psi^{d} \to \nu S_1\,/\,\ell\,\widetilde R_2^{2/3}$ &
    $(0-2)\ell+ 2j + $MET \\
    \bottomrule
    \end{tabular}
    \caption{key decay modes $\mathbb{Z}_2$-odd particles and corresponding final states for H1.}
    \label{tab:h1_decay}
\end{table}

As we can see from \autoref{tab:h1_decay}, most of the decay scenarios following the $\mathbb{Z}_2$-odd particle pair production lead to 0-2 leptons, 2 jets, and MET in the final states. However, for the next-to-lightest dark sector particle i.e. $S_1^{1/3}$ in H1, the leptons and jets are predominantly soft for the parameter space in our consideration. In the other five cases, the $p_T$ of the jets and leptons can be larger depending on the mass gaps. The tiny $\kappa = 10^{-3}$ GeV leads to the $\widetilde{R}_2^{1/3} \to h S_1^{1/3}$ decay mode being highly suppressed. The unmixed up-type VLQ $\Psi^U$ pair-production can lead to a $2t +$ MET signal, which can be exploited. In the reversed hierarchy H2, the modes remain the same as H1, with the $S_1^{1/3} \to h \widetilde{R}_2^{1/3}$ mode opening up, with a tiny BR. 

\subsection{Lightest LQ-DM pair at a muon collider}

The primary feature of our model lies in the relatively compressed LQ-DM pair and their subsequent phenomenology, which in the case of H1 leads to $S_1^{1/3} \to \ell/\nu \, q \,\phi$ decay mode with soft leptons and jets. However, estimating the mass of either $S_1$ or $\phi$ via transverse kinematical variables at the LHC becomes a difficult task for a compressed system. As no single production via quark-gluon fusion is viable for the dark sector LQ, pair production becomes the only option, and with MET emanating from both legs, the transverse mass $M_T$ is not helpful. The \textit{stransverse mass} $M_{T_2}$ \cite{Lester:1999tx, Barr:2003rg}, a dedicated variable for pair-production scenarios like this, may be useful and has been used in both theory and experimental searches aimed at third-generation LQs at the LHC \cite{ATLAS:2019qpq, Gripaios:2010hv}. However, $M_{T_2}$ emerges more suitable for invisible decay products much lighter than the parent, which is not the case in our model. Therefore, we intend to investigate prospects of the compressed system at an experiment where we have a more controlled handle on the energy of the production process, such as a future multi-TeV MuC, where we can look beyond transverse variables to estimate the mass scales of the dark sector particles. At such a collider, the clean initial state allows us to exploit the energy of the visible system in addition to invariant mass observables. We focus on the charged lepton branch of the decay $S_1^{1/3} \to \ell q \phi$, and construct the visible system
\begin{equation}
V \equiv j\ell, \qquad m_V \equiv m_{j\ell}, \qquad E_V \equiv E_{j\ell}=E_j+E_\ell .
\end{equation}
Here, $\ell \equiv e^\pm, \mu^\pm$, and $j$ represents the hadronic jets coming from $u, d, c, s, b$ quarks. For notational simplicity, we write
\begin{equation}
M \equiv M_{S_1}, \qquad m \equiv M_\phi .
\end{equation}
The three-body decay may then be viewed kinematically as an effective two-body decay,
\begin{equation}
S_1 \to V+\phi ,
\end{equation}
where the visible system has a variable invariant mass $(m_V)$. The first useful observable is the invariant mass $(m_{j\ell})$. Since the jet and charged lepton are approximately massless, the maximum visible invariant mass is fixed by the available mass gap:
\begin{equation}
0 \leq m_{j\ell} \leq M_{S_1}-M_\phi .
\end{equation}
Thus,
\begin{equation}
m_{j\ell}^{\max}=M_{S_1}-M_\phi \equiv \Delta M .
\end{equation}
This endpoint is Lorentz invariant and therefore provides a robust handle on the compressed mass splitting. In practice, for pair-produced $S_1^{+1/3} S_1^{-1/3}$ events, we reconstruct the two decay branches and define
\begin{equation}
m_{j\ell}^{\rm max,event} = \max\left(m_{j_1\ell_1},m_{j_2\ell_2}\right).
\end{equation}
For correctly reconstructed signal events, both branches are bounded by $\Delta M$, so this event-level maximum gives a particularly direct probe of the splitting.

The second observable is the visible energy $E_{j\ell}$. Unlike $m_{j\ell}$, this is not Lorentz invariant. However, at a fixed-energy muon collider, this becomes an advantage. For the process $\mu^+ \mu^- \to S_1^{+1/3} S_1^{-1/3}$, not considering beam energy smearing, each $S_1$ carries a fixed lab-frame energy $E_{S_1}=\frac{\sqrt{s}}{2}$. Therefore its boost is fixed by its mass,
\begin{equation}
\gamma_{S_1}=\frac{\sqrt{s}}{2M_{S_1}},
\qquad
\beta_{S_1}=\sqrt{1-\frac{4M_{S_1}^2}{s}} .
\end{equation}
This is the key advantage over the LHC, where the hard partonic collision energy $\sqrt{\hat{s}}$ is unknown event by event. For a fixed value of $m_{j\ell}$, the energy and momentum of the visible system in the $S_1$ rest frame are
\begin{equation}
E_V^* =
\frac{M^2+m_{j\ell}^2-m^2}{2M}, \quad p_V^* =
\frac{\lambda^{1/2}(M^2,m_{j\ell}^2,m^2)}{2M},
\end{equation}
where
\begin{equation}
\lambda(a,b,c)=a^2+b^2+c^2-2ab-2ac-2bc .
\end{equation}
Boosting this visible system to the lab frame gives\footnote{Here, the lab frame is essentially the CM frame of the collision. In refs. \cite{Bandyopadhyay:2020jez, Bandyopadhyay:2020klr, Bandyopadhyay:2020wfv}, the reconstruction of CM frame, which is different from the lab frame, was performed in electron-photon, electron-proton, and proton-proton colliders, but for fully visible final states only.}
\begin{equation}
E_{j\ell} =
\gamma_{S_1}
\left(
E_V^* +\beta_{S_1}p_V^\ast\cos\theta^\ast
\right),
\end{equation}
where $\theta^*$ is the angle between the $j\ell$ system and the boost direction of the parent $S_1$ in the $S_1$ rest frame. Since $-1\leq \cos\theta^* \leq 1$, the allowed region in the $E_{j\ell}-m_{j\ell}$ plane is bounded by
\begin{equation}
E_{j\ell}^{\min/\max}(m_{j\ell}) =
\gamma_{S_1}
\left[
\frac{M^2+m_{j\ell}^2-m^2}{2M}
\mp
\beta_{S_1}
\frac{\lambda^{1/2}(M^2,m_{j\ell}^2,m^2)}{2M}
\right]. \label{eq:ejl_minmax}
\end{equation}
This defines a two-dimensional kinematic envelope:
\begin{equation}
0\leq m_{j\ell}\leq M-m, \quad
E_{j\ell}^{\min}(m_{j\ell})
\leq
E_{j\ell}
\leq
E_{j\ell}^{\max}(m_{j\ell}) .
\end{equation}
The two boundaries encode complementary pieces of information. The vertical edge in the $m_{j\ell}$ direction measures the mass splitting,
\begin{equation}
\Delta M = M_{S_1}-M_\phi .
\end{equation}
The height and curvature of the $E_{j\ell}$ envelope depend on the parent boost,
\begin{equation}
\gamma_{S_1}=\frac{\sqrt{s}}{2M_{S_1}},
\end{equation}
and hence on the absolute mass scale $M_{S_1}$. Therefore, for an unknown benchmark, one may first extract $\Delta M$ from the $m_{j\ell}$ endpoint, then scan over trial values of $M_{S_1}$, with
\begin{equation}
M_\phi=M_{S_1}-\Delta M,
\end{equation}
and compare the resulting envelope with the observed distribution in the $E_{j\ell}-m_{j\ell}$ plane. The best-fit envelope then gives an estimate of the parent mass, while the invisible particle mass follows from the measured splitting. The use of visible-energy endpoints at fixed-energy lepton colliders is well established in slepton-neutralino mass measurements~\cite{Martyn:2004jc,Homiller:2022iax, Battaglia:2005zf} and DM mass inference \cite{Belyaev:2021ngh}, including in compressed spectra at multi-TeV muon colliders \cite{Liu:2022byu}. Related endpoint and cusp methods have also been developed for antler topologies with two invisible particles~\cite{Han:2009ss,Han:2012nm,Christensen:2014yya}, while three-body decay endpoints have been studied in SUSY cascade decays~\cite{Lester:2006cf}, including a system similar to ours but at the LHC instead \cite{Agashe:2015wwa}.

\begin{table}[h]
    \centering
    \begin{tabular}{ccccc}
    \toprule
    $M_{S_1}$ & $M_\phi$ & $Y_{\ell S}$ & $Y_{\phi Q}$ & \makecell{$\sigma(\mu^+ \mu^- \to S_1^{+1/3} S_1^{-1/3})$\\ at 10 TeV MuC} \\
    \midrule
    850 GeV & 800 GeV & 0.50 & 0.21 & 3.87 fb \\
    \bottomrule
    \end{tabular}
    \caption{Relevant parameters of BP-H1.}
    \label{tab:bp-h1}
\end{table}

To illustrate this, we consider a demonstrative benchmark point for hierarchy 1 (BP-H1), and pair-produce the $S_1$ at a 10 TeV MuC. The parameters of the BP-H1 relevant to the discussion are given in \autoref{tab:bp-h1}. With $\Delta M_1 = 50$ GeV, the point satisfies the observed DM relic and evades LHC search limits, as obtained from the \texttt{CheckMATE} analysis in \autoref{sec:dm-h1}. The pair production at the 10 TeV MuC is dominated by the $t$-channel process involving a heavy VLQ, and hence scales as $Y_{\ell S}^4$. The primary SM backgrounds at this energy come from $t\bar{t}$ and $VVV$ ($V \equiv Z, W^\pm$), with leading order cross-sections of 1.73 and 9.58 fb respectively. \footnote{Model backgrounds, especially from $\widetilde{R}_2^{+2/3}$ pair production, can mimic the signal in some phase space. However, with small $Y_{\ell R} = Y_{\phi D} = 0.01$ in H1 coupled with the 1.5 TeV mass, the production rates are much suppressed compared to $S_1$, and hence we do not consider it in our illustrative analysis.} We generate events for the signal and background processes using \texttt{MadGraph5}, shower them with \texttt{Pythia8}, and simulate the proposed MuC detector with \texttt{Delphes} using the \texttt{delphes\_card\_MuonColliderDet.tcl}, and reconstruct jets with the Valencia algorithm with a radius $R = 0.2$ to account for the softness of our signal jets. We do not employ flavour tagging, and hence our jet definition encompasses $u,d,c,s,b$-quarks (and radiated gluons). The detector card also demands a lepton isolation criterion of $\Delta R_{j\ell} \geq 0.1$. 

For BP-H1, the $m_{j\ell}$ endpoint is predicted at $\Delta M_1 = 50$ GeV, while the parent boost at the 10 TeV MuC is:
\begin{equation}
\gamma_{S_1}=\frac{10000}{2\times850}\simeq5.88 .
\end{equation}
The corresponding upper edge of the visible-energy distribution lies at
\begin{equation}
E_{j\ell}^{\max,{\rm global}}\simeq 567 \text{ GeV}. \label{eq:ejlmax_h1bp1}
\end{equation}

\begin{figure}[h]
    \centering
    \includegraphics[width=\linewidth]{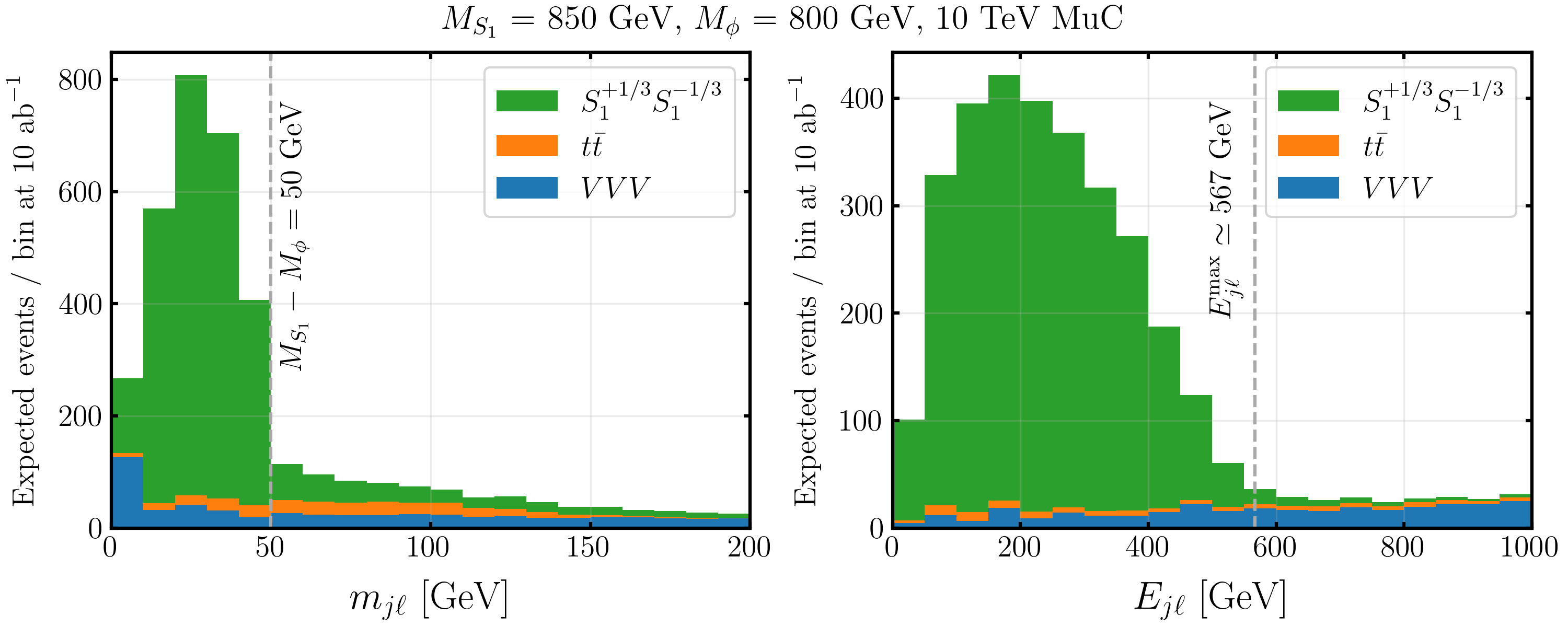}
    \caption{Stacked histograms of $m_{j\ell}$ (left) and $E_{j\ell}$ from three processes: the signal $S_1^{+1/3} S_1^{-1/3}$ (green), and backgrounds $t\bar{t}$ (orange) and $VVV$ (blue), showing expected events at the 10 TeV MuC with 10 ab$^{-1}$ luminosity. }
    \label{fig:stack_bp_h1}
\end{figure}

In \autoref{fig:stack_bp_h1} we present stacked histograms for the distributions of $m_{j\ell}$ (left panel) and $E_{j\ell}$ (right panel) at the 10 TeV MuC, of the expected events at 10 ab$^{-1}$ of luminosity for the signal process $S_1^{+1/3} S_1^{-1/3}$ (green), and the two backgrounds $t\bar{t}$ (orange) and $VVV$ (blue). The distributions are plotted after a pre-selection demanding exactly two charged leptons, and at least two jets, with the leading jet $p_T \leq 2$ TeV to reduce a decent amount of $t\bar{t}$ background. Focusing on the signal, the $j\ell$ pairs are identified using the closest $\Delta R$ values between them. A clear and sharp drop in the $m_{j\ell}$ distribution is observed at 50 GeV, corresponding to the mass splitting between $S_1$ and $\phi$, beyond which only a flat residual tail remains, arising from jet smearing and incorrect $j\ell$ pairing. The background distributions remain much flatter and quite suppressed. On the other hand, we do not observe a sharp drop-off edge in the $E_{j\ell}$ distribution for the signal, but a gradual drop-off from a peak at $\sim 200$ GeV which trickles down into a flat tail at $\sim 570$ GeV, again matching the prediction from \autoref{eq:ejlmax_h1bp1}. 

\begin{figure}[h]
    \centering
    \includegraphics[width=\linewidth]{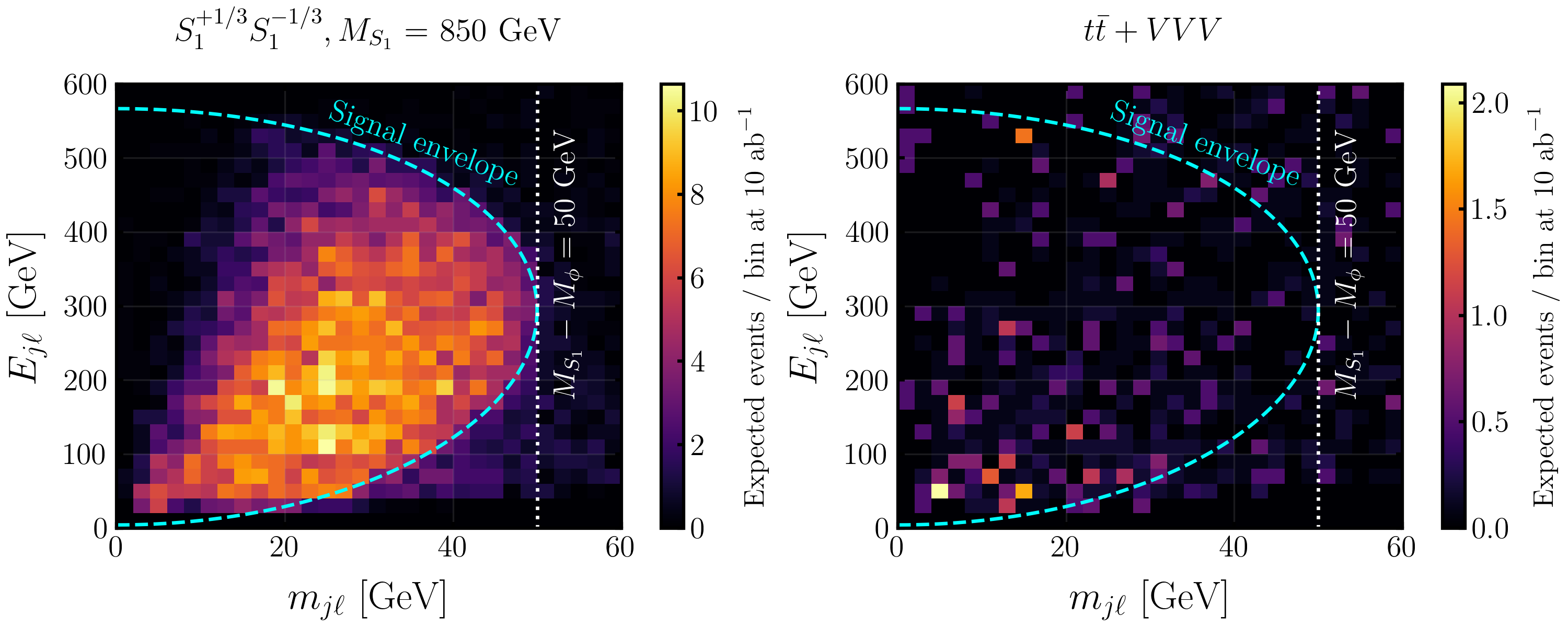}
    \caption{ 2D histograms in the $m_{j\ell} - E_{j\ell}$ plane for the signal $S_1^{+1/3} S_1^{-1/3}$ (left panel) and the combined $t\bar{t} + VVV$ background. The white dotted vertical line shows the predicted $m_{j\ell}^{\rm max} = 50$ GeV boundary, and the cyan dashed curve marks the $E_{j\ell}$ envelope.}
    \label{fig:envelope_bp_h1}
\end{figure}

We next examine the correlated event distribution in the $m_{j\ell} - E_{j\ell}$ plane. Unlike the one-dimensional $E_{j\ell}$ distribution, this plane retains the information on the visible invariant mass in each reconstructed branch and therefore allows us to test the full kinematic hypothesis. For each value of $m_{j\ell}$, the allowed visible energy is bounded by the envelope derived in \autoref{eq:ejl_minmax}: $E_{j\ell}^{\min}(m_{j\ell}) \leq
E_{j\ell}\leq E_{j\ell}^{\max}(m_{j\ell})$, with $0 \leq m_{j\ell} \leq M_{S_1}-M_\phi$. In \autoref{fig:envelope_bp_h1}, we present 2D histograms in the $m_{j\ell} - E_{j\ell}$ plane, with the left panel for the signal $S_1^{+1/3} S_1^{-1/3}$, and the right panel for the combined background. The cyan dashed curves show the ideal fixed-$\sqrt{s}$ kinematic envelope for $S_1\to j\ell\phi$, while the vertical dashed line denotes $m_{j\ell}=M_{S_1}-M_\phi=50$ GeV. In case of the signal, the right edge of the populated region aligns well with the predicted splitting of 50 GeV. The corresponding $E_{j\ell}$ envelope is also closely followed, especially near the upper boundary. The few bins with nonzero events beyond the lower envelope are accounted for by the detector smearing and incorrect $j\ell$-pairing, as mentioned before. The upper left part of the envelope unsurprisingly carries very few events, where the phase space corresponds simultaneously to very small $m_{j\ell}$ and to the visible system being emitted with high boost nearly along $S_1$ direction. For such a boosted system at low $m_{j\ell}$ and high $E_{j\ell}$, the leptons fail isolation, for which the detector-level criteria is $\Delta R_{j\ell} \geq 0.1$. For massless visible objects, one roughly obtains $\Delta R_{j\ell} \sim 2m_{j\ell}/p_{T, j\ell} \sim 2m_{j\ell}/E{j\ell}$ \cite{Altheimer:2013yza}, from which one obtains the depletion region of the envelope. Moreover, the three-body phase space suppresses the distribution near $m_{j\ell}\to0$. Therefore, the dashed curves should be interpreted as kinematic boundaries rather than regions that must be uniformly populated. The background distribution does not exhibit the same correlated structure, as it is not constrained by the same kinematics. Within this envelope, we estimate the signal significance at the 10 TeV MuC with 10 ab$^{-1}$ of target luminosity, counting 2144 signal events ($S$) and a mere 30 total SM background events ($B$), amounting to a large significance of $S/\sqrt{S+B} \simeq 46\sigma$. Hence, even assuming large systematic uncertainties and beam energy smearing, this approach keeps our compressed LQ-DM system largely discoverable at much lower luminosities of the 10 TeV MuC.

\begin{figure}[h]
\centering
\subfigure[]{\includegraphics[width=0.48\linewidth]{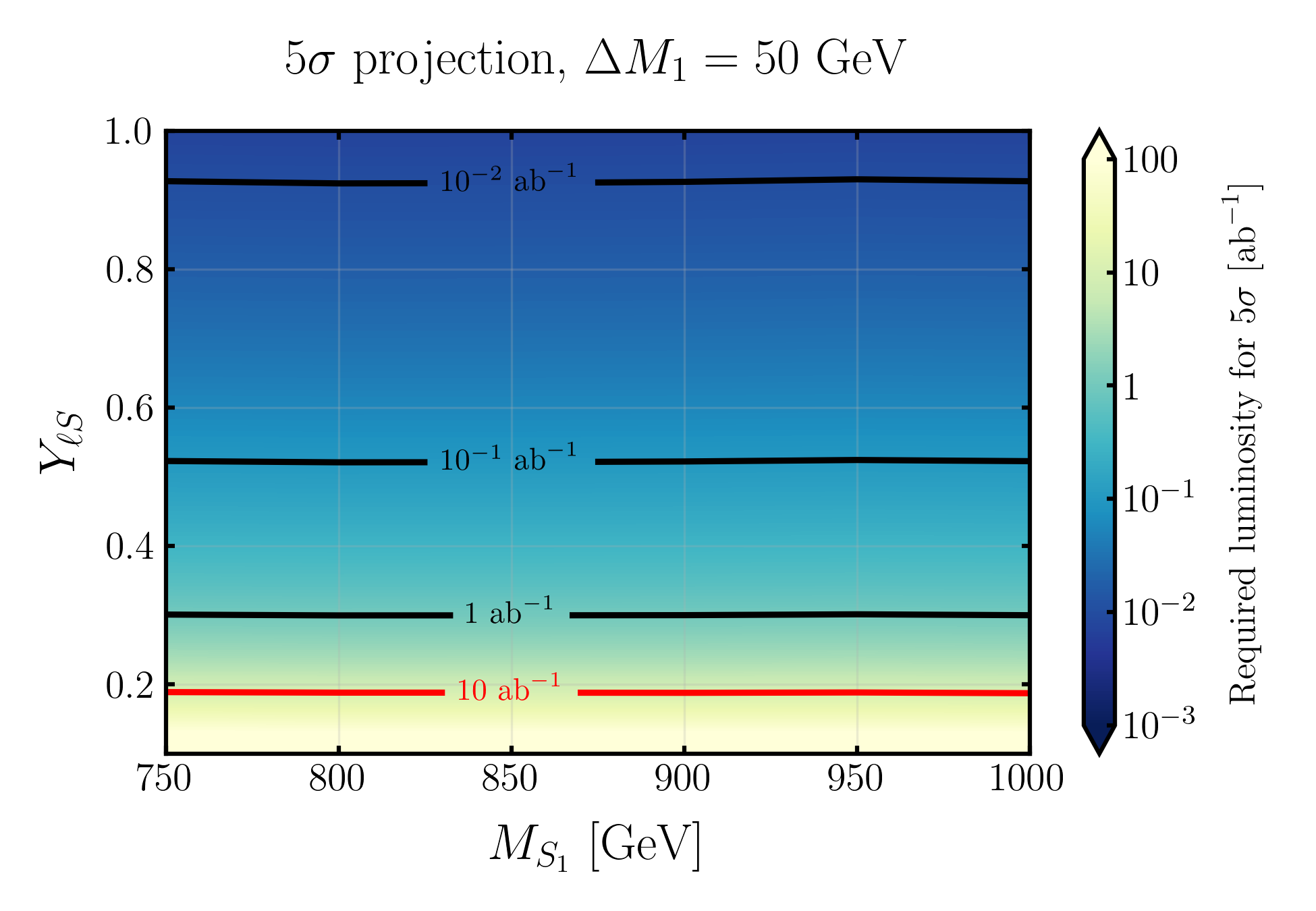}}
\subfigure[]{\includegraphics[width=0.48\linewidth]{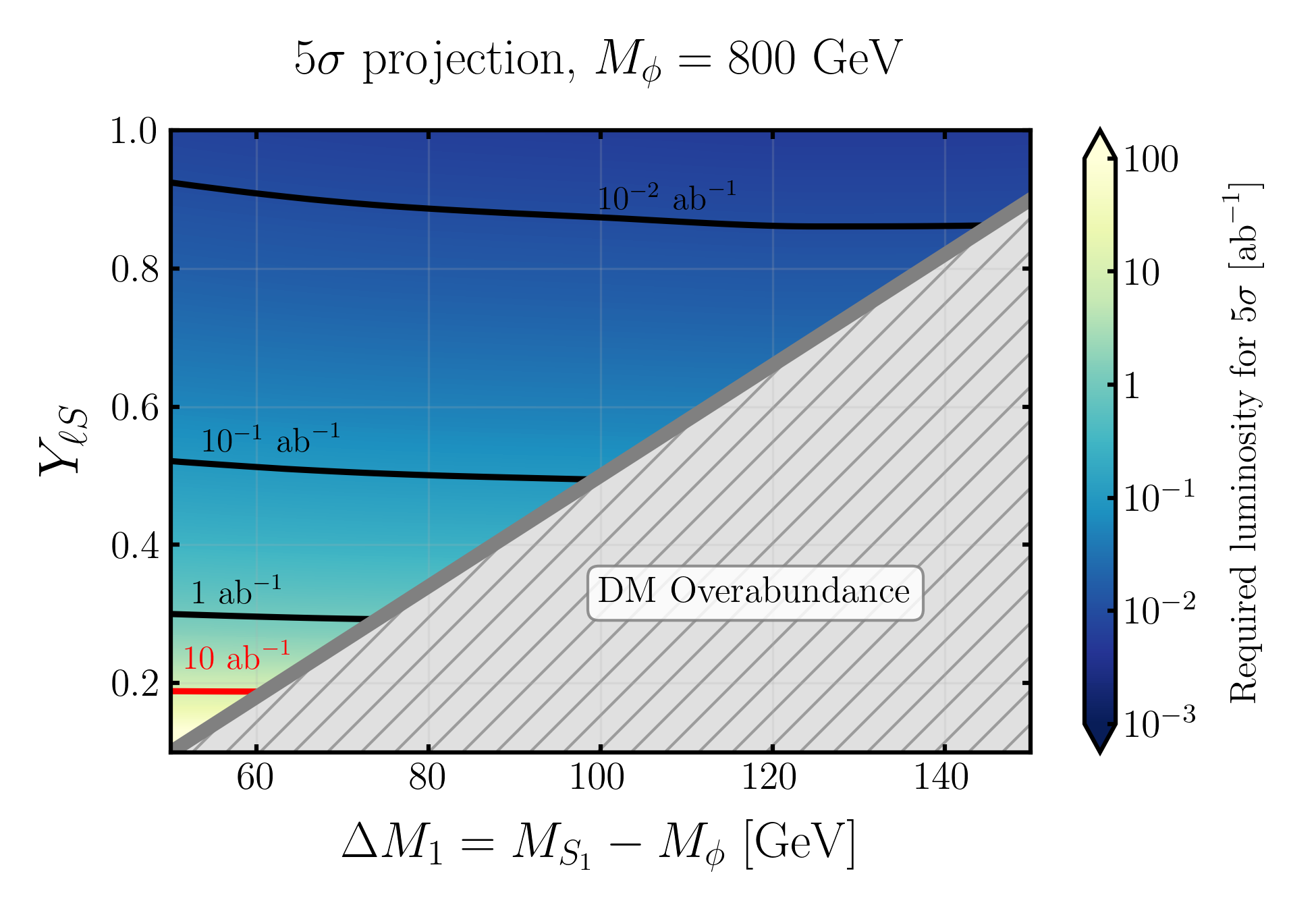}}
\caption{Luminosity projections for 5$\sigma$ probe of the visible energy envelope at the 10 TeV MuC: (a) $M_{S_1}$ vs $Y_{\ell S}$ for a fixed $\Delta M_1 = 50$ GeV, and (b) $\Delta M_1$ vs $Y_{\ell S}$ for a fixed $M_\phi = 800$ GeV.}
\label{fig:h1_proj}
\end{figure}

As mentioned previously, for $Y_{\ell S} \geq \mathcal{O}(0.1)$, the $\mu^+ \mu^- \to S_1^{+1/3} S_1^{-1/3}$ production proceeds dominantly through the $t$-channel exchange of a VLQ, with production rates scaling as $Y_{\ell S}^4$. For our region of interest in H1 i.e. $M_{S_1} \in $[750 GeV, 1 TeV] range where the \texttt{CheckMATE}-derived limits are not applicable, this production rate sees only a minimal drop at the 10 TeV MuC. Hence, this approach can also offer a handle on this key coupling, based on event rates.\footnote{More precisely, the handle is on the component of $Y_{\ell S}$ that couples to the muon, but our work assumes flavour-universal couplings for now.} With variations in $M_{S_1}$ and $\Delta M_1$, the shape and size of the visible energy envelope will change, changing the signal significance accordingly, allowing us to draw projections for a 5$\sigma$ probe. We choose to present this projection in two different settings. Firstly, in \autoref{fig:h1_proj}(a), we keep a fixed $\Delta M_1 = M_{S_1} - M_\phi = 50$ GeV, and vary $M_{S_1} \in $ [750 GeV, 1 TeV], a choice motivated by \autoref{fig:h1_combo_masses}(a) where for $M_\phi$ between $700-900$ GeV, a 50 GeV mass splitting yields the observed DM relic for $Y_{\ell S} \times Y_{\phi Q} \simeq 0.1$. This allows us to calculate the signal and background events in the visible energy envelope for the varied LQ mass points, as well as $Y_{\ell S}$ values between 0.1-1.0, with a free choice of $Y_{\phi Q}$ to balance out the product accordingly. The colour map represents the required luminosity to achieve a $5\sigma$ probe, with the red contour marking the target luminosity of 10 ab$^{-1}$ for the 10 TeV MuC. Fixed $\Delta M_1$ with increasing $M_{S_1}$ actually shrinks the visible energy envelope, leading to smaller total background counts in each step, countering the minimal drop in the production rates, which results in almost horizontal contours for the whole LQ mass range considered, allowing one to probe $Y_{\ell S}$ as low as 0.2 at the target luminosity. \autoref{fig:h1_proj}(b) presents a complementary picture, where we fix $M_\phi = 800$ GeV and vary $\Delta M_1$ between 50-150 GeV by altering the LQ mass accordingly. As per \autoref{fig:h1_combo_masses}(a), not all $Y_{\ell S}$ values are allowed for the entire $\Delta M_1$ range, and hence we keep only those values which, with the correct choice of $Y_{\phi Q} \in [0.1, 1.0]$, can keep the parameter space within the Planck DM relic limits. Increasing $\Delta M_1$ means more leptons pass the reconstruction threshold, increasing the number of raw events that pass the cut, that translate into discovery contours that hit slightly smaller $Y_{\ell S}$ values for higher $\Delta M_1$. This setup also allows 5$\sigma$ probes for $Y_{\ell S} \gtrsim 0.2$, solidifying the approach.

More generally, for an unknown mass point, the same two-dimensional strategy can be used as a mass-template procedure. The endpoint of the $m_{j\ell}$ distribution first determines $\Delta M_1 = M_{S_1}-M_\phi$. 
One can then scan over trial values of $M_{S_1}$, set $M_\phi=M_{S_1}-\Delta M_1$, and compute the corresponding envelope in the $E_{j\ell}-m_{j\ell}$ plane. Since the height and curvature of the envelope are governed by $\gamma_{S_1}=\frac{\sqrt{s}}{2M_{S_1}}$,
the best-matching envelope gives an estimate of the parent mass $M_{S_1}$, while $M_\phi$ follows from the measured splitting. This demonstrates the main advantage of the MuC setup: the fixed initial-state energy converts the visible-energy distribution into a probe of the absolute mass scale, something that is much less direct at the LHC due to the unknown partonic centre-of-mass energy. One can also expand this strategy to the heavier dark sector particles in the model, as well as to H2, owing to the similar decay cascades eventually ending up with a pair each of jets and leptons of varying hardness, alongside the invisible DM scalar. In case of H2, owing to the doublet nature of $\widetilde{R}_2^{2/3}$ and its decay modes consisting exclusively of the $\ell q \phi$ channel, we expect larger event rates for the same kind of signal as H1, potentially offering $5\sigma$ probes of the corresponding Yukawa coupling values smaller than 0.2, a more detailed analysis of which is deferred to future work.

\section{Summary and Conclusion} \label{sec:conc}

In this article, we explore the phenomenology of a novel BSM scenario featuring dark LQs, supplemented by dark VLQs and a dark scalar singlet. The model provides a radiative mechanism for generating Majorana neutrino masses, with the dark LQs and VLQs participating in the scotogenic loop. Compared with the $\mathbb{Z}_2$-even LQ scenario, the neutrino mass constraint requires the trilinear coupling between the doublet and singlet LQs and the Higgs doublet to be highly suppressed, since the light down-type quarks in the neutrino mass loop are replaced by much heavier VLQs.

The $\mathbb{Z}_2$-odd scalar singlet serves as the dark-matter candidate, as the other $\mathbb{Z}_2$-odd states are coloured. Since the LQs are $\mathbb{Z}_2$-odd, their conventional decays into SM quarks and leptons are forbidden, allowing them to evade the stringent LHC constraints on visible LQ signatures and thereby opening the possibility of sub-TeV dark LQs. Moreover, the dark LQs play an important role in achieving the observed dark-matter relic abundance through co-annihilation processes. We investigate the resulting dark-matter phenomenology and assess the prospects for probing this scenario at a future muon collider.

To realise the sub-TeV dark LQ and its effect on the DM phenomenology, we take the DM mass range of $400-900$ GeV, and consider two hierarchies in the dark sector: H1 with the singlet LQ as the next-to-lightest $\mathbb{Z}_2$-odd particle with masses within 10-150 GeV of the DM mass, and H2, where the doublet LQ states play that role. We keep the remaining dark sector particles decoupled in the $\geq 1.5$ TeV mass range. In both cases, the proximity of the LQ to $\phi$ allows efficient co-annihilation and opens up viable dark-matter masses in the few-hundred-GeV range. For H1, the observed relic abundance is typically obtained for a mass splitting $\Delta M_1 \sim 38$--$48$ GeV at moderate $Y_{\ell S}\times Y_{\phi Q}$, while larger Yukawa products allow splittings extending up to $\mathcal{O}(100)$ GeV. In H2, the additional co-annihilation channels associated with the two nearly degenerate components of $\widetilde R_2$ shift the corresponding
low-coupling region to slightly larger splittings, $\Delta M_2 \sim 45$--$55$ GeV. We further restrict ourselves to prompt LQ decays with $c\tau \lesssim 1$ mm.

We confront the relic-compatible parameter space with existing LHC searches using \texttt{CheckMATE}. The conventional $\sim 1.5$ TeV scalar LQ limits are substantially weakened because the dark LQs decay through three-body modes involving $\phi$ and
therefore produce jets, soft leptons and missing energy rather than the standard high-$p_T$ $\ell j$ final state. In H1, singlet LQs with masses around $550$ GeV can remain viable for $M_\phi=500$ GeV, while for $M_\phi=600$ GeV masses above approximately $630$ GeV survive, with $M_\phi \geq 700$ GeV and subsequent combinations of LQ masses remaining unconstrained from the \texttt{CheckMATE} analyses. The constraints are stronger in H2 because of the doublet production rate and its decay structure, although viable sub-TeV LQs remain for the higher DM masses considered. Here, for $M_\phi = 600, 700,$ and 800 GeV, $M_{R_2}\gtrsim 675, 750,$ and 825 GeV remain allowed, while $M_{R_2} \gtrsim 910$ GeV corresponding to 900 GeV DM mass is unconstrained from the {\tt CheckMATE} analyses. This demonstrates the central possibility of our setup: scalar LQs well below
the usual direct-search limits can coexist with a viable scalar DM candidate and participate in radiative neutrino mass generation.

We then explore the prospects of probing the compressed LQ--DM system at a future $10$ TeV muon collider. The fixed centre-of-mass energy allows the three-body decay $S_1\to j\ell\phi$ to be analysed using the correlated $m_{j\ell}$--$E_{j\ell}$ distribution. The invariant mass endpoint, $m_{j\ell}^{\rm max}=M_{S_1}-M_\phi$, directly probes the mass splitting, while the height and curvature of the visible-energy envelope depend on the boost of the parent LQ, and therefore contain information on its absolute mass scale. For the benchmark $M_{S_1}=850$ GeV and $M_\phi=800$ GeV, the predicted $50$ GeV endpoint and visible-energy envelope are clearly reproduced at
detector level. For this benchmark, our illustrative analysis yields 2144 signal events and 30 SM background events at $10~{\rm ab}^{-1}$, corresponding to a statistical significance of approximately $46\sigma$. More generally, the luminosity projections indicate that the target luminosity of the $10$ TeV MuC can probe values of the relevant coupling $Y_{\ell S}$ down to about $0.2$ over the representative mass and relic-compatible splitting ranges considered. The same kinematic strategy can in principle be extended
to H2, where one expects higher cross-sections and hence even finer probes of the relevant Yukawa coupling. 

Overall, our results, as a proof-of-concept, show that placing scalar LQs in a $\mathbb{Z}_2$-odd dark sector can substantially alter both their collider signatures and the phenomenology of singlet
scalar dark matter. The interplay of radiative neutrino masses, LQ-assisted co-annihilation and non-standard LQ decays provides a viable route to sub-TeV LQs, while future high-energy lepton colliders offer a particularly promising environment to discover and reconstruct such compressed dark sector spectra.

\section*{Acknowledgements}
The primary portion of AK's work was carried out under the PRIN 2022 program, funded by the Italian Ministero dell’Universit\`a e Ricerca (MUR) through grant 20227S3M3B (``Bubble Dynamics in Cosmological Phase Transitions''). PB wants to thank IOP for the visit and Phoenix 2025 for the initial phases of the project. SP and DS acknowledge the SAMKHYA High-Performance Computing Facility at the Institute
of Physics (IOP), Bhubaneswar for partial computing aid. SP also acknowledges  the Council of Scientific and Industrial Research (CSIR, File no: 09/1001(0082)/2020-EMR-I), India for funding his research during the initiation of the work.

\appendix

\section{The three-body partial decay widths of the scalar LQs}

The three-body decays of the lightest leptoquarks proceed through an off-shell
vector-like quark and can be written in the generic form
\[
X \to f\,q\,\phi ,
\]
where $X=S_1$ or $\widetilde R_2$, and $f$ denotes a charged lepton or neutrino.
Neglecting the masses of the SM fermions, the partial width for a single flavour
channel is
\begin{equation}
\Gamma_{\rm 1ch}(X\to f q\phi)
=
\frac{|Y_LY_\phi|^2\,M_F^2}
     {512\pi^3 M_X^3}
\int_{M_\phi^2}^{M_X^2} ds\,
\frac{(M_X^2-s)^2(s-M_\phi^2)^2}
     {s^2(s-M_F^2)^2},
\label{eq:LQ_3body_general}
\end{equation}
where $M_F$ denotes the mass of the intermediate VLQ. The correspondence between
the generic parameters and the two leptoquark cases is
\begin{equation}
\begin{array}{c|c|c|c}
X & M_F & Y_L & Y_\phi \\ \hline
S_1 & M_Q & Y_{\ell S} & Y_{\phi Q} \\
\widetilde R_2 & M_D & Y_{\ell R} & Y_{\phi D}
\end{array}.
\end{equation}

In the heavy-mediator limit, $M_F^2\gg M_X^2$, Eq.~\eqref{eq:LQ_3body_general}
reduces to
\begin{equation}
\Gamma_{\rm 1ch}(X\to f q\phi)
\simeq
\frac{|Y_LY_\phi|^2}{1536\pi^3}
\frac{M_X^3}{M_F^2}
F(r),
\qquad
r=\frac{M_\phi^2}{M_X^2},
\end{equation}
with
\begin{equation}
F(r)=1+9r-9r^2-r^3+6r(1+r)\ln r.
\end{equation}

\bibliographystyle{JHEPMod}
\bibliography{darklq_refs.bib}

\providecommand{\href}[2]{#2}\begingroup\raggedright\begin{thebibliography}{10}

\bibitem{ATLAS:2012yve}
{\scshape ATLAS} collaboration, G.~Aad et~al., \textit{{Observation of a new
  particle in the search for the Standard Model Higgs boson with the ATLAS
  detector at the LHC}},
  \href{https://doi.org/10.1016/j.physletb.2012.08.020}{\textit{Phys. Lett. B}
  {\bfseries 716} (2012) 1--29},
  [\href{https://arxiv.org/abs/1207.7214}{{\ttfamily 1207.7214}}].

\bibitem{CMS:2012qbp}
{\scshape CMS} collaboration, S.~Chatrchyan et~al., \textit{{Observation of a
  New Boson at a Mass of 125 GeV with the CMS Experiment at the LHC}},
  \href{https://doi.org/10.1016/j.physletb.2012.08.021}{\textit{Phys. Lett. B}
  {\bfseries 716} (2012) 30--61},
  [\href{https://arxiv.org/abs/1207.7235}{{\ttfamily 1207.7235}}].

\bibitem{Super-Kamiokande:1998kpq}
{\scshape Super-Kamiokande} collaboration, Y.~Fukuda et~al., \textit{{Evidence
  for oscillation of atmospheric neutrinos}},
  \href{https://doi.org/10.1103/PhysRevLett.81.1562}{\textit{Phys. Rev. Lett.}
  {\bfseries 81} (1998) 1562--1567},
  [\href{https://arxiv.org/abs/hep-ex/9807003}{{\ttfamily hep-ex/9807003}}].

\bibitem{Esteban:2024eli}
I.~Esteban, M.~C. Gonzalez-Garcia, M.~Maltoni, I.~Martinez-Soler, J.~P.
  Pinheiro and T.~Schwetz, \textit{{NuFit-6.0: updated global analysis of
  three-flavor neutrino oscillations}},
  \href{https://doi.org/10.1007/JHEP12(2024)216}{\textit{JHEP} {\bfseries 12}
  (2024) 216}, [\href{https://arxiv.org/abs/2410.05380}{{\ttfamily
  2410.05380}}].

\bibitem{Rubin:1970zza}
V.~C. Rubin and W.~K. Ford, Jr., \textit{{Rotation of the Andromeda Nebula from
  a Spectroscopic Survey of Emission Regions}},
  \href{https://doi.org/10.1086/150317}{\textit{Astrophys. J.} {\bfseries 159}
  (1970) 379--403}.

\bibitem{Zwicky:1933gu}
F.~Zwicky, \textit{{Die Rotverschiebung von extragalaktischen Nebeln}},
  \href{https://doi.org/10.1007/s10714-008-0707-4}{\textit{Helv. Phys. Acta}
  {\bfseries 6} (1933) 110--127}.

\bibitem{Jee:2007nx}
M.~J. Jee et~al., \textit{{Discovery of a Ringlike Dark Matter Structure in the
  Core of the Galaxy Cluster Cl 0024+17}},
  \href{https://doi.org/10.1086/517498}{\textit{Astrophys. J.} {\bfseries 661}
  (2007) 728--749}, [\href{https://arxiv.org/abs/0705.2171}{{\ttfamily
  0705.2171}}].

\bibitem{Planck:2018vyg}
{\scshape Planck} collaboration, N.~Aghanim et~al., \textit{{Planck 2018
  results. VI. Cosmological parameters}},
  \href{https://doi.org/10.1051/0004-6361/201833910}{\textit{Astron.
  Astrophys.} {\bfseries 641} (2020) A6},
  [\href{https://arxiv.org/abs/1807.06209}{{\ttfamily 1807.06209}}]. [Erratum:
  Astron.Astrophys. 652, C4 (2021)].

\bibitem{ACT:2020gnv}
{\scshape ACT} collaboration, S.~Aiola et~al., \textit{{The Atacama Cosmology
  Telescope: DR4 Maps and Cosmological Parameters}},
  \href{https://doi.org/10.1088/1475-7516/2020/12/047}{\textit{JCAP} {\bfseries
  12} (2020) 047}, [\href{https://arxiv.org/abs/2007.07288}{{\ttfamily
  2007.07288}}].

\bibitem{Tao:1996vb}
Z.-j. Tao, \textit{{Radiative seesaw mechanism at weak scale}},
  \href{https://doi.org/10.1103/PhysRevD.54.5693}{\textit{Phys. Rev. D}
  {\bfseries 54} (1996) 5693--5697},
  [\href{https://arxiv.org/abs/hep-ph/9603309}{{\ttfamily hep-ph/9603309}}].

\bibitem{Ma:2006km}
E.~Ma, \textit{{Verifiable radiative seesaw mechanism of neutrino mass and dark
  matter}}, \href{https://doi.org/10.1103/PhysRevD.73.077301}{\textit{Phys.
  Rev. D} {\bfseries 73} (2006) 077301},
  [\href{https://arxiv.org/abs/hep-ph/0601225}{{\ttfamily hep-ph/0601225}}].

\bibitem{Bonnet:2012kz}
F.~Bonnet, M.~Hirsch, T.~Ota and W.~Winter, \textit{{Systematic study of the
  d=5 Weinberg operator at one-loop order}},
  \href{https://doi.org/10.1007/JHEP07(2012)153}{\textit{JHEP} {\bfseries 07}
  (2012) 153}, [\href{https://arxiv.org/abs/1204.5862}{{\ttfamily 1204.5862}}].

\bibitem{Restrepo:2013aga}
D.~Restrepo, O.~Zapata and C.~E. Yaguna, \textit{{Models with radiative
  neutrino masses and viable dark matter candidates}},
  \href{https://doi.org/10.1007/JHEP11(2013)011}{\textit{JHEP} {\bfseries 11}
  (2013) 011}, [\href{https://arxiv.org/abs/1308.3655}{{\ttfamily 1308.3655}}].

\bibitem{Hirsch:2013ola}
M.~Hirsch, R.~A. Lineros, S.~Morisi, J.~Palacio, N.~Rojas and J.~W.~F. Valle,
  \textit{{WIMP dark matter as radiative neutrino mass messenger}},
  \href{https://doi.org/10.1007/JHEP10(2013)149}{\textit{JHEP} {\bfseries 10}
  (2013) 149}, [\href{https://arxiv.org/abs/1307.8134}{{\ttfamily 1307.8134}}].

\bibitem{Cai:2017jrq}
Y.~Cai, J.~Herrero-Garc{\'\i}a, M.~A. Schmidt, A.~Vicente and R.~R. Volkas,
  \textit{{From the trees to the forest: a review of radiative neutrino mass
  models}}, \href{https://doi.org/10.3389/fphy.2017.00063}{\textit{Front. in
  Phys.} {\bfseries 5} (2017) 63},
  [\href{https://arxiv.org/abs/1706.08524}{{\ttfamily 1706.08524}}].

\bibitem{Avila:2019hhv}
I.~M. {\'A}vila, V.~De~Romeri, L.~Duarte and J.~W.~F. Valle,
  \textit{{Phenomenology of scotogenic scalar dark matter}},
  \href{https://doi.org/10.1140/epjc/s10052-020-08480-z}{\textit{Eur. Phys. J.
  C} {\bfseries 80} (2020) 908},
  [\href{https://arxiv.org/abs/1910.08422}{{\ttfamily 1910.08422}}].

\bibitem{Karan:2023adm}
A.~Karan, S.~Sadhukhan and J.~W.~F. Valle, \textit{{Phenomenological profile of
  scotogenic fermionic dark matter}},
  \href{https://doi.org/10.1007/JHEP12(2023)185}{\textit{JHEP} {\bfseries 12}
  (2023) 185}, [\href{https://arxiv.org/abs/2308.09135}{{\ttfamily
  2308.09135}}].

\bibitem{Avila:2025qsc}
I.~M. {\'A}vila, A.~Karan, S.~Mandal, S.~Sadhukhan and J.~W.~F. Valle,
  \textit{{Dark matter as the source of neutrino mass: Theory overview and
  experimental prospects}},
  \href{https://doi.org/10.1016/j.physrep.2026.02.003}{\textit{Phys. Rept.}
  {\bfseries 1173} (2026) 1--81},
  [\href{https://arxiv.org/abs/2506.24027}{{\ttfamily 2506.24027}}].

\bibitem{Dorsner:2016wpm}
I.~Dor{\v{s}}ner, S.~Fajfer, A.~Greljo, J.~F. Kamenik and N.~Ko{\v{s}}nik,
  \textit{{Physics of leptoquarks in precision experiments and at particle
  colliders}},
  \href{https://doi.org/10.1016/j.physrep.2016.06.001}{\textit{Phys. Rept.}
  {\bfseries 641} (2016) 1--68},
  [\href{https://arxiv.org/abs/1603.04993}{{\ttfamily 1603.04993}}].

\bibitem{Babu:2019mfe}
K.~S. Babu, P.~S.~B. Dev, S.~Jana and A.~Thapa, \textit{{Non-Standard
  Interactions in Radiative Neutrino Mass Models}},
  \href{https://doi.org/10.1007/JHEP03(2020)006}{\textit{JHEP} {\bfseries 03}
  (2020) 006}, [\href{https://arxiv.org/abs/1907.09498}{{\ttfamily
  1907.09498}}].

\bibitem{Zhang:2021dgl}
D.~Zhang, \textit{{Radiative neutrino masses, lepton flavor mixing and muon g
  {\ensuremath{-}} 2 in a leptoquark model}},
  \href{https://doi.org/10.1007/JHEP07(2021)069}{\textit{JHEP} {\bfseries 07}
  (2021) 069}, [\href{https://arxiv.org/abs/2105.08670}{{\ttfamily
  2105.08670}}].

\bibitem{Parashar:2022wrd}
S.~Parashar, A.~Karan, Avnish, P.~Bandyopadhyay and K.~Ghosh,
  \textit{{Phenomenology of scalar leptoquarks at the LHC in explaining the
  radiative neutrino masses, muon g-2, and lepton flavor violating
  observables}},
  \href{https://doi.org/10.1103/PhysRevD.106.095040}{\textit{Phys. Rev. D}
  {\bfseries 106} (2022) 095040},
  [\href{https://arxiv.org/abs/2209.05890}{{\ttfamily 2209.05890}}].

\bibitem{Dev:2024tto}
P.~S.~B. Dev, S.~Goswami, C.~Majumdar and D.~Pachhar, \textit{{Neutrinoless
  double beta decay from scalar leptoquarks: interplay with neutrino mass and
  flavor physics}},
  \href{https://doi.org/10.1007/JHEP01(2025)004}{\textit{JHEP} {\bfseries 01}
  (2025) 004}, [\href{https://arxiv.org/abs/2407.04670}{{\ttfamily
  2407.04670}}].

\bibitem{ATLAS:2020dsk}
{\scshape ATLAS} collaboration, G.~Aad et~al., \textit{{Search for pairs of
  scalar leptoquarks decaying into quarks and electrons or muons in $ \sqrt{s}
  $ = 13 TeV $pp$ collisions with the ATLAS detector}},
  \href{https://doi.org/10.1007/JHEP10(2020)112}{\textit{JHEP} {\bfseries 10}
  (2020) 112}, [\href{https://arxiv.org/abs/2006.05872}{{\ttfamily
  2006.05872}}].

\bibitem{ATLAS:2019qpq}
{\scshape ATLAS} collaboration, M.~Aaboud et~al., \textit{{Searches for
  third-generation scalar leptoquarks in $\sqrt{s}$ = 13 TeV pp collisions with
  the ATLAS detector}},
  \href{https://doi.org/10.1007/JHEP06(2019)144}{\textit{JHEP} {\bfseries 06}
  (2019) 144}, [\href{https://arxiv.org/abs/1902.08103}{{\ttfamily
  1902.08103}}].

\bibitem{Queiroz:2014pra}
F.~S. Queiroz, K.~Sinha and A.~Strumia, \textit{{Leptoquarks, Dark Matter, and
  Anomalous LHC Events}},
  \href{https://doi.org/10.1103/PhysRevD.91.035006}{\textit{Phys. Rev. D}
  {\bfseries 91} (2015) 035006},
  [\href{https://arxiv.org/abs/1409.6301}{{\ttfamily 1409.6301}}].

\bibitem{Choi:2018stw}
S.-M. Choi, Y.-J. Kang, H.~M. Lee and T.-G. Ro, \textit{{Lepto-Quark Portal
  Dark Matter}}, \href{https://doi.org/10.1007/JHEP10(2018)104}{\textit{JHEP}
  {\bfseries 10} (2018) 104},
  [\href{https://arxiv.org/abs/1807.06547}{{\ttfamily 1807.06547}}].

\bibitem{Mandal:2018czf}
R.~Mandal, \textit{{Fermionic dark matter in leptoquark portal}},
  \href{https://doi.org/10.1140/epjc/s10052-018-6192-3}{\textit{Eur. Phys. J.
  C} {\bfseries 78} (2018) 726},
  [\href{https://arxiv.org/abs/1808.07844}{{\ttfamily 1808.07844}}].

\bibitem{DEramo:2020sqv}
F.~D'Eramo, N.~Ko{\v{s}}nik, F.~Pobbe, A.~Smolkovi{\v{c}} and O.~Sumensari,
  \textit{{Leptoquarks and real singlets: A richer scalar sector behind the
  origin of dark matter}},
  \href{https://doi.org/10.1103/PhysRevD.104.015035}{\textit{Phys. Rev. D}
  {\bfseries 104} (2021) 015035},
  [\href{https://arxiv.org/abs/2012.05743}{{\ttfamily 2012.05743}}].

\bibitem{Belanger:2021smw}
G.~Belanger et~al., \textit{{Leptoquark manoeuvres in the dark: a simultaneous
  solution of the dark matter problem and the $ {R}_{D^{\left(\ast \right)}} $
  anomalies}}, \href{https://doi.org/10.1007/JHEP02(2022)042}{\textit{JHEP}
  {\bfseries 02} (2022) 042},
  [\href{https://arxiv.org/abs/2111.08027}{{\ttfamily 2111.08027}}].

\bibitem{Ghosh:2023xbj}
N.~Ghosh, S.~K. Rai and T.~Samui, \textit{{Search for a leptoquark and
  vector-like lepton in a muon collider}},
  \href{https://doi.org/10.1016/j.nuclphysb.2024.116564}{\textit{Nucl. Phys. B}
  {\bfseries 1004} (2024) 116564},
  [\href{https://arxiv.org/abs/2309.07583}{{\ttfamily 2309.07583}}].

\bibitem{Goncalves:2025snm}
M.~Gon{\c{c}}alves, M.~M{\"u}hlleitner, R.~Santos and T.~Trindade,
  \textit{{Dark matter in multi-singlet extensions of the Standard Model}},
  \href{https://doi.org/10.1007/JHEP03(2026)157}{\textit{JHEP} {\bfseries 03}
  (2026) 157}, [\href{https://arxiv.org/abs/2505.07753}{{\ttfamily
  2505.07753}}].

\bibitem{LZ:2024zvo}
{\scshape LZ} collaboration, J.~Aalbers et~al., \textit{{Dark Matter Search
  Results from 4.2{\,}{\,}Tonne-Years of Exposure of the LUX-ZEPLIN (LZ)
  Experiment}}, \href{https://doi.org/10.1103/4dyc-z8zf}{\textit{Phys. Rev.
  Lett.} {\bfseries 135} (2025) 011802},
  [\href{https://arxiv.org/abs/2410.17036}{{\ttfamily 2410.17036}}].

\bibitem{CMS:2018ncu}
{\scshape CMS} collaboration, A.~M. Sirunyan et~al., \textit{{Search for pair
  production of first-generation scalar leptoquarks at $\sqrt{s} =$ 13 TeV}},
  \href{https://doi.org/10.1103/PhysRevD.99.052002}{\textit{Phys. Rev. D}
  {\bfseries 99} (2019) 052002},
  [\href{https://arxiv.org/abs/1811.01197}{{\ttfamily 1811.01197}}].

\bibitem{CMS:2018lab}
{\scshape CMS} collaboration, A.~M. Sirunyan et~al., \textit{{Search for pair
  production of second-generation leptoquarks at $\sqrt{s}=$ 13 TeV}},
  \href{https://doi.org/10.1103/PhysRevD.99.032014}{\textit{Phys. Rev. D}
  {\bfseries 99} (2019) 032014},
  [\href{https://arxiv.org/abs/1808.05082}{{\ttfamily 1808.05082}}].

\bibitem{ATLAS:2021twp}
{\scshape ATLAS} collaboration, G.~Aad et~al., \textit{{Search for squarks and
  gluinos in final states with one isolated lepton, jets, and missing
  transverse momentum at $\sqrt{s}=13$~ with the ATLAS detector}},
  \href{https://doi.org/10.1140/epjc/s10052-021-09748-8}{\textit{Eur. Phys. J.
  C} {\bfseries 81} (2021) 600},
  [\href{https://arxiv.org/abs/2101.01629}{{\ttfamily 2101.01629}}]. [Erratum:
  Eur.Phys.J.C 81, 956 (2021)].

\bibitem{ATLAS:2023uox}
{\scshape ATLAS} collaboration, G.~Aad et~al., \textit{{Search for pair
  production of third-generation leptoquarks decaying into a bottom quark and a
  $\tau $-lepton with the ATLAS detector}},
  \href{https://doi.org/10.1140/epjc/s10052-023-12104-7}{\textit{Eur. Phys. J.
  C} {\bfseries 83} (2023) 1075},
  [\href{https://arxiv.org/abs/2303.01294}{{\ttfamily 2303.01294}}].

\bibitem{CMS:2024bnj}
{\scshape CMS} collaboration, A.~Hayrapetyan et~al., \textit{{Search for pair
  production of scalar and vector leptoquarks decaying to muons and bottom
  quarks in proton-proton collisions at s=13{\,}{\,}TeV}},
  \href{https://doi.org/10.1103/PhysRevD.109.112003}{\textit{Phys. Rev. D}
  {\bfseries 109} (2024) 112003},
  [\href{https://arxiv.org/abs/2402.08668}{{\ttfamily 2402.08668}}].

\bibitem{CMS:2018kag}
{\scshape CMS} collaboration, A.~M. Sirunyan et~al., \textit{{Search for new
  physics in events with two soft oppositely charged leptons and missing
  transverse momentum in proton-proton collisions at $\sqrt{s}=$ 13 TeV}},
  \href{https://doi.org/10.1016/j.physletb.2018.05.062}{\textit{Phys. Lett. B}
  {\bfseries 782} (2018) 440--467},
  [\href{https://arxiv.org/abs/1801.01846}{{\ttfamily 1801.01846}}].

\bibitem{CMS:2019zmd}
{\scshape CMS} collaboration, T.~C. Collaboration et~al., \textit{{Search for
  supersymmetry in proton-proton collisions at 13 TeV in final states with jets
  and missing transverse momentum}},
  \href{https://doi.org/10.1007/JHEP10(2019)244}{\textit{JHEP} {\bfseries 10}
  (2019) 244}, [\href{https://arxiv.org/abs/1908.04722}{{\ttfamily
  1908.04722}}].

\bibitem{Martyn:2004jc}
H.-U. Martyn, \textit{{Detection of sleptons at a linear collider in models
  with small slepton-neutralino mass differences}},  in \textit{{International
  Conference on Linear Colliders (LCWS 04)}}, 8, 2004,
  \href{https://arxiv.org/abs/hep-ph/0408226}{{\ttfamily hep-ph/0408226}}.

\bibitem{Han:2009ss}
T.~Han, I.-W. Kim and J.~Song, \textit{{Kinematic Cusps: Determining the
  Missing Particle Mass at Colliders}},
  \href{https://doi.org/10.1016/j.physletb.2010.09.010}{\textit{Phys. Lett. B}
  {\bfseries 693} (2010) 575--579},
  [\href{https://arxiv.org/abs/0906.5009}{{\ttfamily 0906.5009}}].

\bibitem{Han:2012nm}
T.~Han, I.-W. Kim and J.~Song, \textit{{Kinematic Cusps With Two Missing
  Particles I: Antler Decay Topology}},
  \href{https://doi.org/10.1103/PhysRevD.87.035003}{\textit{Phys. Rev. D}
  {\bfseries 87} (2013) 035003},
  [\href{https://arxiv.org/abs/1206.5633}{{\ttfamily 1206.5633}}].

\bibitem{Christensen:2014yya}
N.~D. Christensen, T.~Han, Z.~Qian, J.~Sayre, J.~Song and Stefanus,
  \textit{{Determining the Dark Matter Particle Mass through Antler Topology
  Processes at Lepton Colliders}},
  \href{https://doi.org/10.1103/PhysRevD.90.114029}{\textit{Phys. Rev. D}
  {\bfseries 90} (2014) 114029},
  [\href{https://arxiv.org/abs/1404.6258}{{\ttfamily 1404.6258}}].

\bibitem{InternationalMuonCollider:2025sys}
{\scshape International Muon Collider} collaboration, C.~Accettura et~al.,
  \textit{{The Muon Collider}},
  \href{https://arxiv.org/abs/2504.21417}{{\ttfamily 2504.21417}}.

\bibitem{Dey:2025niu}
S.~Dey and S.~K. Rai, \textit{{Exploring fermionic dark matter in the presence
  of scalar leptoquarks}},
  \href{https://doi.org/10.1140/epjc/s10052-026-15682-4}{\textit{Eur. Phys. J.
  C} {\bfseries 86} (2026) 414},
  [\href{https://arxiv.org/abs/2509.02744}{{\ttfamily 2509.02744}}].

\bibitem{Bandyopadhyay:2016oif}
P.~Bandyopadhyay and R.~Mandal, \textit{{Vacuum stability in an extended
  standard model with a leptoquark}},
  \href{https://doi.org/10.1103/PhysRevD.95.035007}{\textit{Phys. Rev. D}
  {\bfseries 95} (2017) 035007},
  [\href{https://arxiv.org/abs/1609.03561}{{\ttfamily 1609.03561}}].

\bibitem{Bandyopadhyay:2021kue}
P.~Bandyopadhyay, S.~Jangid and A.~Karan, \textit{{Constraining scalar doublet
  and triplet leptoquarks with vacuum stability and perturbativity}},
  \href{https://doi.org/10.1140/epjc/s10052-022-10418-6}{\textit{Eur. Phys. J.
  C} {\bfseries 82} (2022) 516},
  [\href{https://arxiv.org/abs/2111.03872}{{\ttfamily 2111.03872}}].

\bibitem{Gonderinger:2009jp}
M.~Gonderinger, Y.~Li, H.~Patel and M.~J. Ramsey-Musolf, \textit{{Vacuum
  Stability, Perturbativity, and Scalar Singlet Dark Matter}},
  \href{https://doi.org/10.1007/JHEP01(2010)053}{\textit{JHEP} {\bfseries 01}
  (2010) 053}, [\href{https://arxiv.org/abs/0910.3167}{{\ttfamily 0910.3167}}].

\bibitem{Elbers:2025vlz}
W.~Elbers et~al., \textit{{Constraints on neutrino physics from DESI DR2 BAO
  and DR1 full shape}}, \href{https://doi.org/10.1103/w9pk-xsk7}{\textit{Phys.
  Rev. D} {\bfseries 112} (2025) 083513},
  [\href{https://arxiv.org/abs/2503.14744}{{\ttfamily 2503.14744}}].

\bibitem{KATRIN:2024cdt}
{\scshape KATRIN} collaboration, M.~Aker et~al., \textit{{Direct neutrino-mass
  measurement based on 259 days of KATRIN data}},
  \href{https://doi.org/10.1126/science.adq9592}{\textit{Science} {\bfseries
  388} (2025) adq9592}, [\href{https://arxiv.org/abs/2406.13516}{{\ttfamily
  2406.13516}}].

\bibitem{Cirelli:2013ufw}
M.~Cirelli, E.~Del~Nobile and P.~Panci, \textit{{Tools for model-independent
  bounds in direct dark matter searches}},
  \href{https://doi.org/10.1088/1475-7516/2013/10/019}{\textit{JCAP} {\bfseries
  10} (2013) 019}, [\href{https://arxiv.org/abs/1307.5955}{{\ttfamily
  1307.5955}}].

\bibitem{Belanger:2013oya}
G.~Belanger, F.~Boudjema, A.~Pukhov and A.~Semenov, \textit{{micrOMEGAs$\_$3: A
  program for calculating dark matter observables}},
  \href{https://doi.org/10.1016/j.cpc.2013.10.016}{\textit{Comput. Phys.
  Commun.} {\bfseries 185} (2014) 960--985},
  [\href{https://arxiv.org/abs/1305.0237}{{\ttfamily 1305.0237}}].

\bibitem{Yu:2024xsy}
Y.~Yu, T.-P. Tang and L.~Feng, \textit{{New constraints on singlet scalar dark
  matter model with LZ, invisible Higgs decay and gamma-ray line
  observations}},
  \href{https://doi.org/10.1016/j.nuclphysb.2025.116910}{\textit{Nucl. Phys. B}
  {\bfseries 1015} (2025) 116910},
  [\href{https://arxiv.org/abs/2410.21089}{{\ttfamily 2410.21089}}].

\bibitem{EscuderoAbenza:2025cfj}
M.~Escudero~Abenza and T.~Hambye, \textit{{The simplest dark matter model at
  the edge of perturbativity}},
  \href{https://doi.org/10.1016/j.physletb.2025.139696}{\textit{Phys. Lett. B}
  {\bfseries 868} (2025) 139696},
  [\href{https://arxiv.org/abs/2505.02408}{{\ttfamily 2505.02408}}].

\bibitem{Alguero:2023zol}
G.~Alguero, G.~Belanger, F.~Boudjema, S.~Chakraborti, A.~Goudelis, S.~Kraml
  et~al., \textit{{micrOMEGAs 6.0: N-component dark matter}},
  \href{https://doi.org/10.1016/j.cpc.2024.109133}{\textit{Comput. Phys.
  Commun.} {\bfseries 299} (2024) 109133},
  [\href{https://arxiv.org/abs/2312.14894}{{\ttfamily 2312.14894}}].

\bibitem{Hisano:2015bma}
J.~Hisano, R.~Nagai and N.~Nagata, \textit{{Effective Theories for Dark Matter
  Nucleon Scattering}},
  \href{https://doi.org/10.1007/JHEP05(2015)037}{\textit{JHEP} {\bfseries 05}
  (2015) 037}, [\href{https://arxiv.org/abs/1502.02244}{{\ttfamily
  1502.02244}}].

\bibitem{Dercks:2016npn}
D.~Dercks, N.~Desai, J.~S. Kim, K.~Rolbiecki, J.~Tattersall and T.~Weber,
  \textit{{CheckMATE 2: From the model to the limit}},
  \href{https://doi.org/10.1016/j.cpc.2017.08.021}{\textit{Comput. Phys.
  Commun.} {\bfseries 221} (2017) 383--418},
  [\href{https://arxiv.org/abs/1611.09856}{{\ttfamily 1611.09856}}].

\bibitem{Baker:2015qna}
M.~J. Baker et~al., \textit{{The Coannihilation Codex}},
  \href{https://doi.org/10.1007/JHEP12(2015)120}{\textit{JHEP} {\bfseries 12}
  (2015) 120}, [\href{https://arxiv.org/abs/1510.03434}{{\ttfamily
  1510.03434}}].

\bibitem{CMS:2021edw}
{\scshape CMS} collaboration, A.~Tumasyan et~al., \textit{{Search for
  supersymmetry in final states with two or three soft leptons and missing
  transverse momentum in proton-proton collisions at $ \sqrt{s} $ = 13 TeV}},
  \href{https://doi.org/10.1007/JHEP04(2022)091}{\textit{JHEP} {\bfseries 04}
  (2022) 091}, [\href{https://arxiv.org/abs/2111.06296}{{\ttfamily
  2111.06296}}].

\bibitem{CMS:2025ttk}
{\scshape CMS} collaboration, V.~Chekhovsky et~al., \textit{{General search for
  supersymmetric particles in scenarios with compressed mass spectra using
  proton-proton collisions at s=13{\,}{\,}TeV}},
  \href{https://doi.org/10.1103/b26z-zmpy}{\textit{Phys. Rev. D} {\bfseries
  112} (2025) 112023}, [\href{https://arxiv.org/abs/2508.13900}{{\ttfamily
  2508.13900}}].

\bibitem{Lester:1999tx}
C.~G. Lester and D.~J. Summers, \textit{{Measuring masses of semiinvisibly
  decaying particles pair produced at hadron colliders}},
  \href{https://doi.org/10.1016/S0370-2693(99)00945-4}{\textit{Phys. Lett. B}
  {\bfseries 463} (1999) 99--103},
  [\href{https://arxiv.org/abs/hep-ph/9906349}{{\ttfamily hep-ph/9906349}}].

\bibitem{Barr:2003rg}
A.~Barr, C.~Lester and P.~Stephens, \textit{{m(T2): The Truth behind the
  glamour}}, \href{https://doi.org/10.1088/0954-3899/29/10/304}{\textit{J.
  Phys. G} {\bfseries 29} (2003) 2343--2363},
  [\href{https://arxiv.org/abs/hep-ph/0304226}{{\ttfamily hep-ph/0304226}}].

\bibitem{Gripaios:2010hv}
B.~Gripaios, A.~Papaefstathiou, K.~Sakurai and B.~Webber, \textit{{Searching
  for third-generation composite leptoquarks at the LHC}},
  \href{https://doi.org/10.1007/JHEP01(2011)156}{\textit{JHEP} {\bfseries 01}
  (2011) 156}, [\href{https://arxiv.org/abs/1010.3962}{{\ttfamily 1010.3962}}].

\bibitem{Bandyopadhyay:2020jez}
P.~Bandyopadhyay, S.~Dutta and A.~Karan, \textit{{Zeros of amplitude in the
  associated production of photon and leptoquark at $e-p$ collider}},
  \href{https://doi.org/10.1140/epjc/s10052-021-09090-z}{\textit{Eur. Phys. J.
  C} {\bfseries 81} (2021) 315},
  [\href{https://arxiv.org/abs/2012.13644}{{\ttfamily 2012.13644}}].

\bibitem{Bandyopadhyay:2020klr}
P.~Bandyopadhyay, S.~Dutta and A.~Karan, \textit{{Investigating the Production
  of Leptoquarks by Means of Zeros of Amplitude at Photon Electron Collider}},
  \href{https://doi.org/10.1140/epjc/s10052-020-8083-7}{\textit{Eur. Phys. J.
  C} {\bfseries 80} (2020) 573},
  [\href{https://arxiv.org/abs/2003.11751}{{\ttfamily 2003.11751}}].

\bibitem{Bandyopadhyay:2020wfv}
P.~Bandyopadhyay, S.~Dutta, M.~Jakkapu and A.~Karan, \textit{{Distinguishing
  Leptoquarks at the LHC/FCC}},
  \href{https://doi.org/10.1016/j.nuclphysb.2021.115524}{\textit{Nucl. Phys. B}
  {\bfseries 971} (2021) 115524},
  [\href{https://arxiv.org/abs/2007.12997}{{\ttfamily 2007.12997}}].

\bibitem{Homiller:2022iax}
S.~Homiller, Q.~Lu and M.~Reece, \textit{{Complementary signals of lepton
  flavor violation at a high-energy muon collider}},
  \href{https://doi.org/10.1007/JHEP07(2022)036}{\textit{JHEP} {\bfseries 07}
  (2022) 036}, [\href{https://arxiv.org/abs/2203.08825}{{\ttfamily
  2203.08825}}].

\bibitem{Battaglia:2005zf}
M.~Battaglia, A.~Datta, A.~De~Roeck, K.~Kong and K.~T. Matchev,
  \textit{{Contrasting supersymmetry and universal extra dimensions at the clic
  multi-TeV e+ e- collider}},
  \href{https://doi.org/10.1088/1126-6708/2005/07/033}{\textit{JHEP} {\bfseries
  07} (2005) 033}, [\href{https://arxiv.org/abs/hep-ph/0502041}{{\ttfamily
  hep-ph/0502041}}].

\bibitem{Belyaev:2021ngh}
A.~Belyaev, A.~Freegard, I.~F. Ginzburg, D.~Locke and A.~Pukhov,
  \textit{{Decoding dark matter at future e+e- colliders}},
  \href{https://doi.org/10.1103/PhysRevD.106.015016}{\textit{Phys. Rev. D}
  {\bfseries 106} (2022) 015016},
  [\href{https://arxiv.org/abs/2112.15090}{{\ttfamily 2112.15090}}].

\bibitem{Liu:2022byu}
J.~Liu, Z.-L. Han, Y.~Jin and H.~Li, \textit{{Unraveling the Scotogenic model
  at muon collider}},
  \href{https://doi.org/10.1007/JHEP12(2022)057}{\textit{JHEP} {\bfseries 12}
  (2022) 057}, [\href{https://arxiv.org/abs/2207.07382}{{\ttfamily
  2207.07382}}].

\bibitem{Lester:2006cf}
C.~G. Lester, M.~A. Parker and M.~J. White, \textit{{Three body kinematic
  endpoints in SUSY models with non-universal Higgs masses}},
  \href{https://doi.org/10.1088/1126-6708/2007/10/051}{\textit{JHEP} {\bfseries
  10} (2007) 051}, [\href{https://arxiv.org/abs/hep-ph/0609298}{{\ttfamily
  hep-ph/0609298}}].

\bibitem{Agashe:2015wwa}
K.~Agashe, R.~Franceschini, D.~Kim and K.~Wardlow, \textit{{Mass Measurement
  Using Energy Spectra in Three-body Decays}},
  \href{https://doi.org/10.1007/JHEP05(2016)138}{\textit{JHEP} {\bfseries 05}
  (2016) 138}, [\href{https://arxiv.org/abs/1503.03836}{{\ttfamily
  1503.03836}}].

\bibitem{Altheimer:2013yza}
A.~Altheimer et~al., \textit{{Boosted Objects and Jet Substructure at the LHC.
  Report of BOOST2012, held at IFIC Valencia, 23rd-27th of July 2012}},
  \href{https://doi.org/10.1140/epjc/s10052-014-2792-8}{\textit{Eur. Phys. J.
  C} {\bfseries 74} (2014) 2792},
  [\href{https://arxiv.org/abs/1311.2708}{{\ttfamily 1311.2708}}].

\end{thebibliography}\endgroup

\end{document}